%% file: BrownDwarfsWithGaia.tex
\documentclass[structabstract,longauth]{aa}

\usepackage{graphicx}
\usepackage{lscape}
\usepackage{amsmath}
\usepackage{txfonts}
\usepackage{url}
\usepackage[usenames,dvipsnames,svgnames,table]{xcolor}
\usepackage{ulem}
\usepackage{tabularx}
\usepackage{longtable}
\usepackage{multirow}
\usepackage{subcaption} 

\usepackage{arydshln} 

\usepackage{array}    
\usepackage{booktabs} 

\usepackage{hyperref}

\usepackage{xcolor}
\newcommand\EDIT[1]{{\color{black}#1}}						

\def\gaia{\textit{Gaia}\xspace}

\begin{document} 


\title{Gaia predicted brown dwarf detection rates around FGK stars\\in astrometry, radial velocity, and photometric transits}

\titlerunning{Predicted BD detection rates by \gaia around FGK stars}

\author{
B.~Holl\inst{\ref{inst1},\ref{inst2}}\fnmsep\thanks{Corresponding author: B. Holl (\href{mailto:berry.holl@unige.ch}{\tt berry.holl@unige.ch})}, 
M.~Perryman\inst{\ref{inst3}}, L.~Lindegren\inst{\ref{inst4}}, D.~Segransan\inst{\ref{inst1}}, \and M. Raimbault\inst{\ref{inst1}}
} \authorrunning{Holl et al.} 
\institute{
Department of Astronomy, University of Geneva, Ch. Pegasi 51, CH-1290 Versoix, Switzerland\label{inst1}
\and 
Department of Astronomy, University of Geneva, Ch. d'Ecogia 16, CH-1290 Versoix, Switzerland\label{inst2}
\and 
School of Physics, University College Dublin, Ireland\label{inst3}
\and
Lund Observatory, Department of Astronomy and Theoretical Physics, Lund University, Box 43, SE-22100 Lund, Sweden\label{inst4}
}

\date{Received ; accepted}

\abstract
   {
   After more than two decades of relevant radial velocity surveys, the current sample of known brown dwarfs (BDs) around FGK-stars is only of the order of a hundred, limiting our understanding of their occurrence rate, properties, and formation. 
   The ongoing ESA mission \gaia has already collected more than its nominal 5~years of mission data, and is expected to operate for up to 10~years in total. Its exquisite astrometric precision allows for the detection of (unseen) companions down to the Jupiter-mass level, hence allowing for the efficient detection of large numbers of brown dwarfs. Additionally, its low-accuracy multi-epoch radial velocity measurements for $G_{\text{RVS}}<12$ can provide additional detections or constraints for the more massive BDs, while a further small sample will have detectable transits in the \gaia photometry.
   }
%
   {
Using detailed simulations, we provide an assessment of the number of brown dwarfs that could be discovered by \gaia astrometry, radial velocity, and photometric transits around main sequence (V) and subgiants (IV) FGK host stars for the nominal 5-yr and extended 10-yr mission.  }
%
   { 
   Using a robust $\Delta \chi^2$ statistic we analyse the BD companion detectability from the Besan\c con Galaxy population synthesis model for $G=10.5-17.5$\,mag, complemented by Gaia DR2 data for $G<10.5$, using the latest \gaia performance and scanning law, and literature-based BD-parameter distributions.
    }
%
   {
We report here only reliable detection numbers with $\Delta \chi^2>50$ for a 5-year mission and between square brackets for a 10-year mission. For astrometry alone we expect  28\,000--42\,000 [45\,000--55\,000] detections out to several hundred~pc [up to more than a kiloparsec]. The majority of these have $G\sim14-15$ [$14-16$] and periods above 200~d, extending up to the longest simulated periods of 5~year. 
\gaia radial velocity time series for $G_{\text{RVS}}<12$ ($G\lesssim12.7$), should allow the detection of 830--1100 [1500--1900] BDs, most having orbital periods <10~days, and being amongst the most massive BDs ($55-80M_{\text{J}}$), though several tens will extend down to the `desert' and lowest BD masses. Systems with at least three photometric transits having S/N>3 are expected for 720--1100 [1400--2300] BDs, averaging at 4--5 [5--6] transits per source.
The combination of astrometric and radial velocity detections should yield some 370--410 [870--950] detections. Perhaps 17--27 [35--56] BDs have both transit and radial velocity detections, while both transits and astrometric detection will lead to a meagre 1--3 [4--6] detection(s).

}
%
   {
Though above numbers have $\pm 50\%$ uncertainty due to the uncertain occurrence rate and period distribution of BDs around FGK host stars, detections of BDs with \gaia will number in the tens-of-thousands, enlarging the current BD sample by at least two orders of magnitude and thus allowing to investigate the BD fraction and orbital architectures as a function of host stellar parameters in greater detail than every before.
}

\keywords{
-- astrometry 
 -- Techniques: radial velocities 
 -- Techniques: photometric 
-- planets and satellites: general
-- space vehicles: instruments
      }
\maketitle

%
\section{Introduction\label{sec:introduction}}
\input{Sec_Introduction}


\section{Detection criteria \label{sec:gaiaSensitivity}}
\input{Sec_GaiaSensitive}

\section{Method\label{sec:method}}
\input{Sec_Method}

\subsection{Host star distribution as observed by \gaia \label{sec:hostStarDistribution}}
\input{Sec_HostStarDistribution}



\section{Results\label{sec:results}}
\input{Sec_Results}

\section{Discussion\label{sec:discussion}}

\input{Sec_Discussion}
\section{Conclusions
\label{sec:conclusions}}
\input{Sec_Conclusions}

\input{Sec_Acknowledgements}

\bibliographystyle{aa}
\raggedbottom
\bibliography{local}

\clearpage

\begin{appendix}
\input{Sec_GaiaDetectionLimits}
\input{Sec_Appendices}

\clearpage

\end{appendix}

\end{document}

%% file: Sec_Introduction.tex



Brown dwarf companions with orbital period $\lesssim10$\,yr are within the range of sensitivity and time-span of various long-running radial velocity surveys 
\citep[e.g.][]{2016ASSL..428....3H,2018ASSP...49..199D,2018ASSP...49..181F,2018ASSP...49..239O}. But despite the many thousands of stars surveyed to date, less than a hundred brown dwarf companions around solar-type stars are currently known \cite[e.g.][]{2014MNRAS.439.2781M,Grieves:2017aa}.
This has reinforced the evidence for a `brown dwarf desert', i.e.\ the general absence of companions in the mass range $10-80M_\text{J}$, being most pronounced between $30-55M_\text{J}$ \citep{Marcy:2000ab, Grether:2006aa}. 

The low-mass BD distribution seems to be the high-mass tail of the planetary distribution function \citep[e.g.][]{Sahlmann:2011yq}, while the high mass BD distribution seems to be the low-mass tail of the stellar binary distribution function \citep[][Sect.~2.7.1]{1991A&A...248..485D,1992A&A...256..121T,2018exha.book.....P}. This points to formation scenarios that populate the adopted BD mass range having diminishing efficiency toward the BD desert 
\citep{Grether:2006aa,2010A&A...509A.103S,Sahlmann:2011yq,2014MNRAS.439.2781M,2014prpl.conf..619C,2016A&A...585A..46B,2016A&A...587A..64S,2016AJ....151...85T,2016A&A...588A.144W,2017A&A...599A..57B,Grieves:2017aa}. 
 \cite{Sahlmann:2011yq} provide an upper limit on the occurrence rate of tight-orbit ($P\lesssim300$\,d) BD companions around solar-type stars of 0.3--0.6\%, illustrating that to significantly increase the statistics of BDs, one needs to survey millions of host stars.

So far, the ongoing \gaia ESA mission \citep{2016A&A...595A...1G} has provided the scientific community with three data releases \citep{2016A&A...595A...2G,2018A&A...616A...1G, 2020arXiv201201533G} containing information on over a billion sources on the sky based on data spanning 13, 22, and 33 months of data, respectively. 
For all sources \gaia measures (near) simultaneously the astrometric position, three-band (spectro)photometry, and radial velocities \cite[for $G_{\text{RVS}}\leq16$~mag][]{Sartoretti2018aa}\footnote{Note that epoch radial velocities will only be published for sources with $G_{\text{RVS}}\leq12-13$~mag, for fainter sources only mean radial velocities will be published.}. For the nominal five-year mission this averages to about about 70 epochs, increasing linearly with time during its extended mission. 

These three measurements in principle provide three distinct discovery methods for any orbiting companions to main-sequence stars, with the astrometric and radial velocity measurements probing the dynamical reflex motion of the host star, and the photometric measurements yielding transit information. We will consider detection yields from these approaches separately, and also in combination.

The estimation of orbital parameters of astrometric BDs by \gaia is roughly limited to companions with orbital periods less than the mission length, thus a possible extension providing up to ten years of \gaia data would be greatly beneficial.


No orbital solutions have yet been published by the \gaia consortium; the first of these are expected to be part of the upcoming full DR3 release in 2022 based on 33~months of data (not to be confused with the \textit{early} Data Release 3 of Dec 2020). In that release some radial velocity orbital solutions are also expected to be included, though no radial velocity nor astrometric time series will be provided; these will become available with DR4. Although photometric time series have been released for a 550\,000 sample of variable stars in DR2 \citep{2018A&A...618A..30H, 2018A&A...616A...4E, 2018A&A...616A...3R}, none of these were targeted to photometric transits of unseen companions.

Release of these orbital and time series data will have a major impact on the brown dwarf field, as it will provide a nearly magnitude-complete survey of host stars to at least $G=18$ \citep[DR2,][]{2018A&A...616A..17A}, although extending to a limiting magnitude of $G<20.7$\,mag. As we will show, and largely as a result of \gaia's unprecedented astrometric precision, BDs should be found in large numbers.

Only a few previous publications have made any estimate of the number of brown dwarfs discoverable by \gaia. \cite{Bouchy:2014aa} estimated a number of 20\,000 BD companions for $G_{\text{RVS}}\leq12$ having both \gaia radial velocity and astrometric measurements.
{\cite{2014MmSAI..85..643S} estimated at least several thousands BD companions to $G=16$\,mag extrapolating from known radial velocity samples.
\citet[][Sect.~2.10.5]{2018exha.book.....P} considered that `\ldots many tens of thousands of [brown dwarf] desert occupants might be expected'. The prediction of several tens of BDs mentioned in \cite{2019ApJ...886...68A} was mainly related to their goal of tightly constraining the mass function, which requires more signal than that for a detection or even an orbital fit. They also examined predictions of unseen companions using simulated \gaia radial velocities.  
 
 In this paper we will not address the issue of detecting BDs around binary stars \citep[see e.g.][]{2015MNRAS.447..287S}, nor \textit{direct} detection of \EDIT{isolated brown dwarfs by \gaia as theoretically discussed in \cite{2013A&A...550A..44S} and \cite{2014arXiv1404.3896D} (the latter predicting `several thousand' with measured parallax) and DR2 data estimates in e.g. \cite{2018ApJ...868...44F} and \cite{2018A&A...619L...8R}}.

This paper provides up-to-date estimates of the number of brown dwarf companions around FGK host stars expected to be detected in the \gaia astrometric, radial velocity, and photometric transit data. These numbers are derived for the nominal \mbox{5-year} mission (approximately corresponding to DR4), and for an extended 10-yr mission. 
We limit our analysis to FGK main sequence and sub-giant host stars because only for this restricted population are meaningful occurrence rates, required for our analysis, currently known.

We base our study on the latest \gaia scanning law and instrument models. We derive brown dwarf population distributions, a necessary input to our simulations, from the literature. Below G<10.5 we use Gaia DR2 and above G>10.5 the Besan\c con population synthesis model as a proxy of the \gaia observed FGK host star distribution, although for the lower and higher mass ends of the stellar mass function \gaia will eventually \textit{define} the observational distribution.

In this paper we will use the term `brown dwarf' to specify a substellar companion in the range $10-80M_{\text{J}}$, despite its overlap with the giant-\text{planet} regime, and without any assumptions on composition or consideration of deuterium burning \citep[e.g.][]{2014prpl.conf..619C}.  We assume that the brown dwarf is sufficiently faint that its light does not influence the astrometric measurement of its FGK-type host star, nor is it considered to be visible in the spectral lines (i.e.\ not a spectroscopic binary). 

Sect.~\ref{sec:gaiaSensitivity} provides an overview of the adopted \gaia detection criteria for astrometric, radial velocity, and photometric transits.  In Sect.~\ref{sec:method} we outline our method of simulating the expected number of brown dwarfs observed by \gaia. The predicted BD numbers and distributions are presented in Sect.~\ref{sec:results}, with discussion of the results in Sect.~\ref{sec:discussion}. Sect.~\ref{sec:conclusions} summarises our conclusions. 

In Appendix~\ref{sec:gaiaDetectionLimits} we illustrate the \gaia astrometric and radial velocity detection limits.
Appendices~\ref{sec:gaiaError}, \ref{sec:gaiaErrorRv}, and \ref{sec:gaiaErrorPhot} contain details of the derived error models. Details of how the data samples were prepared from both the Besan\c con and \gaia DR2 data are given in Appendices~\ref{appendix:besancon} and~\ref{appendix:gaiaGt10p5}.

%% file: Sec_GaiaSensitive.tex
The most fundamental factor in estimating the expected number of brown dwarfs by \gaia is to quantify what we call a `detection'. We define this for astrometry in Sect.~\ref{sec:gaiaDetectMethod}, for radial velocity in Sect.~\ref{sec:gaiaDetectMethodRv}, and for photometric transits in Sect.~\ref{sec:gaiaDetectMethodTransit}.
We do not attempt to quantify false-detection rates for the various thresholds, as this will be strongly dependent on the noise properties and remaining `features' in the final data.
 

\subsection{Astrometric detection criteria \label{sec:gaiaDetectMethod}} 
The high accuracy astrometric measurements of \gaia \citep{2001A&A...369..339P,Lindegren:2012aa, Lindegren:2016qy, Lindegren:2018aa, 2020arXiv201203380L} are key to the astrometric detection of brown dwarf companions, and indeed unseen companions in general. 
For a circular orbit and small mass ratio \cite[][Eq.~3.2]{2018exha.book.....P}
\begin{equation}
\label{eq:astrometricSignal}
\alpha =
\left( \frac{M_\text{BD}}{M_{*}} \right)
\left( \frac{a}{1~\text{au}} \right)
\left( \frac{1~\text{pc}}{d} \right) 
\text{arcsec} \ .
\end{equation}
This astrometric signal increases linearly with semi-major axis $a$ [au] of the companion, such that wider orbits, and longer periods, yield a larger signal. However, the signal decreases linearly with increasing distance, so that detections will be dominated by relatively nearby stars.

The astrometric detection method used here is adopted from \cite{2014ApJ...797...14P}, where it was used to assess the number of planets detectable with \gaia. It is based on the  astrometric $\Delta \chi^2 = \chi^2_{\mathrm{min}}\hbox{(5 parameter)} - \chi^2_{\mathrm{min}}\hbox{(12 parameter)}$, i.e., the difference in $\chi^2$ between a single-star astrometric model fit to the observations (having 5 parameters) and a Keplerian orbital fit to the observations (having 12 parameters). As $\Delta \chi^2$ is sensitive to the number and distribution of observations, the eccentricity and inclination of the system, and orbital period in relation to the total length of the observations, it can serve as a precise estimate of detectability once the relevant thresholds have been calibrated.

Following the findings resulting from the simulations in \cite{2014ApJ...797...14P}, we adopt their three thresholds for $\Delta \chi^2$:
\begin{enumerate}
\item $\Delta \chi^2 \simeq 30$ is considered a marginal detection, 
\item $\Delta \chi^2 > 50$ is generally a reliable (`solid') detection,
\item $\Delta \chi^2 > 100$ generally results in orbital parameters determined to 10\% or better. 
\end{enumerate}
This procedure can be applied to both real and simulated data. 
However, the drawback of this $\Delta \chi^2$ detection metric is that it is computationally expensive to find the best 12-parameter Keplerian solution for millions of astrometric time series. For simulation work, Sect.~5 of \cite{2014ApJ...797...14P} provides however an elegant shortcut through the introduction of the \textit{noncentrality} parameter $\lambda$, which is simply the 5-parameter $\chi^2_{\mathrm{min}}\hbox{(5 parameter)}$ fit to the \textit{noiseless} data, and thus is applicable to simulations only. 

For the astrometric $\Delta \chi^2$ the difference in number of parameters between the 5- and 12-parameter model is~7;  therefore $\lambda+7$ approximates the expectation value of $\Delta \chi^2$ \citep[][Appendix~B]{2014ApJ...797...14P}. Note that this estimate does not include any statistical variance that would be induced by observation noise. This can easily be re-introduced by randomly drawing from the expected variance distribution around the predicted $\Delta \chi^2$. 

For the adopted thresholds in this paper these distributions are largely symmetric, and thus any biases when omitting this `noisification' of our large number of (and randomly initialised) samples is expected to be negligible. In this paper, as in \cite{2014ApJ...797...14P}, this noisification step was omitted, and for each star in our simulation we therefore only need to generate a noiseless astrometric signal representing the true orbit, compute the $\chi^2$ of a 5-parameter fit, and add~7 to determine the (noiseless) $\Delta \chi^2$ astrometric detection statistic\footnote{Though some computational cost is involved in simulating the orbital signal of a Keplerian system and making a 5-parameter solution to it, it is \textit{much cheaper} than taking the average of a large number of Keplerian 12-parameter solutions to Monte Carlo noise realisations of the same signal, while providing basically the same information.}.

It is worth pointing out the study of \cite{2018A&A...614A..30R} which shows that though the adopted $\Delta \chi^2$ thresholds for 
reliable detection are still valid, the false-detection rate is higher than expected from the
naively expected $\chi^2$ with 7~degrees of freedom, and actually is closer to a 11--16 degrees-of-freedom 
distribution. 

The adopted per-FoV astrometric precision as function of magnitude is derived in  Appendix~\ref{sec:gaiaError}.

\subsubsection*{Validity of the $\Delta \chi^2$ thresholds}
The quality indicators associated with $\Delta \chi^2 \simeq30$, 50, and 100 are only valid when the period $P\lesssim$ simulated mission length $T$, i.e., for (somewhat) longer periods the orbital parameters will no longer be recovered, even when $\Delta \chi^2 > 100$. This is because for $P<T$ the orbital signal has typically a low correlation with any of the 5~astrometric parameters because of its oscillating (periodic) nature. Consequently, it does not get absorbed in (and biasses) any of the positions, parallax\footnote{An exception to this are systems with an orbital period very close to the 1 year parallactic motion, see \cite{2000fepc.conf..479L,Butkevich:2018aa,   2010IAUS..265..416S} for more detailed discussions.}, or proper motions efficiently.  Instead, the orbital signal remains largely unmodelled, and ends up as an increased variance in the residuals. It increases the $\chi^2$, allowing us to use it as a measure of the difference (and hence detectability) between the 5-parameter single star solution, and the 12-parameter orbital solution. 

In the regime $P>T$, the longer the period, the smaller the section of the orbit that is probed. For longer periods, this curved orbital segment more closely resembles a contribution to the proper motion, resulting in biassed estimates of the proper motion. This then reduces the residual of the 5-parameter solution, and hence the $\chi^2$. Orbital fitting tests show that the period recovery starts to drop significantly beyond $P\sim T$, hence the $\Delta \chi^2$ thresholds no longer correspond to a well-defined quality of the parameter fits. Results from planetary mass companion fitting for a 5-year mission shows that periods are not properly recovered beyond 6--6.5~years \citep[][Fig.~6]{Casertano:2008aa}, though for BDs or heavier unseen companions  this may be extended to some 10~years or more. In this paper we will generally not consider periods beyond the simulated mission lengths of 5 and 10~years, as this would require actual orbital fitting. As will be discussed in Sect.~\ref{sec:bdPriors}, the BD period distributions used in this study are limited to $P<5$~yr (due to poor literature constraints for longer periods), which justifies the use of our detection method to give meaningful statistics for the whole simulated sample.
For the recovery of an acceleration model instead of an orbital model, the  $\Delta \chi^2 > 100$ might still be considered a good criterion for $P>T$.

In summary, for $P\lesssim T$, higher values of $\Delta \chi^2$ give both more reliable detections and higher precision of the estimated orbit. 
In this paper all noiseless $\Delta \chi^2$ values are approximated by the use of the non-centrality parameter $\lambda$  via the relation $\Delta \chi^2=\lambda+7$. 


\subsection{Radial velocity detection criterion \label{sec:gaiaDetectMethodRv}} 
\EDIT{The maximum observable radial velocity signal at inclination $i$ is characterised by the radial velocity semi-amplitude $K$ 
 \cite[e.g.][Eq.~2.28]{2018exha.book.....P}:
\begin{eqnarray}
\label{eq:rvSignal}
      \left( \frac{K}{\text{m\,s}^{-1}} \right) &=& 203.3 \,  \sin i \ (1-e^2)^{-1/2}    \nonumber \\
    &&  \left( \frac{P}{\text{days}} \right)^{-1/3} \left( \frac{M_\text{BD}  }{M_\text{J}} \right) \left( \frac{M_* + M_\text{BD}}{M_\odot} \right)^{-2/3}
\end{eqnarray}
}
where $K$ scales as $a^{-1/2}$, i.e., decreasing for wider orbits. The radial velocity signal has no direct dependence on distance, apart from the radial velocity precision which will depend on the apparent magnitude of the star. We note that inclination cannot be deduced from one-dimensional radial velocity data alone, while it can be derived from (2-dimensional sampled) astrometric data.

The Radial Velocity Spectrometer (RVS) instrument on-board \gaia \citep{2018A&A...616A...5C, 2019A&A...622A.205K} records a spectrum of every transiting object with $G\lessapprox17$ when the source transits any of rows 4--7 in the field of view (FoV). As there are 7~rows of CCDs in the focal plane, this means that on average 4 out of 7 field-of-view transits will have RVS measurements (Fig.~\ref{fig:FovNumObs}). For the majority of sources the accumulated RVS spectra will be stacked to boost the signal-to-noise and derive an averaged radial velocity. However, for $G_{\text{RVS}}\lessapprox12$ the signal-to-noise ratio will allow the derivation of meaningful radial velocities for each transit, and hence will allow detailed transit-by-transit modelling and the search for velocity variations due to a companion.
The per-FoV radial velocity (RV) precision is of the order of 1\,km\,s$^{-1}$ for the bright stars (Appendix~\ref{sec:gaiaErrorRv}). Despite this being only comparable to the signal amplitude of a $90^\circ$ inclined BD with $80 \mathrm{M}_\text{Jup}$ mass orbiting at a Jupiter distance from its host star,
with an average of $\sim40$ transits over a 5-year mission it will nonetheless provide orbital constraints on the more massive BDs accompanying bright host stars (see also the detection thresholds in Appendix~\ref{sec:gaiaDetectionLimits}).

To explore the detectability of BD using \textit{only} \gaia radial velocities, we derive the approximate per-transit precision in Appendix~\ref{sec:gaiaErrorRv}, 
and compute a $\Delta \chi^2$ criteria for the same levels (30, 50, and 100) as adopted for the astrometric detections. 
For the radial velocity time series of stars with $G_{\text{RVS}}<12$ we compute $\Delta \chi^2 = \chi^2_{\mathrm{min}}\hbox{(1 param)} - \chi^2_{\mathrm{min}}\hbox{(6 parameter)}$, i.e., the difference in $\chi^2$ between a constant radial velocity model fit to the observations (having 1 parameter) and a radial velocity orbital fit to the observations (having 6 parameters). In radial velocity fitting, of the 7 Keplerian parameters the $\Omega$ cannot be determined, and the two components of $a_{*}\sin i$ cannot be separated, hence effectively 5 parameters are left, plus one for the system's barycentric radial velocity. The difference in number of parameters is thus 5 (6--1) and we therefore compute the (noiseless) $\Delta \chi^2=\lambda+5$ for the radial velocity non-centrality parameter $\lambda$ for the threshold levels 25, 45, and 95 to represent the 30, 50, and 100 $\Delta \chi^2$ criteria. The non-centrality parameter in this case is simply the weighted mean of the noiseless observations that include the radial velocity signal of the orbital motion.


\subsection{Photometric transit detection criterion
\label{sec:gaiaDetectMethodTransit}} 
To decide whether a system is detectable based on photometric transits we require $\geq3$ transits with a signal-to-noise ratio $\geq3$
\begin{equation}\label{eq:transitSn}
  S/N = 1.086\frac{ \left(R_\text{BD}/R_{*}\right)^2}{\sigma(G)}	
\end{equation}
where $\left(R_\text{BD}/R_{*}\right)^2$ is the transit depth, and $\sigma(G)$ the photometric uncertainty as function of $G$-band magnitude (derived in Appendix~\ref{sec:gaiaErrorPhot}). 
The  requirement of a minimum of three transits allows to estimate the periodicity to some degree (which is required for efficient follow-up observations) and is also a reasonable requirement to reduce false detections due to noise or potential outliers in the real \gaia data.
It turns out that the majority of transiting systems have only 1 or 2 transits, thus significantly reducing the numbers of reported detectable transits in this study (for more  details see Table~\ref{tab:fracTransiting} and associated discussion).
Note that \gaia has three (almost) simultaneously taken photometric band measurements: the broad band $G$ magnitude 
and the two narrower photometric bands $G_\text{BP}$ and $G_\text{RP}$. All bands can potentially detect photometric eclipses, 
although here we will only consider the most precise $G$-band observations. 

Our simulations do not require us to {\it search\/} for potential transits. Instead, we rigorously compute for each observation whether it is transiting, or partially transiting. We have

\begin{eqnarray}\label{eq:transitEq}
      \Delta^\text{*}_{\alpha_*} (t) \text{ [mas]} &=& B  X(t) + G  Y(t) \nonumber \\
      \Delta^\text{*}_{\delta} (t) \text{ [mas]} &=& A  X(t) + F  Y(t)  \nonumber \\
          \Delta^\text{*} (t) \text{ [mas]} &=& \sqrt{ \left( \Delta^\text{*}_{\alpha_*} (t) \right)^2 + \left(\Delta^\text{*}_{\delta} (t) \right)^2 } \nonumber \\
	d(t) \ \text{[au]} &=& (1 + 1/q) \ \Delta^\text{*} (t) \ /  \ \varpi \nonumber \\
	\text{in transit} &=&\begin{cases}
   \text{full transit:} & d(t)< R_\odot - R_\mathrm{BD} \\
    \text{partial transit:} &d(t)< R_\odot + R_\mathrm{BD} 
     \end{cases} \nonumber \\
     \theta_T &=& \text{atan2}(Y,X) \nonumber \\
     z_\text{sign}     & =& \text{sign}(\  \sin( \theta_T + \omega  ) \sin( i) \ ) \nonumber \\
     	\text{transit} &=&\begin{cases}
   \text{primary:} & z_\text{sign} = 1 \\
    \text{secondary:} & z_\text{sign}= -1
     \end{cases}
\end{eqnarray}
where $A,B,F,G$ are the Thiele--Innes orbital parameters in [mas], and $X(t)$ and $Y(t)$ the solution to the Kepler equation at observation time $t$.
The $\Delta^\text{*}_{\alpha_*} (t)$ and $\Delta^\text{*}_{\delta} (t)$ are the time-dependent sky-projected barycentric offsets of the host star in $\alpha$ and $\delta$, respectively.
The $\Delta^\text{*}(t)$ is the total sky-projected barycentric offset of the host star in mas, which can be converted into the sky-projected distance
$d(t)$  in au between the host star and the BD companion using the mass ratio $q$ and parallax $\varpi$ in mas. To yield a detectable transit signal, we will only consider the primary transits (where the BD passes in front of the host star, i.e., the host is `behind' the system barycentre as seen from \gaia) corresponding to $z_\text{sign}=1$ in Eq.~\ref{eq:transitEq}, which is computed from the true anomaly $\theta_T$, longitude of periastron $\omega$, and inclination~$i$.

The transit counts provided in this paper include all partial and full primary transits, and assuming $R_\mathrm{BD}=R_\text{J}$ for all simulated BD masses. For completeness we note that  Eq.~\ref{eq:transitSn} overestimates the signal-to-noise for partial transits, as well as for transits close to the outer rim of the stellar disk, as it does not take into account any limb-darkening effects. We ignore these second-order effects as they only affect a small fraction of the data, and will not drastically alter our results.
The $R_\odot$ for each host star is given for the Besan\c con model simulation data ($G>10.5$, see Appendix~\ref{appendix:besancon}) and estimated from the \gaia DR2 data ($G<10.5$, see Appendix~\ref{appendix:gaiaGt10p5}).

%% file: Sec_Method.tex
The number of \gaia detectable BDs can be estimated as

\begin{eqnarray}\label{eq:LL1}
N_\text{det}&=&N_* \int \text{d}\vec{\theta}_* \int \text{d}\vec{\theta}_\text{BD}\,  \nonumber \\
&& f_\text{obs}(\vec{\theta}_*) f_\text{BD}(\vec{\theta}_*) 
f_\text{det}(\vec{\theta}_\text{BD},\vec{\theta}_*) 
p_*(\vec{\theta}_*) p_\text{BD}(\vec{\theta}_\text{BD}|\vec{\theta}_*)
\end{eqnarray}
with the following dependencies:
\begin{itemize}
\item  $N_{\mathrm{*}}$ is the number of F, G, and K (main sequence and subgiant) host stars in our Galaxy;
\item $f_\text{obs}(\vec{\theta}_*)$ is the fraction of observable host stars with stellar parameters $\vec{\theta}_*$;
\item  $f_{\mathrm{BD}}(\vec{\theta}_{\mathrm{*}})$ is the fraction of BDs as function of host star stellar parameters $\vec{\theta}_{\mathrm{*}}$, see Sect.~\ref{sec:bdFractHostStars};
\item $f_{\mathrm{det}}(\vec{\theta}_\text{BD},\vec{\theta}_*)$ is the fraction of detectable BDs by \gaia for a host star with parameters $\vec{\theta}_{\mathrm{*}}$ and BD  parameters $\vec{\theta}_{\mathrm{BD}}$, we will assume in this paper that all systems with $\Delta \chi^2>30$, 50, and 100 are \textit{detected} to the degree as defined in Sect.~\ref{sec:gaiaSensitivity}. An illustration of the astrometric and radial velocity detection limits is provided in Appendix~\ref{sec:gaiaDetectionLimits};
\item $p_*(\vec{\theta}_*)$ is the (true) prior normalised probability density function of host stars;
\item  $p_\text{BD}(\vec{\theta}_\text{BD}|\vec{\theta}_*)$ is the (true) prior normalised probability density function of BDs given certain host star stellar parameters $\vec{\theta}_{\mathrm{*}}$, see Sect.~\ref{sec:bdFractHostStars}.
\end{itemize}
Note that the  $f$ fractions 
are not normalised, and correspond to a number typically much smaller than one when integrated over. 

We use the Besan\c con model and \gaia DR2 data (see Sect.~\ref{sec:hostStarDistribution}) to predict the number of stars observed by \gaia per stellar parameter (of which stellar type, $G$-band magnitude, and sky position are the most relevant)
\begin{equation}\label{eq:LL2}
N_\text{obs}(\vec{\theta}_*)=N_* f_\text{obs}(\vec{\theta}_*) p_*(\vec{\theta}_*)		\ ,
\end{equation}
reducing Eq.~\ref{eq:LL1} into:
\begin{equation}\label{eq:LL3}
N_\text{det}=\int \text{d}\vec{\theta}_* \int \text{d}\vec{\theta}_\text{BD}\, \
N_\text{obs}(\vec{\theta}_*) f_\text{BD}(\vec{\theta}_*)
f_\text{det}(\vec{\theta}_\text{BD},\vec{\theta}_*) 
p_\text{BD}(\vec{\theta}_\text{BD}|\vec{\theta}_*)
\end{equation}
The following sections define the various components in detail, and summarised as follows:
\begin{itemize}
\item Sect.~\ref{sec:bdFractHostStars} introduces the assumption of a constant BD fraction, i.e., $f_\text{BD}(\vec{\theta}_*)=f_\text{BD}$ (= 0.6\%);
\item Sect.~\ref{sec:bdPriors} justifies our assumption that the distribution of BD parameters can be taken to be independent of the stellar parameters, i.e.\ 
$p_\text{BD}(\vec{\theta}_\text{BD}|\vec{\theta}_*)=p_\text{BD}(\vec{\theta}_\text{BD})$, and then goes on to specify that distribution; 
\item finally Sect.~\ref{sec:hostStarDistribution} gives $N_\text{obs}(\vec{\theta}_*)$ from the Besan\c con model and \gaia DR2 data.

\end{itemize}

\subsection{BD fraction as a function of host stellar parameters\label{sec:bdFractHostStars}} 
Estimates of the numbers of BD detectable with \gaia rest on the presently known occurrence rates as function of host star and companion properties.
For the CORALIE planet search sample containing 1600 Sun-like stars within 50 pc, \cite{Sahlmann:2011yq} derived a range of 0.3--0.6\% for the frequency of close-in brown-dwarf ($P\lesssim300$\,d), which is consistent with the $<0.5\%$ upper limit of \cite{Marcy:2000ab}, and the recent 0.56\% for $P\lesssim300$\,d of \cite{Grieves:2017aa}. \cite{Grether:2006aa} find a similar occurrence rate of  $<1\%$ when considering BD companions around solar-type stars with periods $P\lesssim5$\,yr.

For AF-type main sequence stars, \citet[][Table 4]{Borgniet:2017aa} found that for the lower mass sample ($M_{*} \leq 1.5 M_{\odot}$, i.e., F-type stars) the occurrence rate of BDs with $1<P<1000$~d is $\leq 2\%$ (similar for both $1<P<100$~d and $1<P<10$\,d), though with a $1\sigma$ confidence interval between 0--7\%, hence consistent with, but not more constraining than, the previously mentioned occurrence rates. At the lower end, \cite{2016A&A...587A..64S} found a rate of $0.29\pm0.17\%$ for BD having a Kepler transit with $P<400$\,d \cite{2010Sci...327..977B}, while at the high end \cite{2019A&A...631A.125K} used SOPHIE to study FGK stars within 60~pc in the northern hemisphere, finding
12 new BD, and concluding a BD rate of $2.0 \pm 0.5 \%$ for $P<10\,000$\,d. The latter is likely overestimated due to its derivation from RV-data.

\cite{Sahlmann:2011yq} conclude that the CORALIE sample was too small to make any inferences on a dependence on the metallicity distribution of the host stars, as found for planet host stars \citep[e.g.][]{Santos:2001aa}.

We therefore adopt a constant BD fraction for this study, i.e., $f_\text{BD}(\vec{\theta}_*)=f_\text{BD} = 
0.6\pm0.3 \%$, where the lower value is based on the lower estimate of $0.3\%$ by \cite{Sahlmann:2011yq} for $P\lesssim300$\,d and consistent with \cite{2016A&A...587A..64S}, and the upper value of $0.9\%$ by the upper limit of $1\%$ from \cite{Grether:2006aa} for $P\lesssim5$\,yr. Though this 50\% uncertainty will dominate the accuracy of our estimate of the number of expected BD discoverable, it is the most realistic value to use given the current data available. We will limit the longest simulated periods to 5\,yr, as discussed in the following section.

While we have focused our predictions on FGK stars for which estimates of occurrence frequencies are available, we stress that the Gaia results will define these distributions over the entire range of spectral types and luminosity class.

\subsection{BD prior orbital parameters for simulations \label{sec:bdPriors}} 

\paragraph{Mass distribution}
As discussed in the introduction, there is clear evidence for a general absence of sub-stellar companions in the BD mass range (the so-called BD `desert'), being most pronounced between $ 30-55M_{\text{J}}$. Examining the cumulative mass distribution function from Fig.~18 of \cite{Sahlmann:2011yq} and Fig.~3 of \cite{2014MNRAS.439.2781M}, it appears that about half of the BD candidates reside on one side of this desert. Given their paucity, the relative occurrence rate as a function of mass is currently rather ill-constrained.
Adopting the power-law fits of \cite{Grether:2006aa}, as shown in their Figs~8 or 9, would result in zero BDs between 13--55 or 20--57 $M_{\text{J}}$, respectively, which seems too restrictive as a generative function for this study in which we focus on the $10-80M_{\text{J}}$ range.

\begin{figure}[h]
  \includegraphics[width=0.5\textwidth]{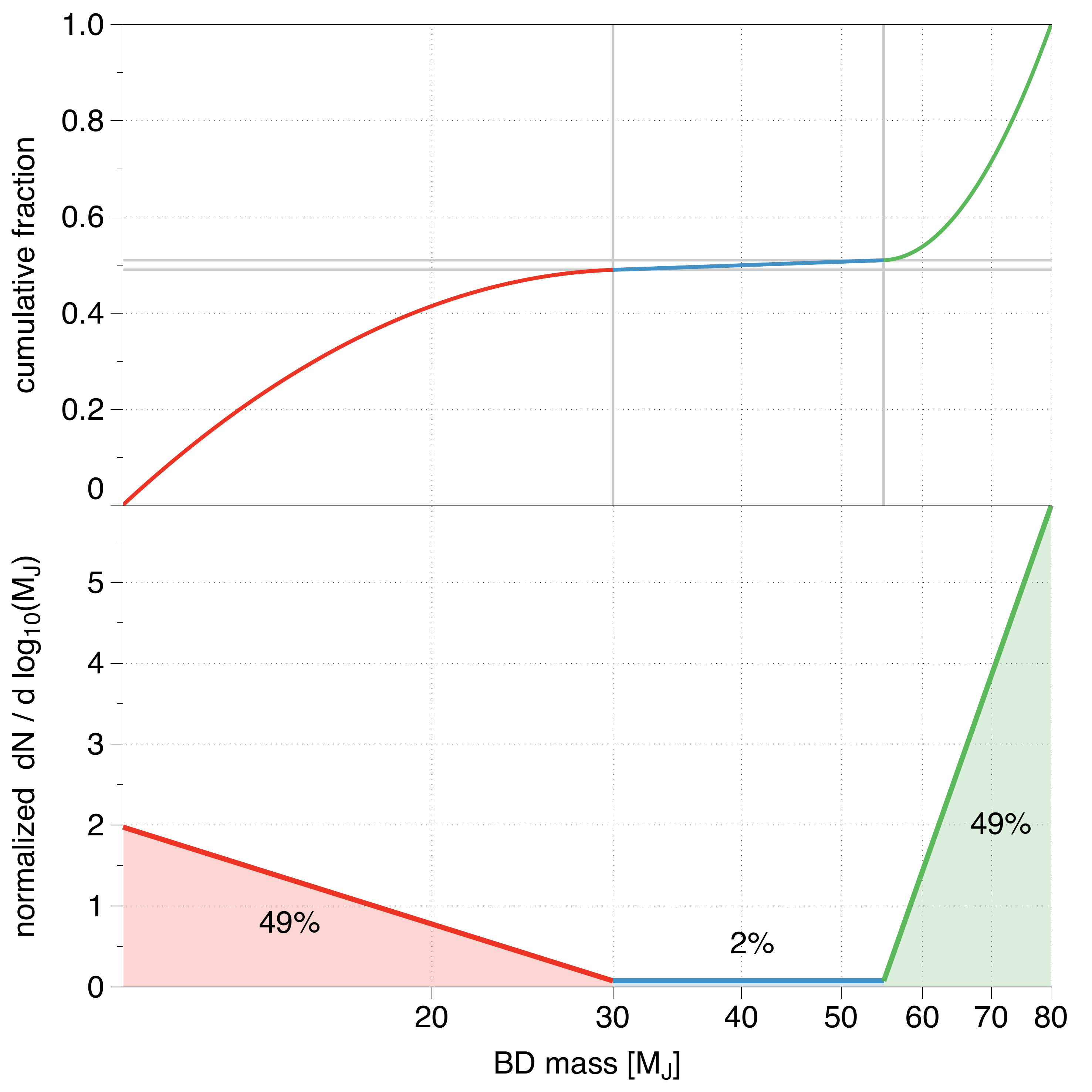} 
 \caption{Adopted BD mass distribution in this study, see text for details.
}
\label{fig:hollMassFunction}
\end{figure}

Based on the above, we therefore create a customised mass distribution function suited for this study:
\begin{itemize}
\item all companions have masses in the range $10-80M_{\text{J}}$,
\item the BD `desert' occurs at half the cumulative mass distribution \citep[conforming to Fig.~18 of][]{Sahlmann:2011yq},
\item we assume a linear relation in $\log_{10}(M_{\text{BD}})$ \citep[as in][]{Grether:2006aa} on both sides of the desert: one with a negative slope between $10-30M_\text{J}$ and one with a positive slope between $55-80M_{\text{J}}$.
\item we add a constant level in $\log_{10}(M_{\text{BD}})$ of $2\%$ of the BD companions in the $30-55M_{\text{J}}$ range, which is a rate that could easily lead to non-detection in the previous studies given their sample sizes.
\end{itemize}
The result is shown in Fig.~\ref{fig:hollMassFunction}. The cumulative density function is given in quadratic form $c + bx + ax^2$, with $x=\log_{10}(M_\text{J})$:
\begin{equation}
    {\rm CDF}_M(x)=\begin{cases}
    -3.974 + 5.969x -1.995x^2, & \text{$10\leq M_\text{J} <30$} \\
   0.3778 + 0.0760x, & \text{$30\leq M_\text{J} <55$}\\
    55.052 -62.753x + 18.05x^2, & \text{$55\leq M_\text{J} \leq 80$}
    \label{eq:massDCF}
  \end{cases}
\end{equation}
The normalized histogram values shown in Fig.~\ref{fig:hollMassFunction} are simply the first derivative of CDF$_M(x)$.
In the simulations the BD mass $(M_{\text{BD}})$ was generated as a random number using the inverse CDF (Appendix~\ref{sec:cdfMass}).

\paragraph{Period distribution}
Though radial velocity and transit observations are biassed towards the detection of short-period objects, observations show that the number of BDs increases with orbital period, which is consistent with direct imaging surveys \citep{Diaz:2012aa, 2014MNRAS.439.2781M}.  
Because \gaia 's astrometric sensitivity increases with increasing period up to the mission duration (Appendix~\ref{sec:gaiaResponseComp}), it is important to have a good representation of the period distribution up to at least the nominal 5~yr mission length. Most literature data have poor statistics beyond several years orbital period (and often limited to only 300--500\,d), so we have limited our adopted period distributions to 5~years.

\begin{figure}[h]
  \includegraphics[width=0.5\textwidth]{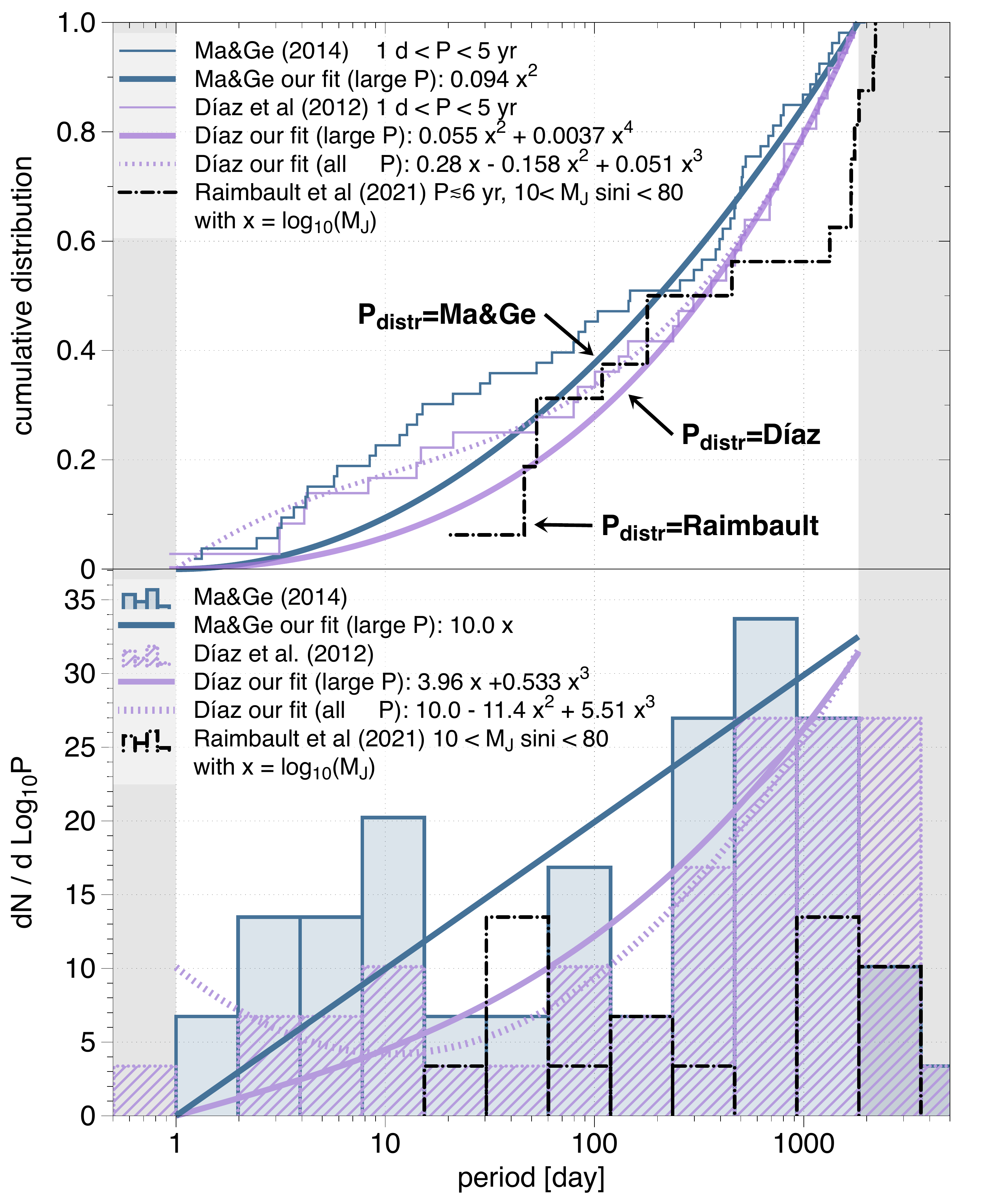} 
\caption{Period distribution of the BD sample from \cite{2014MNRAS.439.2781M} and \cite{Diaz:2012aa} having masses between 9 and 90 $M_\text{J}$ orbiting solar-type stars. 
Top: the cumulative distribution function for  BDs, the `large P' fits of Ma\,\&\,Ge and D\' iaz are used to generate the BD period distributions and will be referred to as $\textrm{P}_\textrm{distr}$=Ma\,\&\,Ge and $\textrm{P}_\textrm{distr}$=D\' iaz, respectively. The bottom figure shows the log-binned histogram with the corresponding fit-curves (bin size $\log_{10}(1.98)$). Our simulations are limited to the indicated (non-shaded) period range  1\,d\,$<P<$\,5\,yr. The \cite{raimbault19} data is used for an additional bootstrap sample estimate (Sect.~\ref{sec:raimbaultBootstrap}).
}
\label{fig:cdfHistPeriodFit}
\end{figure}

To satisfy our need to sample beyond more than 1--2~years, we derived the period distribution from the samples of \cite{2014MNRAS.439.2781M} and \cite{Diaz:2012aa} consisting of (partially overlapping) BDs with masses between 9 and 90 $M_\text{J}$ and orbiting solar-type stars with a period shorter than $\sim 10^4$~d. Unfortunately these studies contain an aggregation of identified objects without the kind of bias sample corrections applied in, e.g., \cite{Grether:2006aa},
Accordingly, their period distributions should probably be used as approximate at best, which is already illustrated by their differences in the cumulative distribution functions (Fig.~\ref{fig:cdfHistPeriodFit}).
Since we need to represent the longest periods most accurately, because they have the highest detectability, we bias our fits towards them. Hereafter these are referred to as the `Ma\,\&\,Ge' and `D\' iaz' fits and labelled `our fit (large $P$)' in Fig.~\ref{fig:cdfHistPeriodFit}. For illustration we also fit the `whole' \cite{Diaz:2012aa} curve with a more complex model (see `D\' iaz our fit (all $P$)' in Fig.~\ref{fig:cdfHistPeriodFit}), though the comparison with the histogram seems to suggest such a model does not seem to really represent the underlying distribution better, and it will not be further used for this study. Because the period distribution is amongst the most important factors in determining our predicted numbers, we will asses (and combine) the result of both the Ma\,\&\,Ge and D\' iaz fits, because they nicely represent two rather different steepnesses of the assumed period distribution towards larger periods.

In both the Ma\,\&\,Ge and D\' iaz fits we set the lower limit to 1\,d, based on the apparent cutoff in the population (as these systems should have a high detectability in the radial velocity studies). The lower panel of Fig.~\ref{fig:cdfHistPeriodFit} suggest that the distributions peak around 2.5--5\,yr, though it is unclear if this could be ascribed to a selection effect in the surveys (due to the survey lengths) or is really a feature of the underlying BD distribution.  In case a significant population exists with $P>5$\,yr, \gaia will be able to detect a large fraction of those, especially when the (currently ongoing) 10-year extended mission will be completed. 

\paragraph{Period--eccentricity distribution\label{sec:eccVsPeriod}} 

\cite{2014MNRAS.439.2781M} found that there seems to be a statistically significant different distribution of eccentricities for BDs below and above $42.5M_{\mathrm{J}}$ (the `driest' part of the BD desert). For the BDs with (minimum) mass above $42.5M_{\mathrm{J}}$ they identify the short period distribution to be consistent with a circularisation limit of $\sim$12~d, similar to that of nearby stellar binaries \citep{Raghavan:2010aa}. Also, these massive BDs seem to have an  apparent absence of eccentricities below $e<0.4$ for periods $300<P<3000$~d, while these eccentricities are populated by BDs with (minimum) masses below $42.5M_{\mathrm{J}}$. Additionally, the latter sample suggests a reduction in maximum eccentricity toward higher (minimum) mass, while such trend seems absent for the sample with (minimum) mass above $42.5M_{\mathrm{J}}$.

\begin{figure}[h]
 \includegraphics[width=0.5\textwidth]{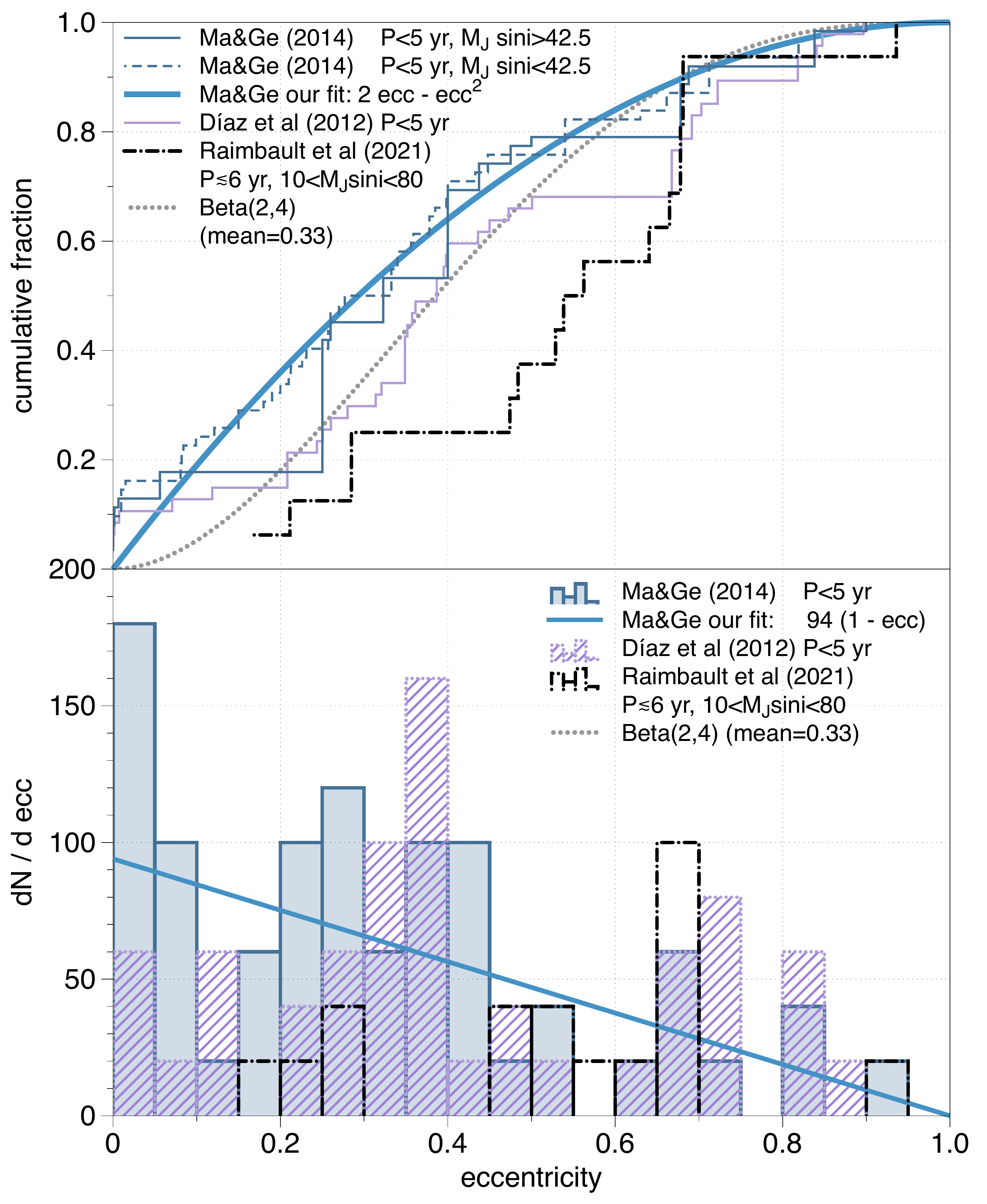} 
\caption{Eccentricity distribution of the BD sample from \cite{2014MNRAS.439.2781M} and \cite{Diaz:2012aa}. Top: the cumulative distribution function for  BDs. The `Ma\,\&\,Ge' fit is used for the simulations in this paper. The bottom figure shows the binned histogram with the corresponding fit-curve (bin size 0.05). The \cite{raimbault19} data is used for an additional bootstrap sample estimate (Sect. ~\ref{sec:raimbaultBootstrap}).
}
\label{fig:eccDist}
\end{figure}

As for the period distribution, we inspect the eccentricity distribution of \cite{2014MNRAS.439.2781M} and \cite{Diaz:2012aa} in Fig.~\ref{fig:eccDist}. Generally there is not a clear distinction between the lower mass and higher mass regimes in the \cite{2014MNRAS.439.2781M} data, and we simply adopt a single fit that allows us to draw eccentricities in the range 0--1 in a continuous fashion. 
%
The \cite{Diaz:2012aa} sample has a slightly more eccentric distribution than our fit, though this is in the interval $e=0-0.6$, for which our detection efficiency varies only by about 10\% (Fig.~\ref{fig:gaiaDetecEcc}). Accordingly, if the true brown dwarf eccentricity distribution follows the D\' iaz curve, it would have only a marginal effect on the predictions.  

\cite{Kipping:2013aa} suggested that the orbital eccentricity distribution for extrasolar planets is well described by the Beta distribution. We did not perform an extensive fitting, but for illustration plot the Beta distribution $B(2,4)$ in Fig.~\ref{fig:eccDist}, having the rough mean of the distributions shown (0.30~for  \cite{2014MNRAS.439.2781M}, and 0.41 for \cite{Diaz:2012aa} for $P<5$\,yr). This illustrates that there in no good match with  \cite{2014MNRAS.439.2781M}, although it resembles the \cite{Diaz:2012aa} reasonably well. The good fit found for planetary data, $B(0.867,3.03)$, provides a poor match to these data, heavily under-predicting the observed non-zero eccentricity distribution.

\begin{figure}[h]
    \includegraphics[width=0.5\textwidth]{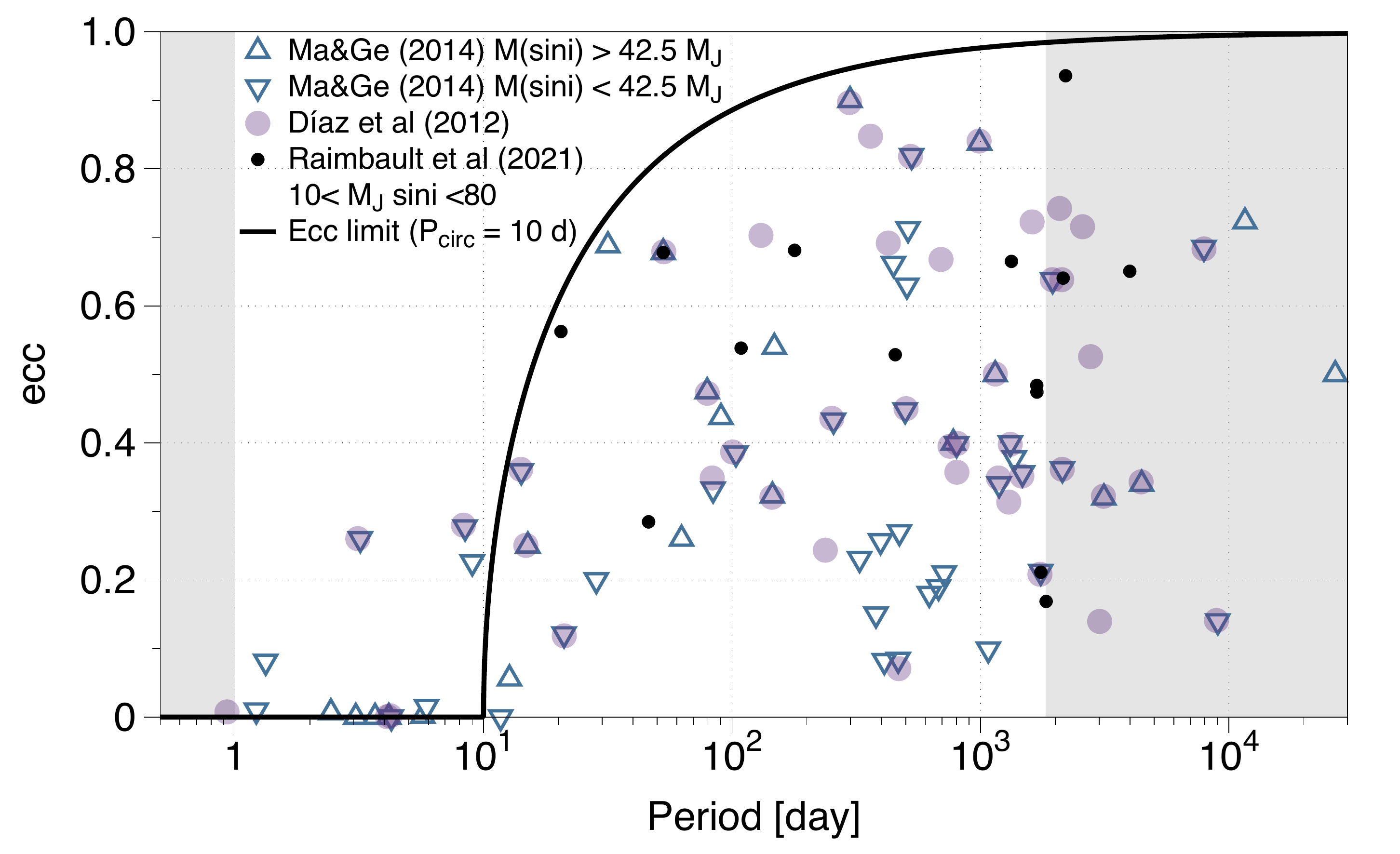} 
\caption{Period--eccentricity distribution of the BD sample from \cite{2014MNRAS.439.2781M} split in low-mass and high-mass samples, \cite{Diaz:2012aa}, and from \cite{raimbault19} only the companions in the $10-80M_\text{J}\sin i$ range. In our simulations, a circularisation limit of 10\,d is adopted, resulting in the eccentricity limit indicated by the black line.
We limit (most of) our derived models to the interval between the shaded areas, i.e., 1\,d\,$<P<$\,5\,yr. 
}
\label{fig:pEcc}
\end{figure}

Having chosen our independent distribution functions, we can inspect the period--eccentricity distribution of both \cite{Diaz:2012aa} and \cite{2014MNRAS.439.2781M} in Fig.~\ref{fig:pEcc}. 
We adopted a circularisation period of $P_{\mathrm{circ}}=10$\,d, using the formulation of \cite{Halbwachs:2005aa}, being an obvious feature in the combined parameter space. We did not adopt a specific mass-dependent eccentricity distribution, in part because the effect is only very marginal in Fig.~\ref{fig:eccDist}), but also because \cite{Grieves:2017aa} find some new candidates (their Fig.~9) which do not follow the regimes introduced by \cite{2014MNRAS.439.2781M}, although they do still support a two-population trend.

The way the samples for our simulations are constructed is by first drawing a period from either the Ma\,\&\,Ge or D\' iaz period-distribution fit. Then we use this period to compute the maximum eccentricity allowed due to the circularisation criterion, and use this to limit the range of the drawing of the eccentricity, effectively cutting the PDF from which we draw the eccentricity, from zero to the maximum eccentricity allowed. This means that the effective eccentricity distribution that is used has an increased rate of low eccentricity systems with respect to the CDF fit-line shown in Fig.~\ref{fig:eccDist}.

%% file: Sec_HostStarDistribution.tex

\begin{figure}[!htp]
   \centerline{\includegraphics[width=0.49\textwidth]{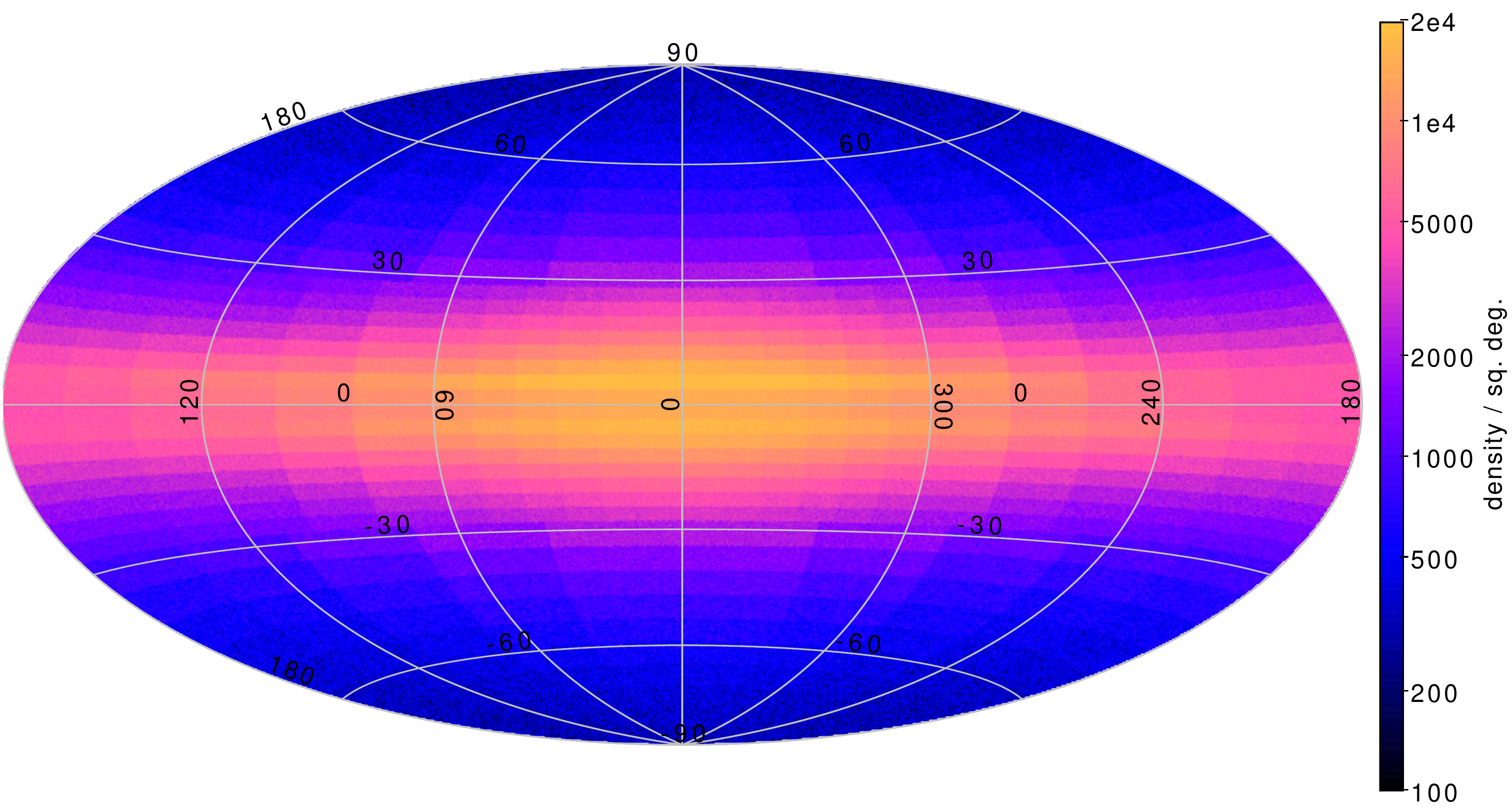} }
   \centerline{\includegraphics[width=0.49\textwidth]{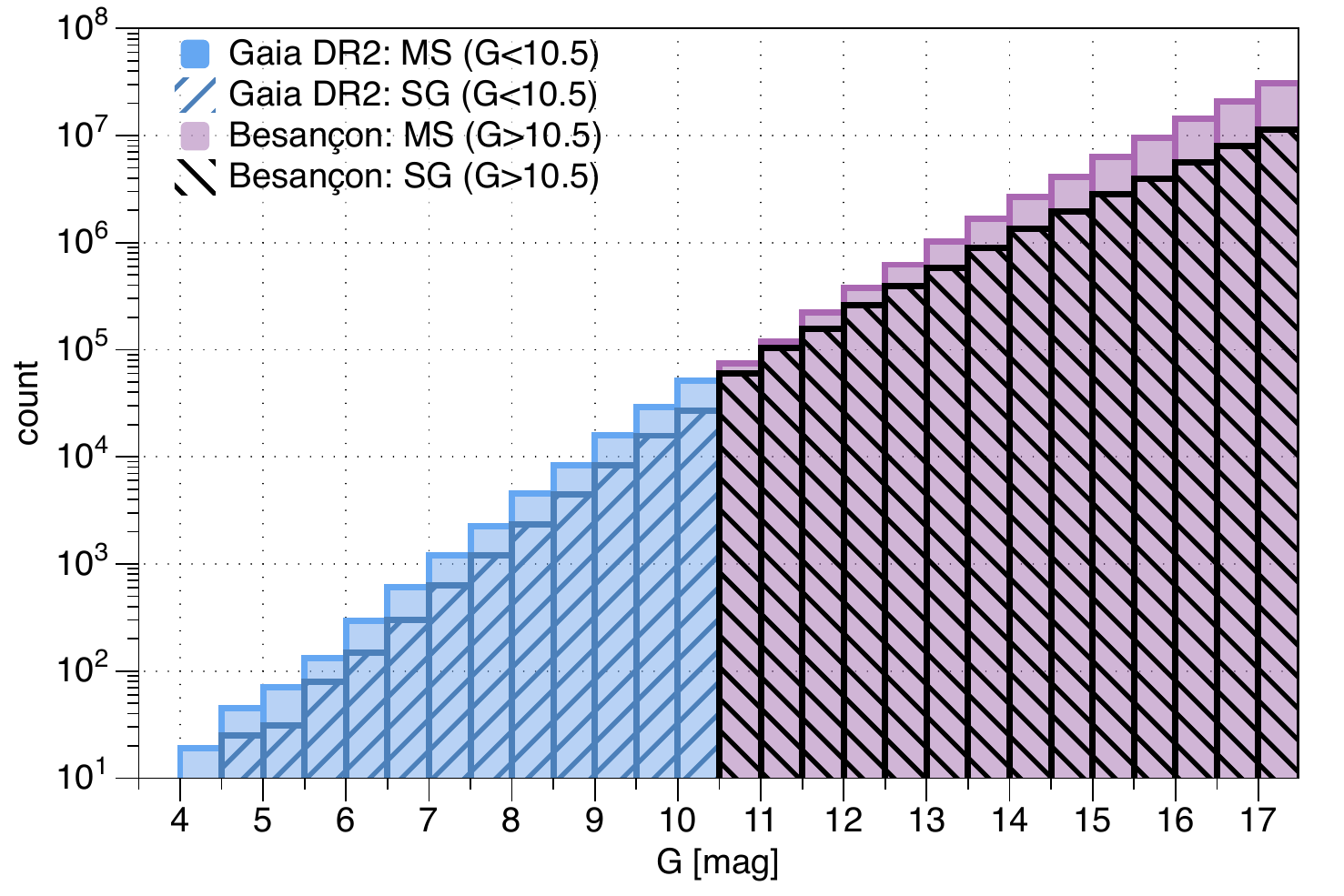} } 
              \vspace{-0.1cm}
    \centerline{ \includegraphics[width=0.49\textwidth]{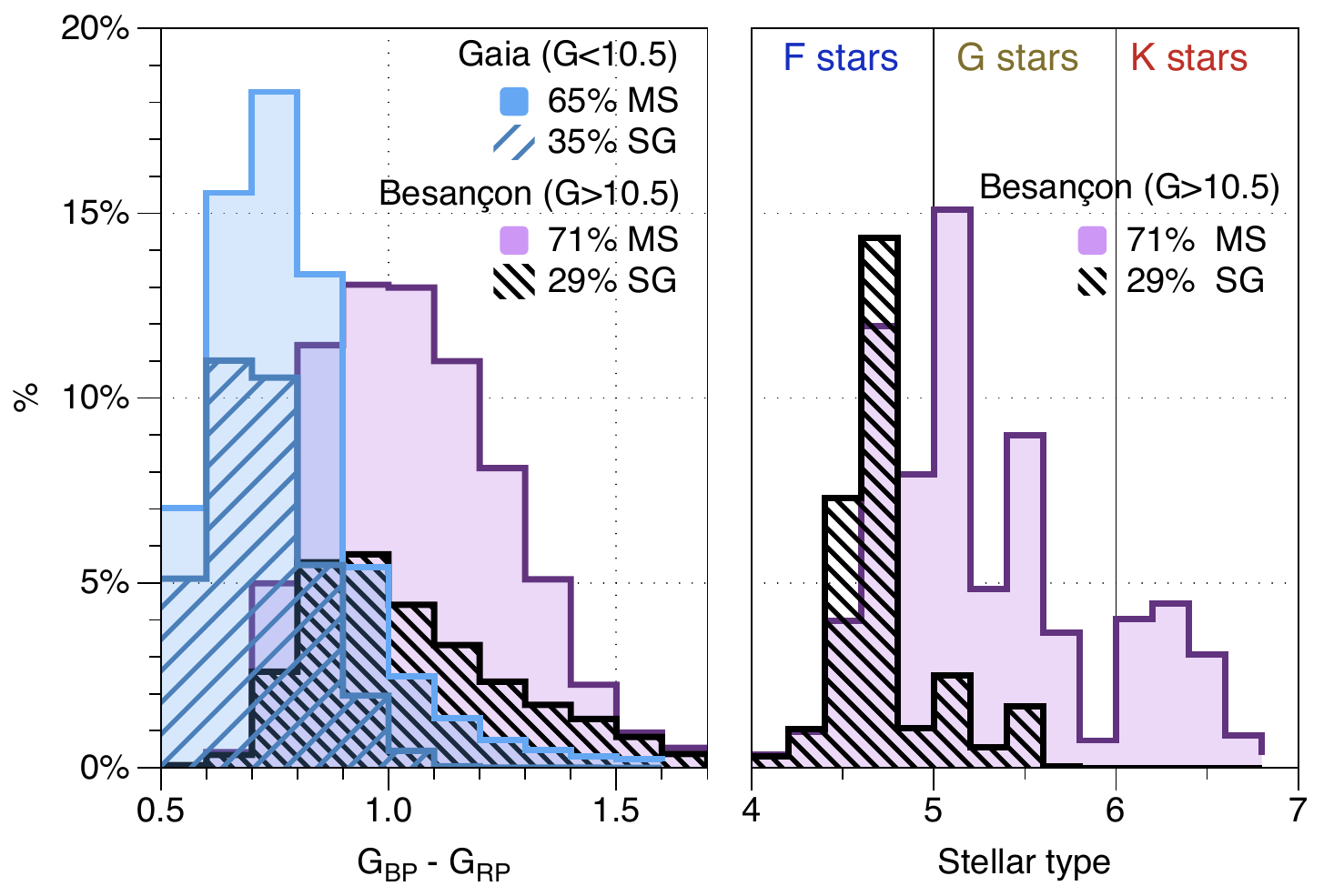} } 
    
           \vspace{-0.1cm}
     \centerline{  \includegraphics[width=0.49\textwidth]{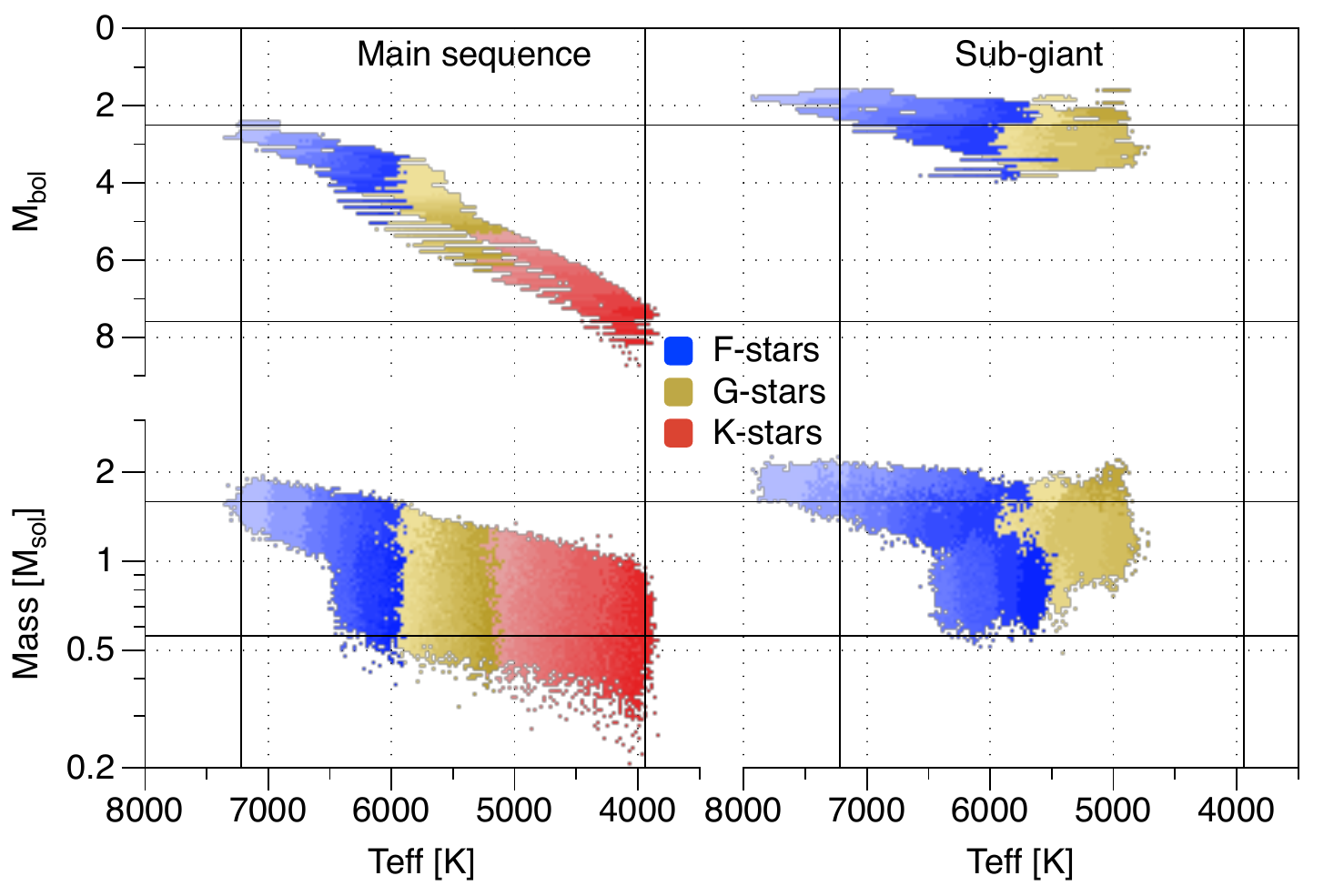} }
       \vspace{-0.2cm}
\caption{Distribution of the 130~million simulated FGK main sequence (MS) and sub-giants (SG) host stars with $G<17.5$. Top: logarithmic Galactic sky density (Besan\c con data only), second: histogram of $G$-band counts (bin size: 0.5 mag), third:  histogram of colour (bin size: 0.1) and type (bin size: 0.2), bottom: colour-absolute magnitude diagram (Besan\c con data only). 
}
\label{fig:bscFig1}
\end{figure}

The number of host stars observed by \gaia as a function of stellar parameter is given by our function $N_\text{obs}(\vec{\theta}_*)$. This is taken from the Besan\c con population synthesis model \citep{2003A&A...409..523R, 2012A&A...538A.106R,2014A&A...569A..13R} for stars with $G>10.5$, and is derived from \gaia DR2 data for $G<10.5$, as described in Appendices~\ref{appendix:besancon} and \ref{appendix:gaiaGt10p5}, respectively. The total number of stars brighter than $G=17.5$\,mag is 130~million.
Fig.~\ref{fig:bscFig1} summarises the adopted distribution of host stars, both on the sky, as a function of colour and stellar type, and according to their position in the HR-diagram.
The sky distribution of the \gaia data is not shown, but varies between roughly 2--6 stars per square degree, with the densest regions in the Galactic plane.

Given the adopted constant 0.6\% occurrence rate of BDs around any FGK star (Sect.~\ref{sec:bdFractHostStars}) we can simulate BD systems for a random sample of the 130 million Besan\c con and \gaia DR2 stars and scale the resulting counts. For the main results we simulated a 1\% random sample (1.4 million BD systems), scaled by 0.6 to produce the BD counts used to compute the statistics for the first four rows of result Table~\ref{tab:simResults}. For the $G_{\text{RVS}}<12$ selection (see Sect.~\ref{sect:resultsRv}) we simulated BD systems on the full set of main sequence and sub-giant stars (in our model 1.9~million), and then scaled the simulation counts by 0.006 to get a detailed estimate of the detection rate.

Throughout this paper we have made the assumption that all \gaia and Besan\c con (host) stars are not in a binary system, apart from the unseen BD component we  simulate. The rigorous inclusion of binarity will most likely reduce the number of detectable BDs, as it complicates the modelling. Also, the BD orbital parameter distributions adopted in this paper are likely to be quite different for those orbiting a binary host system.

%% file: Sec_Results.tex
The total number of BDs around \gaia observable host stars is about 780\,000, i.e.\ 0.6\% of the 130~million considered stars with G<17.5, listed on the top row of Table~\ref{tab:simResults}. 
Because the number of \gaia detections is highly sensitive to the period distribution, which itself is amongst the most ill-defined priors, we express the final numbers as the range of the counts resulting from two models we derived in Sect.~\ref{sec:bdPriors} (Fig.~\ref{fig:cdfHistPeriodFit}):  $P_\textrm{distr}$\,=\,Ma\,\&\,Ge and $P_\textrm{distr}$\,=\,D\' iaz. Their labels are found in many of the result tables and figures.
As seen from Table~\ref{tab:simResults}, the D\' iaz period distribution predicts about 20--30\% more BDs than Ma\,\&\,Ge.
Table~\ref{tab:simResults} allows us directly to draw some overall conclusions: 
\begin{itemize}
    \item the number of astrometric BDs detectable with \gaia will be in the tens of thousands;
    \item the radial velocity time-series for $G_{\text{RVS}}<12$ will be able to detect about one to two thousand BDs;
    \item the total number of stars with detectable BD transits will be about a thousand;
    \item stars with both good astrometric and radial velocity BD detections will be in the hundreds;
    \item the number of detectable transiting exoplanets in combination with radial velocity detection will be in the tens;
    \item detections of photometric transits together with astrometry (whether or not also including radial velocity) will be countable on one hand, and only after an extended 10~year mission.
\end{itemize}

\begin{figure*}[!htp]
\begin{tabular}{l@{\kern0em} r}          %
    5 year (nominal mission) & 10 year ( extended mission) \\
    \includegraphics[width=0.49\textwidth]{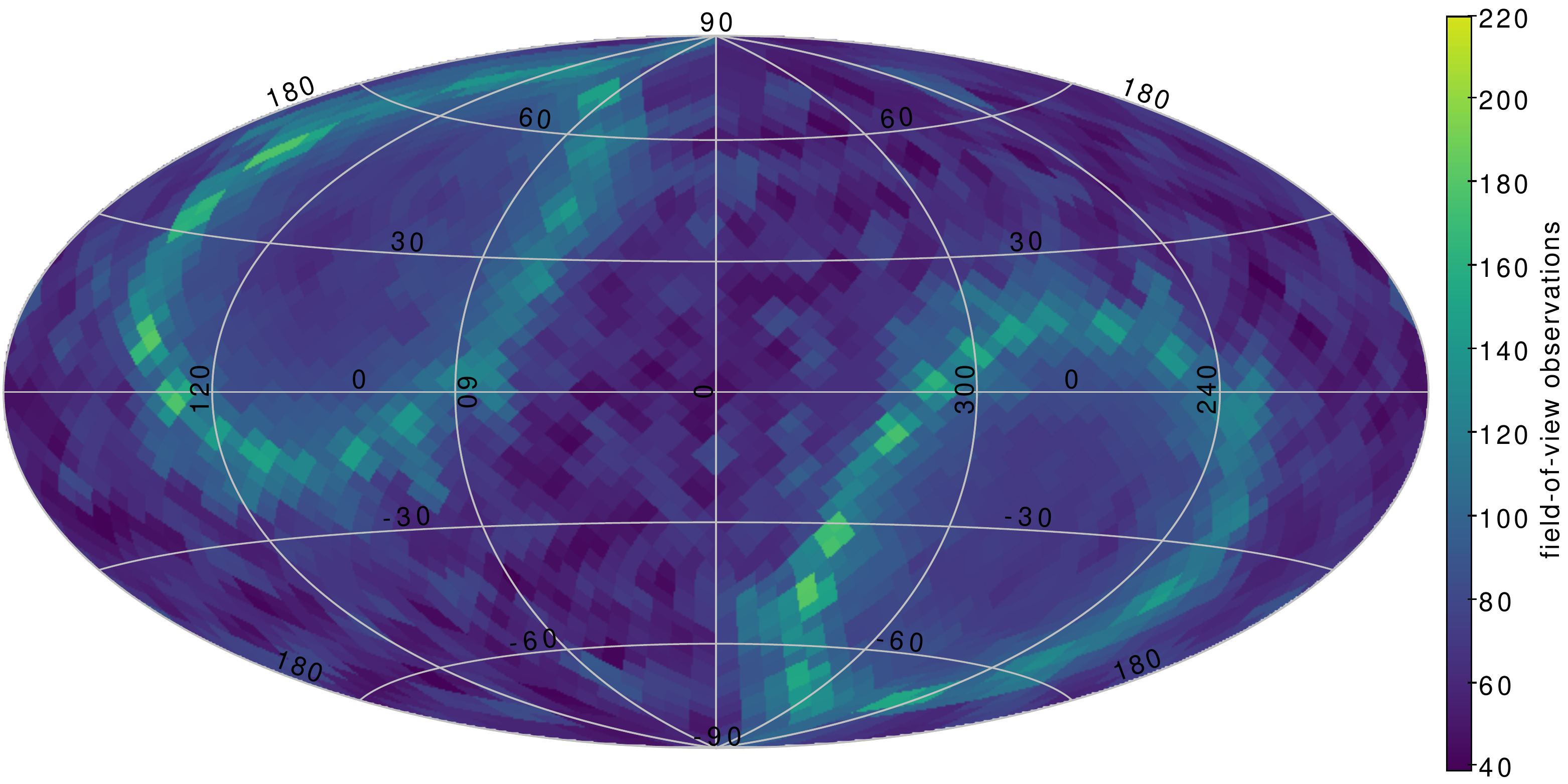}
    & 
    \includegraphics[width=0.49\textwidth]{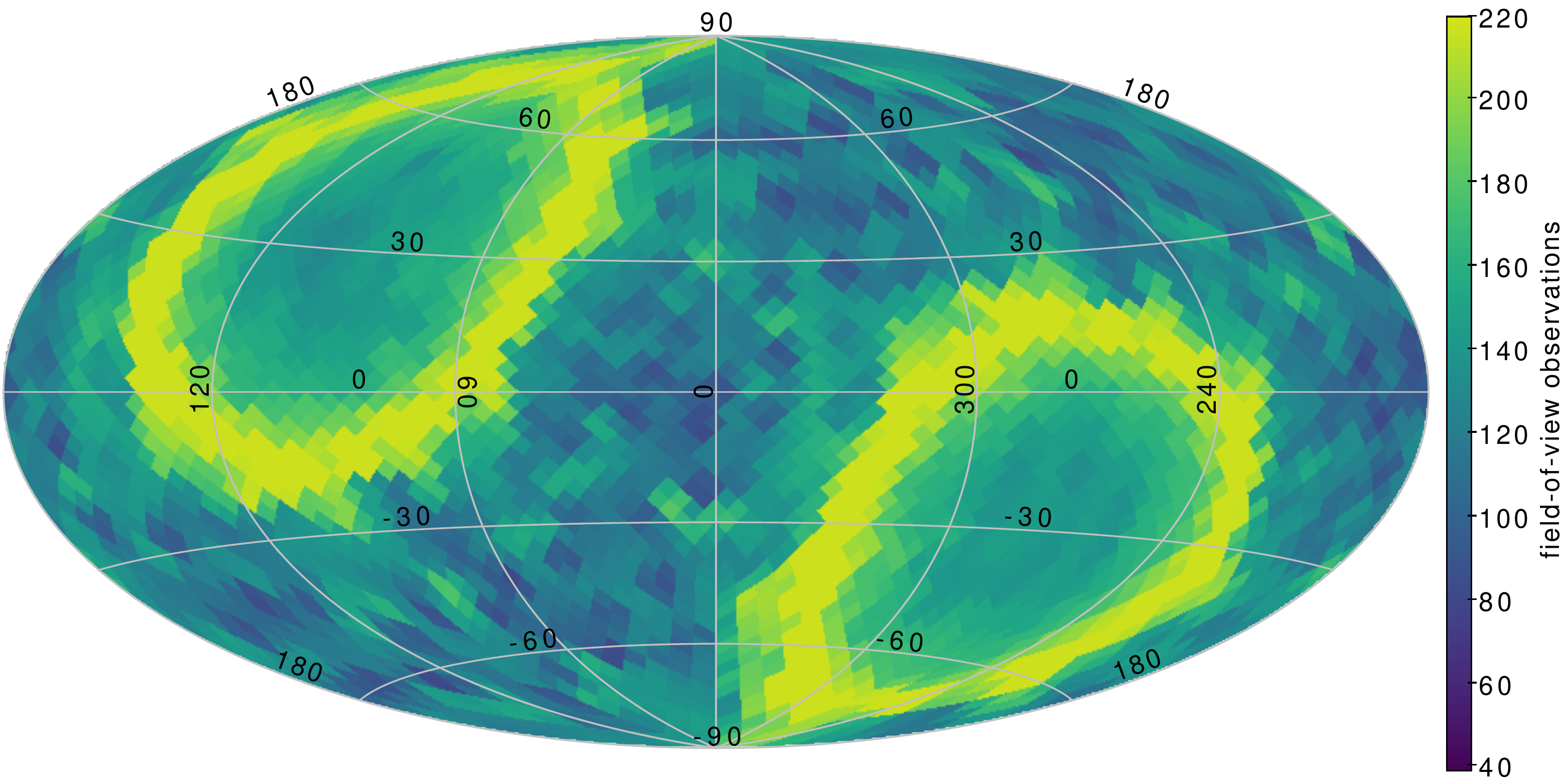} 
    \\

   \includegraphics[width=0.49\textwidth]{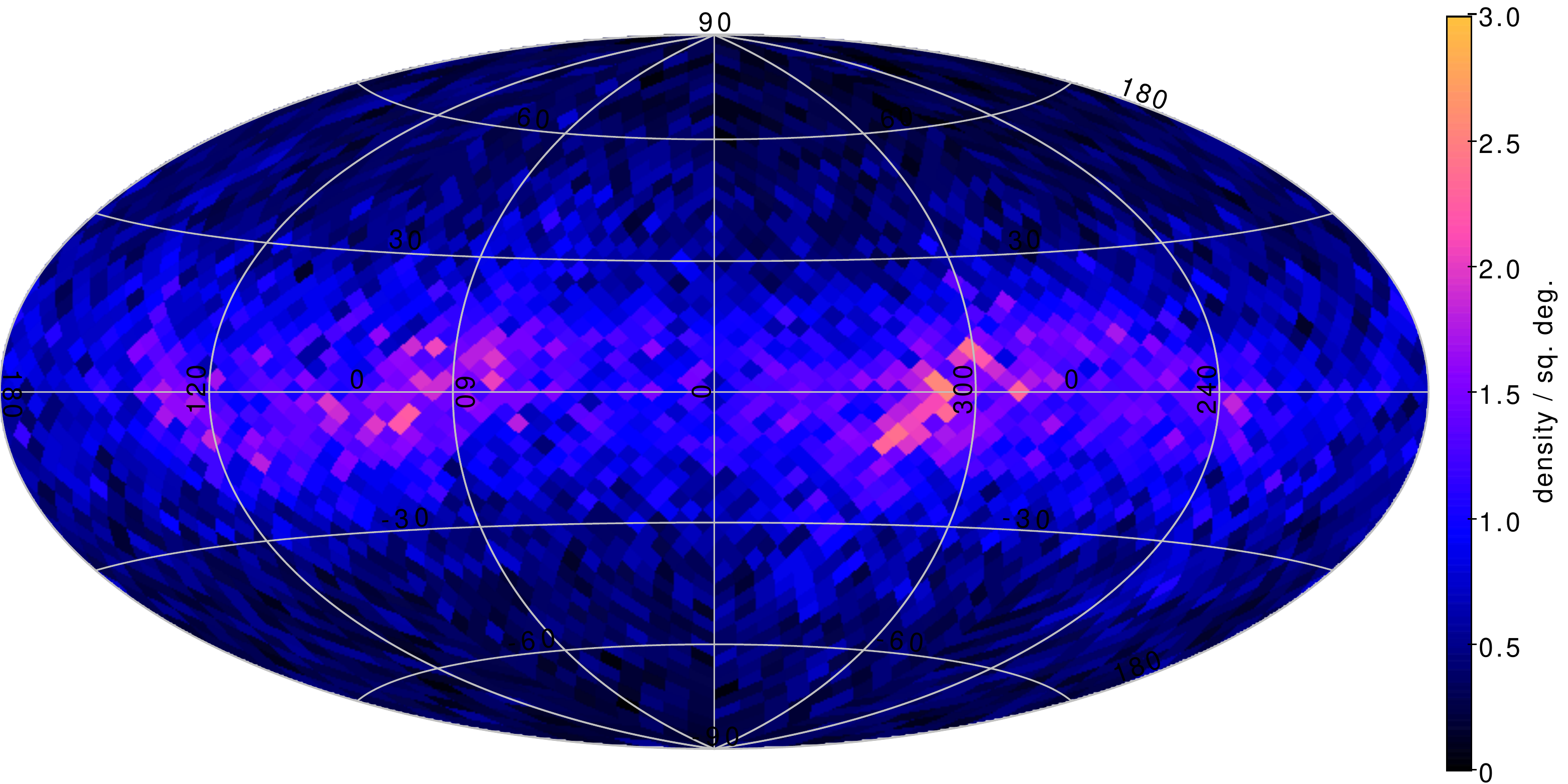}
    &   \hspace{0.2cm}
    \includegraphics[width=0.49\textwidth]{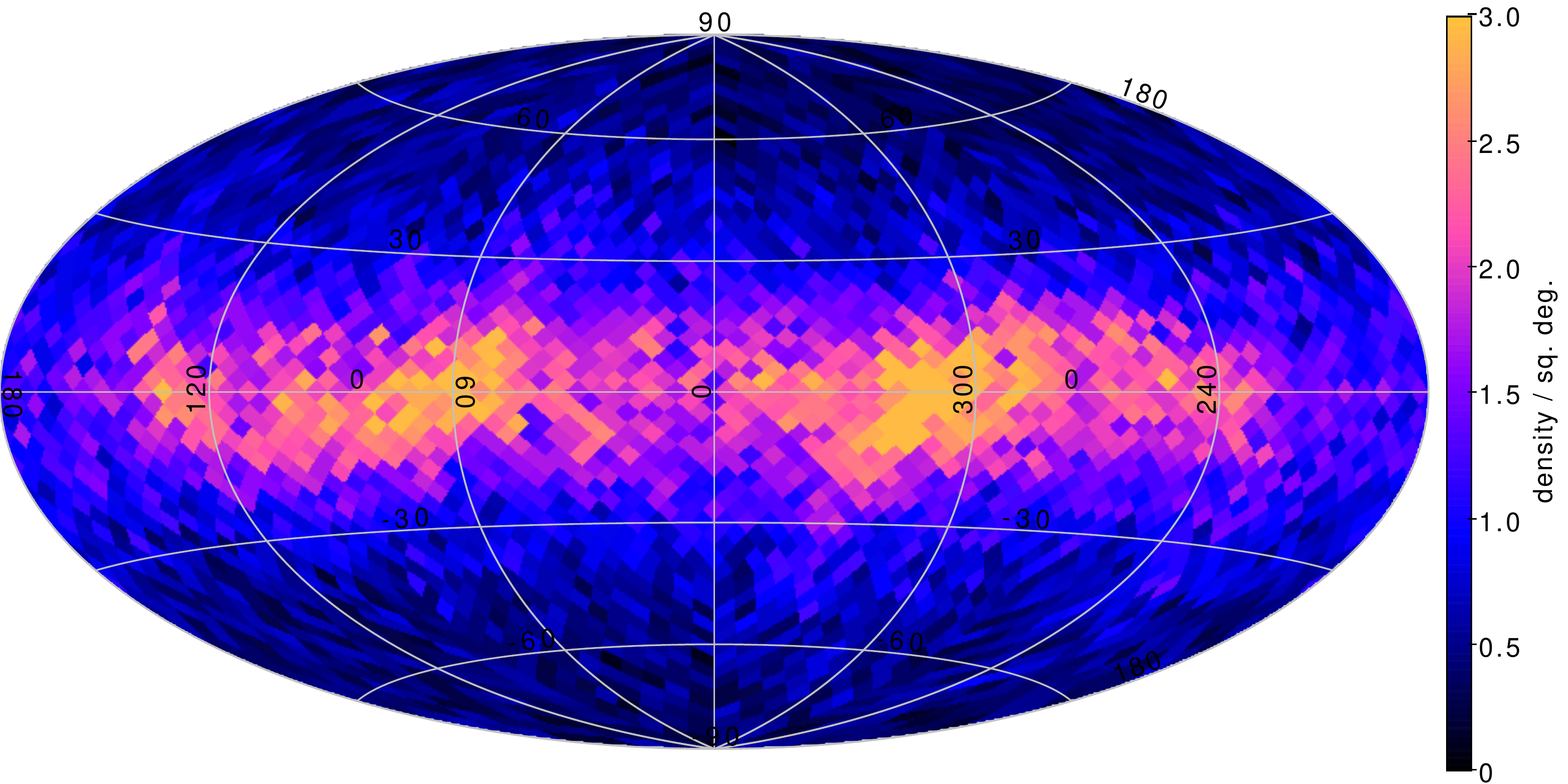} \\
    
    \includegraphics[width=0.49\textwidth]{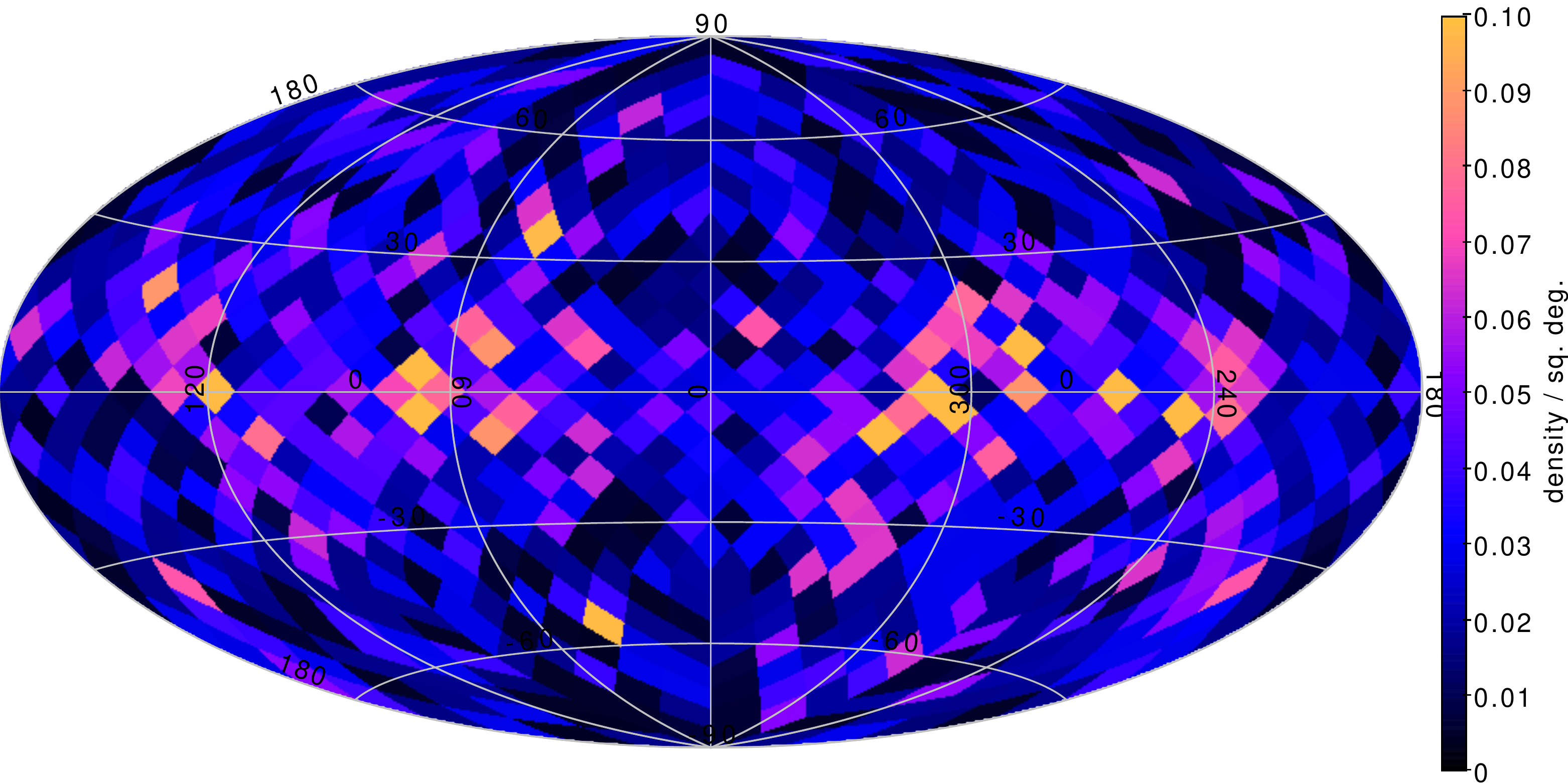} 
     & 
    \includegraphics[width=0.49\textwidth]{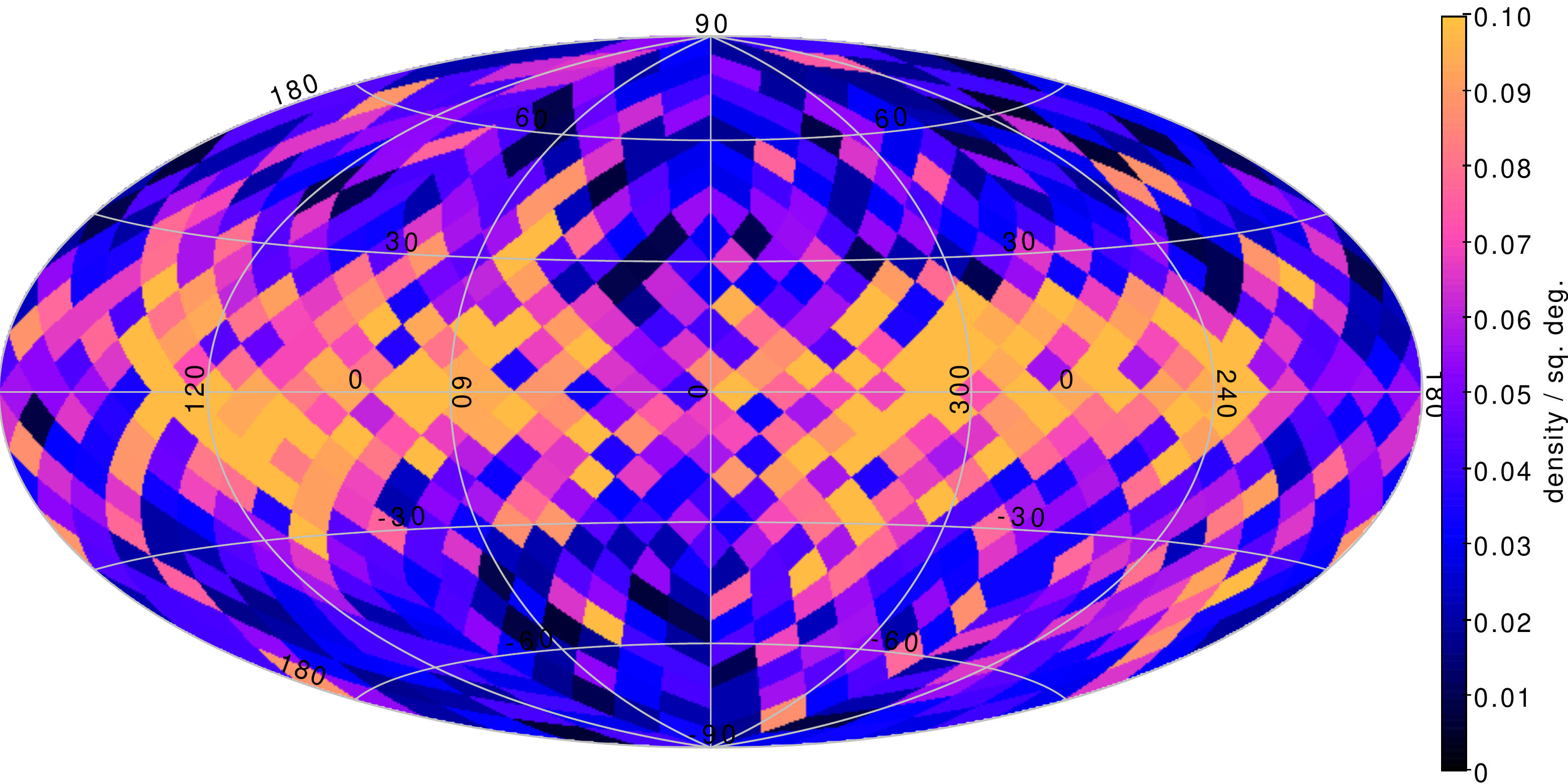} \\
    
    \multicolumn{2}{l}{ \includegraphics[width=0.96\textwidth]{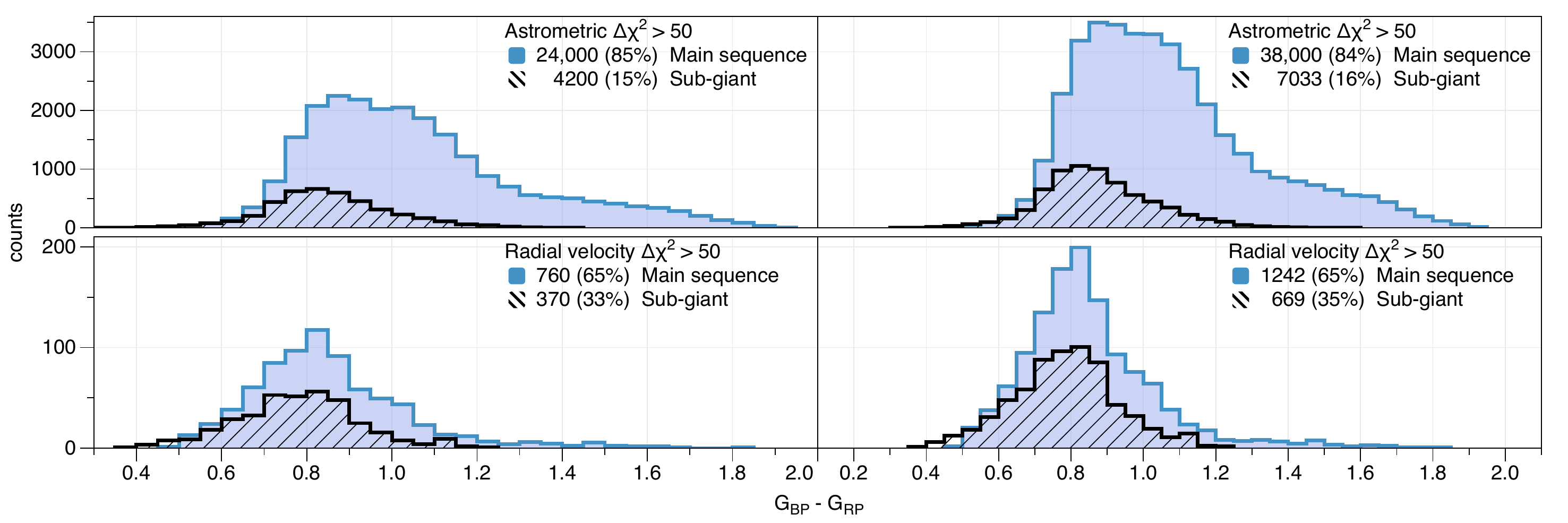} }
      \vspace{-0.2cm}\\
\end{tabular}
\caption{\label{fig:resultsTypeRelated}Distributions of the astrometeric and radial velocity solid detections ($\Delta \chi^2 >50$) resulting from the $\textrm{P}_\textrm{distr}$= `Ma\,\&\,Ge'  simulation. Top: the number of field-of-view transits for comparison with lower panels, second: astrometric detections, third: radial velocity detections,
bottom: histogram of counts as function of colour (bin size:~0.05) separated by evolutionary stage. 
 }
\end{figure*}

The sky distributions of solid detections ($\Delta \chi^2>50$) of BDs is shown in the second and third panels of Fig.~\ref{fig:resultsTypeRelated}, showing that it generally follows the Galactic host-star distribution, `weighted' by the available number of observations shown in the top panel. The bottom panel shows the detection rates as function of colour, showing that our recovery peaks around $\textrm{G}_\textrm{BP}-\textrm{G}_\textrm{RP}=0.8-0.9$.  For astrometry the vast majority of detections (85\%) is around main-sequence stars, with only 15\% coming from the sub-giants, while for the radial velocity this ration is 65\% and 33\% (similar to the ratios in the input data for the bright magnitudes shown in Fig.~\ref{fig:bscFig1}).


\subsection{Detections by astrometry\label{sect:resultsAstro}}
\begin{figure*}[!htp]
\begin{tabular}{l@{\kern0em} r}          %
    \hspace{1.1cm}5 year (nominal mission) & 10 year ( extended mission)\hspace{0.95cm} \\
    \multicolumn{2}{c}{ \includegraphics[width=0.99\textwidth]{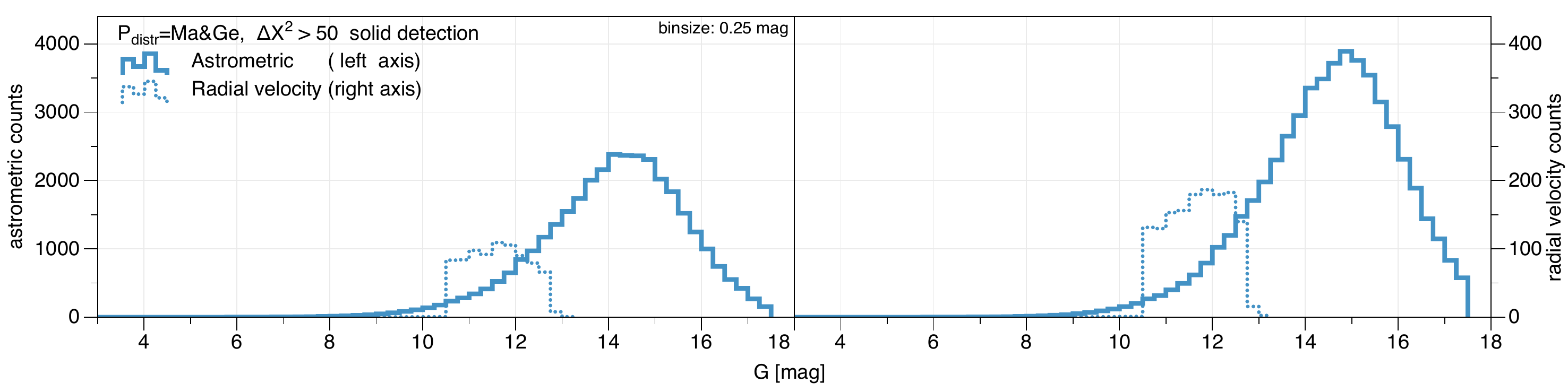} }\\
    \multicolumn{2}{c}{ \includegraphics[width=0.99\textwidth]{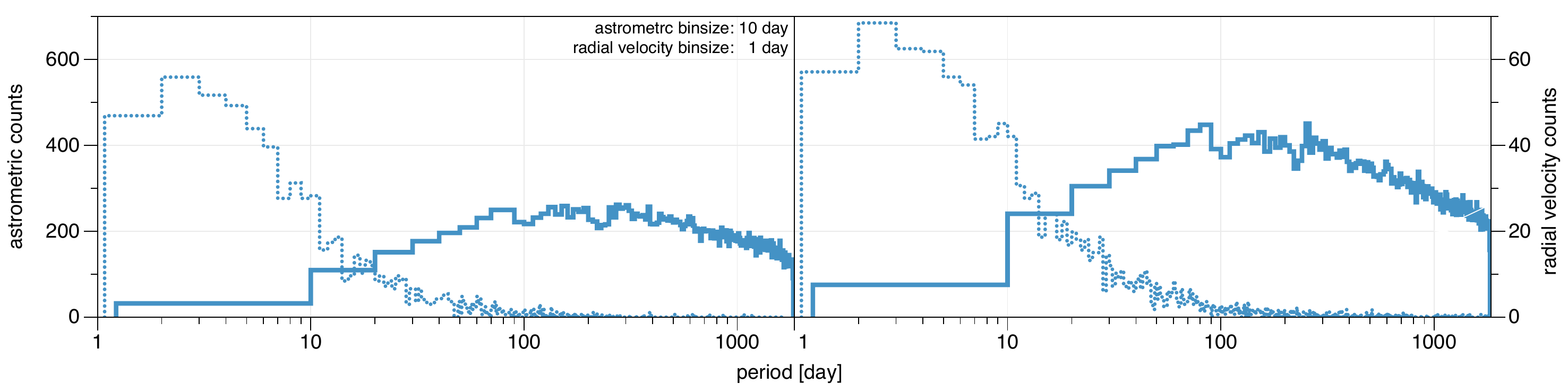} }\\
    
    \multicolumn{2}{c}{ \includegraphics[width=0.99\textwidth]{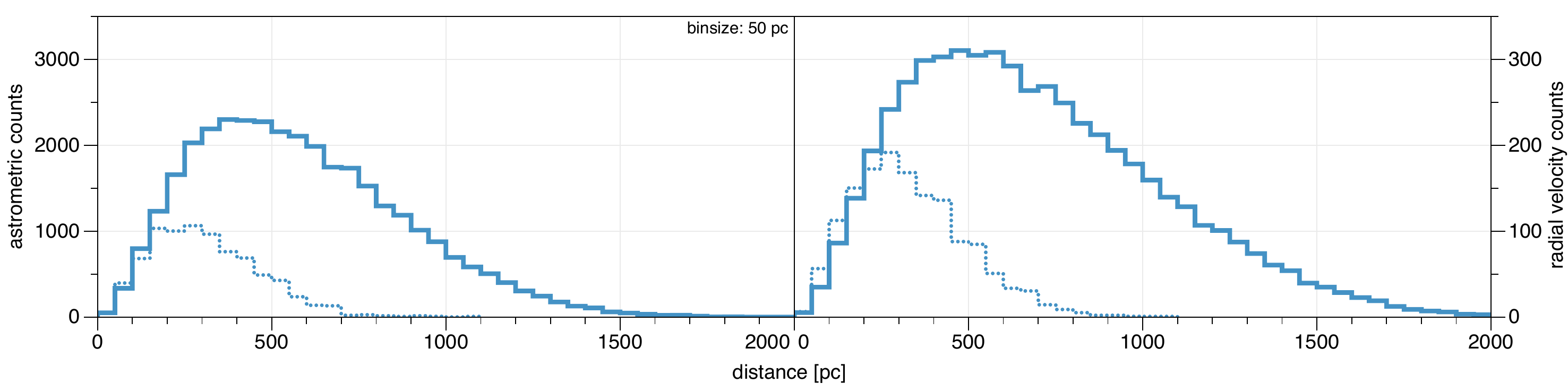} }\\
    \multicolumn{2}{c}{ \includegraphics[width=0.99\textwidth]{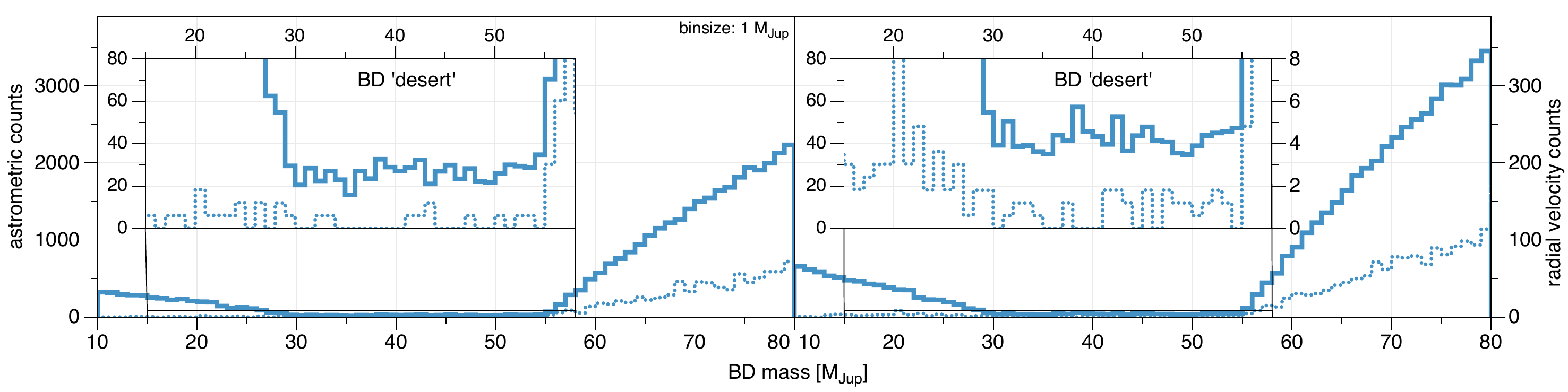} }\\
    \multicolumn{2}{c}{ \includegraphics[width=0.99\textwidth]{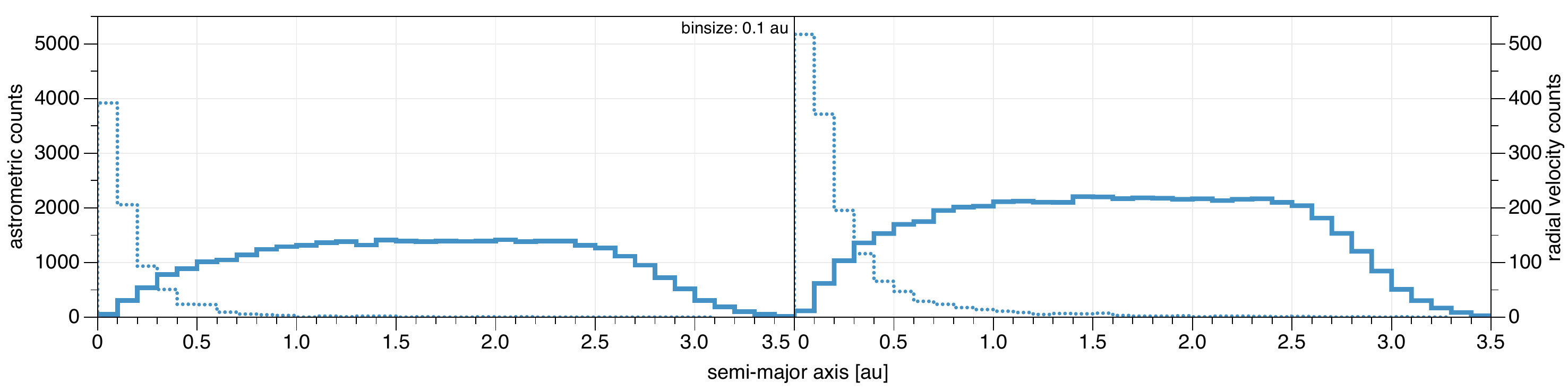} }\\

\end{tabular}
\caption{\label{fig:resultHistograms}Histograms of the astrometeric and radial velocity solid detections ($\Delta \chi^2 >50$) resulting from the $\textrm{P}_\textrm{distr}$=`Ma\&Ge'  simulation. Top panel: host star magnitude, second: period, third: distance, fourth: BD mass, and fifth panel: semi-major axis $\overline{a}$.}

\end{figure*}

For the adopted BD mass range $10-80M_{\text{J}}$ we conclude from  Table~\ref{tab:simResults} and Fig.~\ref{fig:resultHistograms} that using astrometry alone, \gaia should detect between 41\,000--50\,000 ($\pm$23\,000) BDs, of which 28\,000--34\,000 ($\pm$15\,000) BDs 
should be reliably detected out to several hundred~pc with semi-major axis typically in the $0.25 - 2.7$~au range. The majority will have $P>200$\,d,  extending up to the 5~yr adopted upper period-limit. Good orbital solutions will be possible for about 17\,000--21\,000 ($\pm$10\,000) systems. 
Note that the detection sensitivity for short astrometric periods $P \lesssim 100$~days is rather non-uniform, see  Appendix~\ref{sec:gaiaDetectionLimits}.

In Fig.~\ref{fig:resultHistograms} we see that the distribution peaks around $G\sim14-15$ where the stellar number density (Fig.~\ref{fig:bscFig1}) and still relatively small astrometric error (Fig.~\ref{fig:FovAccLaw}) produce the highest gain. Also it show the majority of recoveries are for the more massive BDs, as expected from Eq.~\ref{eq:astrometricSignal}. Note that even though we depleted the mass-prior distribution of the `desert' region to contain only 2\% of the simulated BD, there are still tens of confident detections.

\begin{figure*}[h]
  \includegraphics[width=0.98\textwidth]{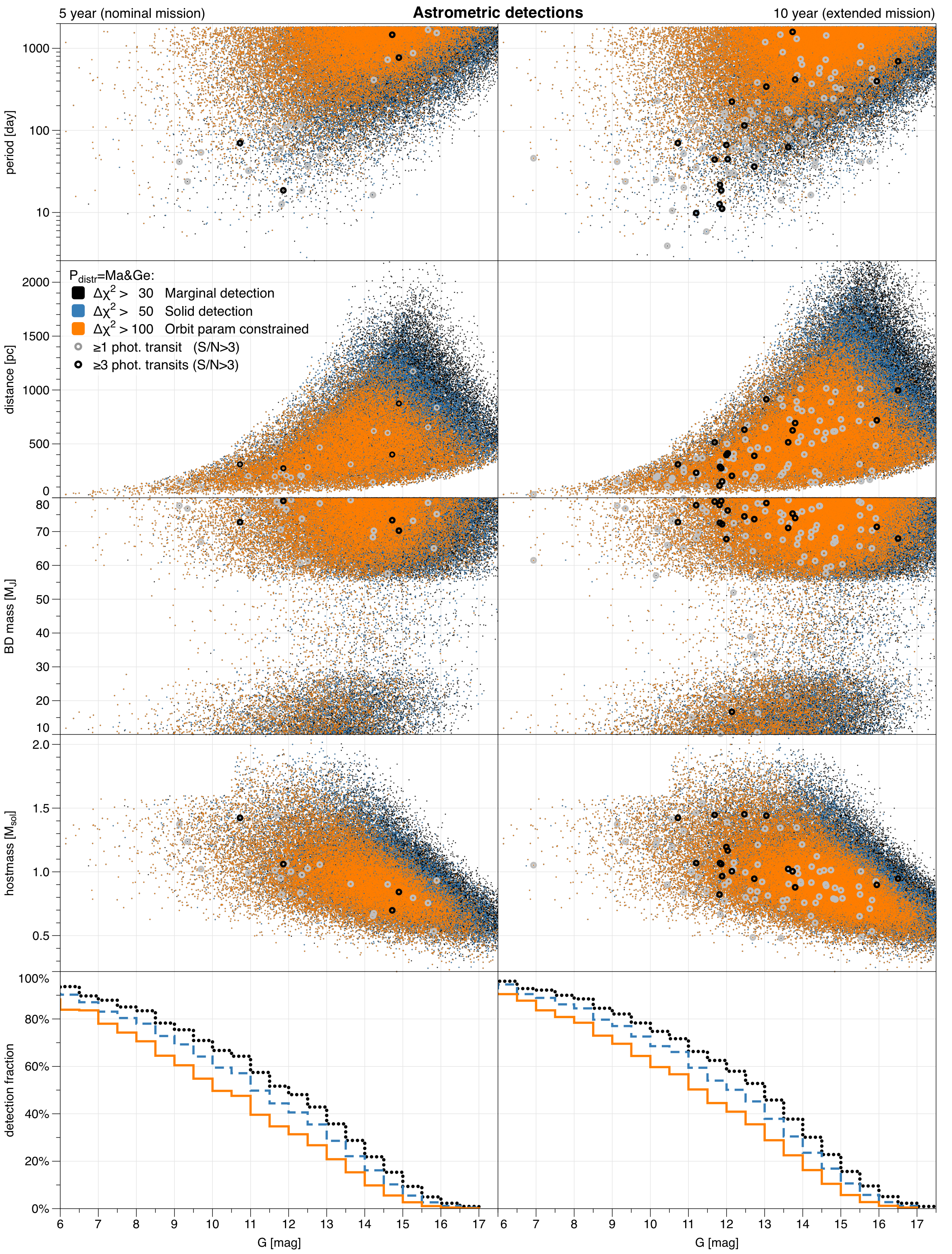}  			
  \vspace{-0.3cm}
\caption{Distributions of the BDs resulting from the 5-yr (left) and 10-yr (right) \gaia simulation using $\textrm{P}_\textrm{distr}=$Ma\,\&\,Ge for the three adopted \textbf{astrometric} detection thresholds: $\Delta~\chi^2 >30$~(41\,000, 64\,000), $\Delta~\chi^2~>50$~(28\,000, 45\,000), and $\Delta~\chi^2 >100$~(17\,000, 28\,000). (The point-cloud data are from 1\% simulations (Sect.~\ref{sec:hostStarDistribution}), hence scale by 0.6 to get the BD rates of Table~\ref{tab:simResults}.)}
\label{fig:results1page}
\end{figure*}

Fig.~\ref{fig:results1page} shows the distributions as function of magnitude for all detection threshold for various parameters. The bottom panel show that the detection fraction (completeness) steadily increases towards the bright end, though for the most populated magnitudes it reaches only 10--30\%. For a 10-yr mission, this number rises to 20--40\% and the mode of the distribution is extended by about 0.3 mag and 20-30\% in distance.

For the $G>10.5$ Besan\c con model data (not plotted) we have the detailed type information available, telling us that for the solid detections 40\% are G-type host stars, 32\% are host K-stars, and 28\% are host F-stars. The recovery numbers are declining at the edges of our adopted range of stellar types, i.e.\ towards the more massive F-stars and less-massive K-stars, confirming our assumption in the introduction that the majority of BDs detections is expected around FGK-stars.


\subsection{Detections by radial velocity\label{sect:resultsRv}}
\begin{figure*}[h]
  \includegraphics[width=0.98\textwidth]{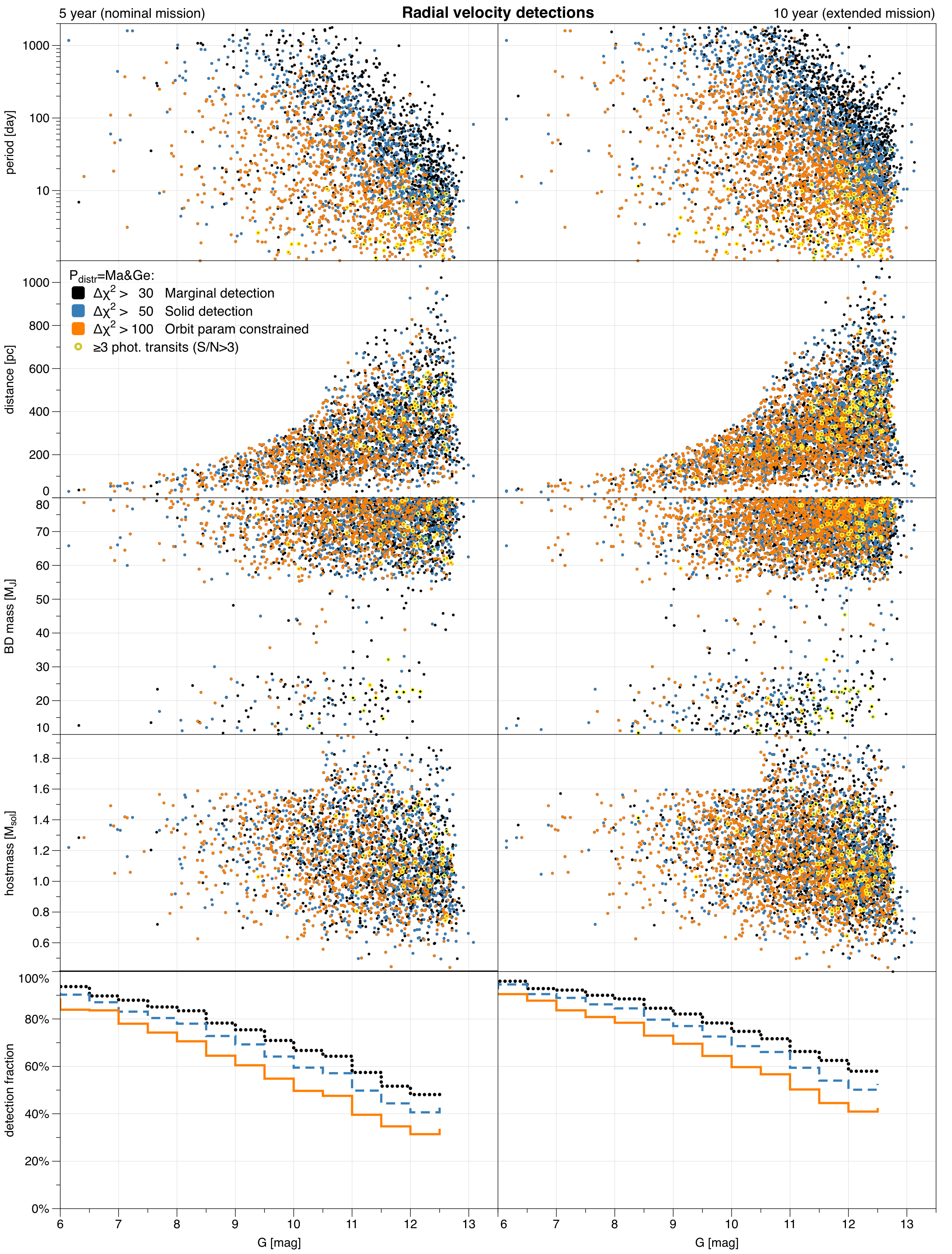} 				
  \vspace{-0.3cm}
\caption{Same as Fig.~\ref{fig:results1page} but now for the three adopted \gaia \textbf{radial velocity} detection thresholds (i.e., limited to stars with $G_\textrm{RVS}<12$): $\Delta~\chi^2 >30$~(1700, 2700), $\Delta~\chi^2~>50$~(1100, 1900), and $\Delta~\chi^2 >100$~(570, 1100). (The point-cloud data are from 1\% simulations (Sect.~\ref{sec:hostStarDistribution}), hence scale by 0.6 to get the BD rates of Table~\ref{tab:simResults}.)}
\label{fig:results1pageRv}
\end{figure*}

Here we discuss only sources with \gaia radial velocity time series, selected by $G_{\text{RVS}}<12$ ($G\lessapprox 12.7$). From Table~\ref{tab:simResults} 
we see that we can expect 1700--1300 ($\pm$800) BD detections from \gaia radial velocity measurements, of which 1100--830 ($\pm$500) robustly, and 570--410 ($\pm$250) with good orbital parameters. As expected, shown in Figs.~\ref{fig:resultHistograms} and \ref{fig:results1pageRv}, the largest numbers are found at short orbital periods and for the most massive BDs ($55-80M_{\text{J}}$), although several tens of those will extend to the lowest BD masses. 
The largest number of detections is around $G=11-12$, where the detection fraction is around 30--60\%. A 10-year mission boosts this to 40--70\%.

Fig.~\ref{fig:resultHistograms} shows the distance limit is a few hundred pc at maximum with a semi-major axis typically below 0.25~au. Mainly periods  below 10 days are recovered, and hence the eccentricity distribution (not shown) is peaking at 0 due to the adopted eccentricity limit (which is 0 for periods below 10~days, see Fig.~\ref{fig:eccDist}). Recovery as function of inclination (not shown) is roughly limited between $20^\circ< i < 160^\circ$ and being rather flat between $60^\circ< i < 120^\circ$, as expected from Eq~\ref{eq:rvSignal} and the random orientation input distribution.
 
\begin{figure*}[h]
  \includegraphics[width=0.98\textwidth]{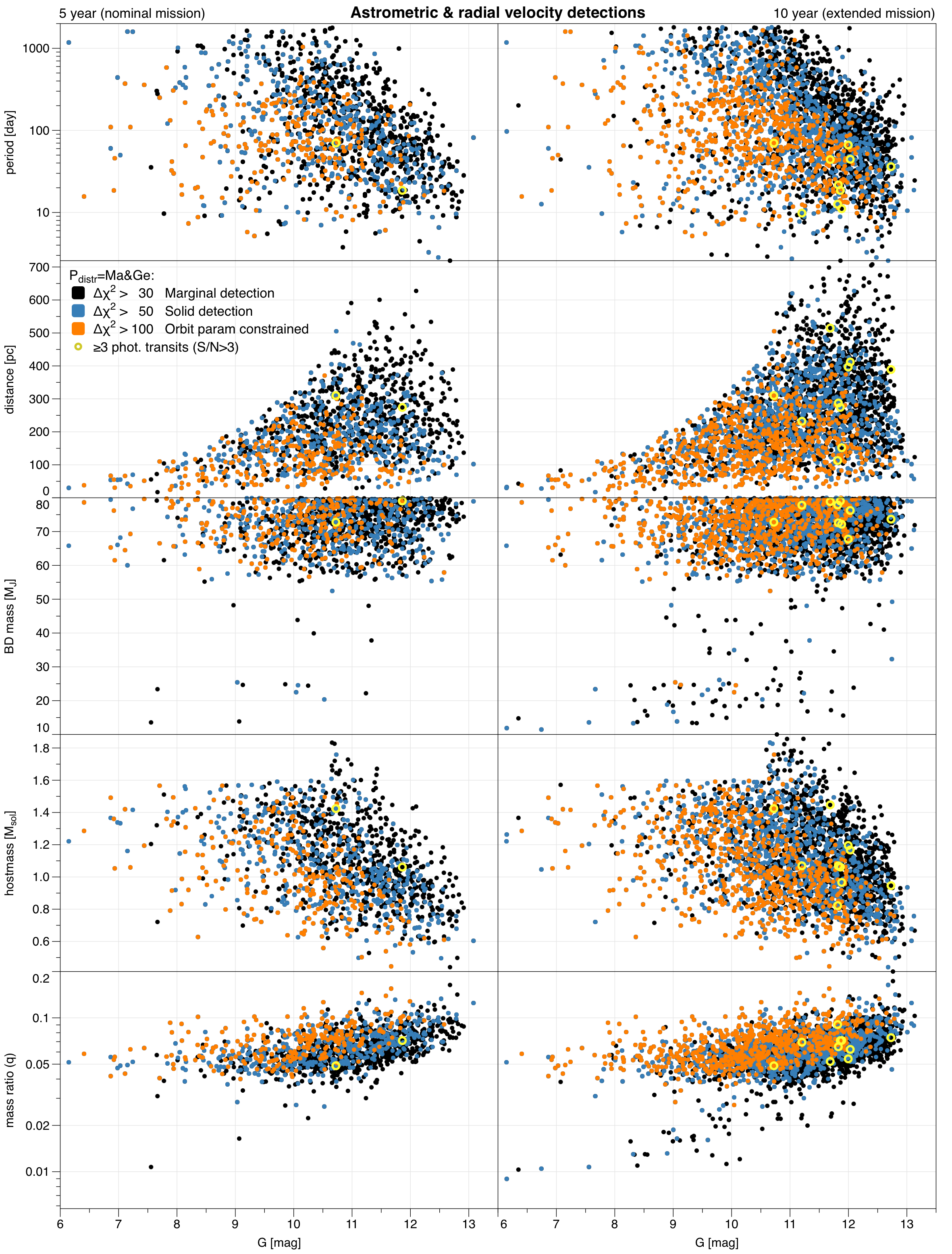} 		
  \vspace{-0.3cm}
\caption{Intersection of Figs.~\ref{fig:results1page} and \ref{fig:results1pageRv}, i.e., where the same detection threshold is passed for both \gaia \textbf{astrometry} and \textbf{radial velocity} (hence limited to stars with $G_\textrm{RVS}<12$): $\Delta~\chi^2 >30$~(810,1600), $\Delta~\chi^2~>50$~(410, 950), and $\Delta~\chi^2 >100$~(140, 400). (The point-cloud data are from 1\% simulations (Sect.~\ref{sec:hostStarDistribution}), hence scale by 0.6 to get the BD rates of Table~\ref{tab:simResults}.)}
\label{fig:results1pageAstroRv}
\end{figure*}

The combination of astrometric and radial velocity detections (i.e. the last three rows of Table~\ref{tab:simResults} and shown in Fig.~\ref{fig:results1pageAstroRv}) yields  810--750 ($\pm$400) systems, of which 410--370 ($\pm$200) have good orbital solution from either data set. This will be an important sample, since a joint solution would better constrain all system parameters. Additionally, since all sources with \gaia RV time-series are relatively bright, this set will be ideal for follow-up observations.
Most noticeable is that the number of systems in this sample doubles for an extended mission of 10 years!

\subsection{Detections by photometric transits\label{sec:transit}}
\begin{figure*}[!htp]
  \includegraphics[width=0.98\textwidth]{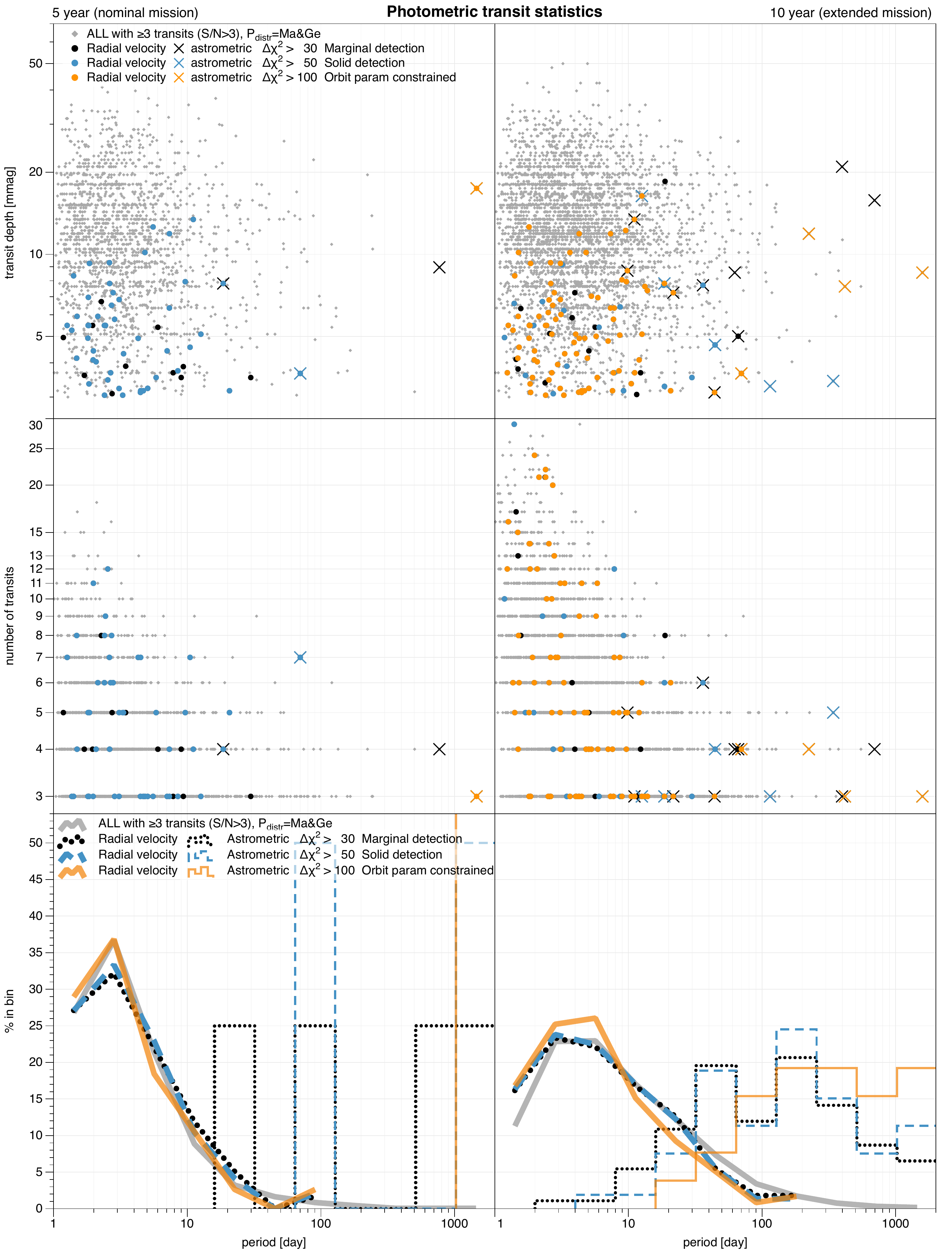} 
    \vspace{-0.3cm}
\caption{Transit statistics from the $\textrm{P}_\textrm{distr}$= Ma\,\&\,Ge simulation for a 5-year and 10-year mission of 1.5~million randomly oriented BDs. These BD number counts are scaled by 0.6 to represent a 0.6\% BD occurence rate in Table~\ref{tab:simResults}, see Sect.~\ref{sec:hostStarDistribution}. Detailed period-binned statistics of the `ALL' part are shown in Table~\ref{tab:fracTransiting}. 
}
\label{fig:numInTransit}
\end{figure*}

For each simulation we list the number of sources that have at least three detectable transits (S/N\,>\,3) in parenthesis in Tables~\ref{tab:simResults} and~\ref{tab:RbsimResults}, together with their average number of observation in transit. The transit depth and period distribution of the Ma\,\&\,Ge 5-year sample are shown in Fig.~\ref{fig:numInTransit}.

As seen from the top row of Table~\ref{tab:simResults}, detectable photometric transits are expected for some 720-1100 ($\pm$450) mainly short-period BDs with periods mainly below 10 days (Fig.~\ref{fig:numInTransit}), each yielding around four to five transits on average.
Direct identification of these objects as BDs would be difficult from the \gaia photometry alone, but this would yield a very interesting follow-up sample.

For $G_{\text{RVS}}<12$ this number drops to 34--53 BDs each with typically five observable transits. Because both radial velocity and photometric transit detection methods are sensitive to large orbital inclination and short orbital period, out of this small sample about half should be detectable with both methods, growing to perhaps 40--60 BDs for a 10-yr mission.

For astrometrically detectable BDs with at least three detectable photometric transits the numbers only reach in the few, and drop to basically zero below $G_{\text{RVS}}<12$ (whether or not we check for coincidence with radial velocity detection). A 10-yr mission could yield at most a few of these `triple method' detections.

Let us go for a moment in a bit more detail.
The largest main sequence stars\footnote{Using radii from reference in footnote~\ref{footnoteMamajek}.} have diameters around $1.79R_{\odot}$ (F0V) and the smallest around $0.65R_{\odot}$ (K9V). For an adopted BD radius of $R_\text{J}\sim0.1R_{\odot}$, this means that the predicted transit depth $(R_{\mathrm{BD}}/R_*)^2$ ranges between 0.3--2.3\%, corresponding to $\Delta m\simeq3-25$\,mmag. This is around the \gaia noise limit (see Appendix~\ref{sec:gaiaErrorPhot}) for most of the considered magnitude range and with a S/N>3 threshold it causes a drop of a factor 3-4 in detected number of sources with at least one observable transit (not shown).
In Table~\ref{tab:fracTransiting} we list the total transit statistics for the all magnitude simulation using $\text{P}_\text{distr}$ = Ma\,\&\,Ge for different period bins. When only requiring at least one transit with a S/N>3 (left column): for a 5-year simulation 64--128~day bin we can see that of all observations of all random oriented sources there is 1 in  56\,000 observations in transit ($1/f^\text{obs}_\text{tr}$), and 1 in 1100 sources that have non-zero transits ($1/f^\text{src}_\text{tr}$). For those sources that have non-zero transit, the mean number of transits is $\langle N^\text{obs}_\text{tr} \rangle=1.6$.  When we select a minimum number of three transits with a S/N>3 on the photometric transits (as we demand in this study) the source occurrence rates of all period bins drop by a significant factor of 2.5 to 8. For a ten-year mission the fraction of observations in transit does not change much (as this is mainly due to the inclination of the systems), but the number of sources with transits is increasing by up to a factor of 2 for the longer periods, though hardly changes for those with short orbital periods, as they were likely to transit already in 5-year data. For a ten-year mission the average number of transits per source increases by up to a factor of 1.5 with the largest increase mainly for the short period objects as expected. 



In the Ma\,\&\,Ge 5- and 10-year transit `ALL' sample shown in Fig.~\ref{fig:numInTransit}, 71\% of the identified transiting systems are around main sequence stars, reflecting the simulated input distribution. 

\begin{table*}[t]
\caption{Simulation results. The number of detected BDs with in parentheses the number of host stars having $\geq3$ photometric transits (S/N>3) in \gaia data together with their mean number of transits. 
The two period distributions $\textrm{P}_\textrm{distr}$ `Ma\,\&\,Ge'  and `D\' iaz' are detailed in Sect.~\ref{sec:bdPriors} (Fig.~\ref{fig:cdfHistPeriodFit}). All counts could be up to $50\%$ higher or lower due to the uncertainty on the assumed $0.6\%$ BD occurrence rate (see Sect.~\ref{sec:bdFractHostStars}).
} 
\label{tab:simResults}     
\centering    
\begin{tabular}{c c c | r @{\ }l r@{\ } l | r@{\ } l  r@{\ } l}          %
\hline\hline      
Magnitude & Astro & RV & \multicolumn{4}{c|}{5-yr nominal} & \multicolumn{4}{c}{10-yr extended} \\
selection & $\Delta \chi^2$ & $\Delta \chi^2$ & \multicolumn{2}{c}{$\text{P}_\text{distr}$ = Ma\,\&\,Ge }   & \multicolumn{2}{c|}{$\text{P}_\text{distr}$ = D\' iaz} & \multicolumn{2}{c}{$\text{P}_\text{distr}$ = Ma\,\&\,Ge }   & \multicolumn{2}{c}{$\text{P}_\text{distr}$ = D\' iaz}  \\    
\hline                                 
--                 & -- 	        & --	& 780\,000 & (1100: 4.5)        &   780\,000 	&( \ 720: 4.4)  & 780\,000 &(2300: 5.9) 	    & 780\,000 &(1400: 5.8) \\
\hline
--                 & $>30$         & --	    & 41\,000 & ( \quad \ 3: 4.5)   &   50\,000 & ( \quad \  4: 3.0)& 64\,000 & ( \ \ \ 12: 3.9)    & 77\,000 & ( \quad \  6: 3.5)  \\
--                 & $>50$         & --	    & 28\,000 & ( \quad \ 1: 4.9)   &   34\,000 & ( \quad \  3: 3.0)& 45\,000 & ( \quad \ 6: 3.7)   & 55\,000 & ( \quad \  4: 3.4)  \\
--                 & \ \ $>100$    & --	    & 17\,000 & ( \quad \ 1: 3.3)   &	21\,000 & ( \quad \  2: 3.0)& 28\,000 & ( \quad \ 3: 3.7)   & 35\,000 & ( \quad \  2: 3.4)  \\
\hline
\hline
$G_{\text{RVS}}<12$ & -- 		    & --	& 12\,000 & ( \ \ \   53: 4.8)     & 12\,000 & ( \ \ \ 34: 4.8)    & 12\,000 & ( \ \ \  94: 6.7)      & 12\,000 & ( \ \ \ 60: 6.5)	 \\ 
\hline
$G_{\text{RVS}}<12$ & $>30$ 		& --	& 6200 & ( \quad  \ 1: 3.9)     &  7100 & ( \quad  \ 1: 3.6)    &  7300 & ( \quad  \ 6: 4.4)      & 8100 & (  \quad  \ 5: 4.3)	 \\ 
$G_{\text{RVS}}<12$ & $>50$ 		& --	& 5300 & ( \quad  \ 1: 3.9)     &  6100 & ( \quad  \ 1: 3.7)    & 6500 & ( \quad  \ 4: 4.3)       & 7300 & (  \quad  \ 3: 4.1)	 \\ 
$G_{\text{RVS}}<12$ & \ \  $>100$ 	& --	& 4300 & ( \quad  \ 0\quad \ \ \ )     &  5000 & ( \quad  \ 0\quad \ \ \ )    &  5400 & ( \quad  \ 2: 4.1)    & 6200 & (  \quad  \ 2: 4.0)	 \\ 
\hline
$G_{\text{RVS}}<12$ & --	        & $>30$     & 1700 & ( \ \ \  32: 4.9)  &  1300 & ( \ \ \  20: 4.9)     & 2700 & ( \ \ \  64: 7.0)      & 2200 & ( \ \ \  40: 6.8)  \\
$G_{\text{RVS}}<12$ & --	        & $>50$     & 1100 & ( \ \ \ 27: 5.0)  &  830 & ( \ \ \  17: 4.9)      & 1900 & ( \ \ \  56: 7.0)      & 1500 & ( \ \ \  35: 6.9)  \\
$G_{\text{RVS}}<12$ & --	        & \ \ $>100$& 570 & ( \ \ \  21: 5.1)   &  410 & ( \ \ \  13: 5.0)      & 1100 & ( \ \ \  46: 7.1)      & 810 & ( \ \ \  29: 7.0) \\
\hline
$G_{\text{RVS}}<12$ & $>30$	        & $>30$     & 810 & ( \quad  \ 1: 3.9) & 750 & ( \quad  \  1: 3.7)     & 1600 & (  \quad \   6: 4.5)      & 1500 & ( \quad  \  4: 4.4)  \\
$G_{\text{RVS}}<12$ & $>50$	        & $>50$     & 410 & ( \quad  \ 1: 4.0)  &  370 & ( \quad  \  0\quad \ \ \ )    & 950 & (  \quad \  3: 4.4)       & 870 & ( \quad  \  2: 4.3)  \\
$G_{\text{RVS}}<12$ & \ \ $>100$	& \ \ $>100$& 140 & ( \quad  \ 0\quad \ \ \ )  &  120 & ( \quad  \ 0\quad \ \ \ )     & 400 & ( \quad \ 1: 4.3)     & 360 & ( \quad  \  1: 4.2)  \\

\hline                                             
\end{tabular}
\end{table*}

\begin{table*}[t]
\caption{Photometric transit statistics as function of period for the all-magnitudes simulation using $\text{P}_\text{distr}$ = Ma\,\&\,Ge  (irrespective of astrometric and radial velocity detectability and sky orientation). Only sources with $\geq3$ transits having S/N>3 are considered in the final analyses of this paper (Table~\ref{tab:simResults}). Here, 
$f^\text{obs}_\text{tr}$ is the fraction of selected transiting observations w.r.t. to the total number of observations, and 
$f^\text{src}_\text{tr}$ is the fraction of sources having the selected transiting observations, both being for randomly oriented systems. For sources that have selected transiting observations, $\langle N^\text{obs}_\text{tr}\rangle$ is the mean number of transits per source. E.g. see the top row on the left: for a nominal mission with all simulated orbital periods (between 1 to 1826 days): one in 7800 observations is a transit, and one in 250 sources has transiting observations with on average 2.5 transits per source. Following rows provide numbers in specific period ranges showing that transits become increasingly less likely for longer orbital periods.
 The numbers in the bottom three period bins are particularly poorly defined due to their low occurrence rates in the simulated data.
}. 
\label{tab:fracTransiting}     
\centering                        
\begin{tabular}{c | r@{\kern0.7em} r@{\kern0.7em} r | r@{\kern0.em} r@{\kern0.7em} r | r@{\kern0.7em} r@{\kern0.7em} r | r@{\kern0.7em} r@{\kern0.7em} r}        
\hline\hline  
 &  \multicolumn{6}{c|}{5-yr nominal ($\text{P}_\text{distr}$ = Ma\,\&\,Ge)} & \multicolumn{6}{c}{10-yr extended ($\text{P}_\text{distr}$ = Ma\,\&\,Ge)} \\
\hline
period bin &  \multicolumn{3}{c|}{$\geq1$ transit (S/N > 3)} &  \multicolumn{3}{c|}{$\geq3$ transits (S/N > 3)} &  \multicolumn{3}{c|}{$\geq1$ transit (S/N > 3)} &  \multicolumn{3}{c}{$\geq3$ transits (S/N > 3)} \\
 $[$day$]$ &  $1/f^\text{obs}_\text{tr}$ & $1/f^\text{src}_\text{tr}$ & $\langle N^\text{obs}_\text{tr} \rangle$ &  $1/f^\text{obs}_\text{tr}$ & $1/f^\text{src}_\text{tr}$ & $\langle N^\text{obs}_\text{tr} \rangle$
 &  $1/f^\text{obs}_\text{tr}$ & $1/f^\text{src}_\text{tr}$ & $\langle N^\text{obs}_\text{tr} \rangle$ &  $1/f^\text{obs}_\text{tr}$ & $1/f^\text{src}_\text{tr}$ & $\langle N^\text{obs}_\text{tr} \rangle$\\
\hline                                             
1--1826 	& 7800 & 250 & 2.5 		& 12\,000 & 690 & 4.5		& 7800 & 190 & 4.0		& 9300 & 350 & 5.9		\\
\hline
1--2 		& 300 & 15 & 4.1 		& 330 & 22 & 5.3		& 300 & 15 & 7.9		& 310 & 16 & 8.4 		\\
2--4 		& 640 & 25 & 3.1 		& 840 & 49 & 4.6		& 640 & 22 & 5.5		& 680 & 28 & 6.5 		\\
4--8 		& 1500 & 43 & 2.2		& 2700 & 140 & 3.9		& 1500 & 36 & 3.8		& 1800 & 57 & 5 		\\
8--16 		& 4100 & 97 & 1.9 		& 9700 & 460 & 3.8		& 4000 & 68 & 2.7		& 5600 & 150 & 4.2		\\
16--32 		& 9500 & 200 & 1.6 		& 37\,000 & 1600 & 3.5		& 9400 & 120 & 2.1		& 19\,000 & 460 & 3.8		\\
32--64 		& 21\,000 & 440 & 1.6		& 82\,000 & 3900 & 3.8		& 22\,000 & 250 & 1.8		& 58\,000 & 1300 & 3.6		\\
64--128 	& 56\,000 & 1100 & 1.6		& 200\,000 & 9000 & 3.6		& 56\,000 & 630 & 1.8		& 160\,000 & 3800 & 3.8		\\
128--256 	& 140\,000 & 2700 & 1.6 	& 500\,000 & 21\,000 & 3.3	& 150\,000 & 1500 & 1.5		& 1\,100\,000 & 28\,000 & 3.8	\\
\hdashline
256--512 	& 590\,000 & 12\,000 & 1.6	& 3\,600\,000 & 180\,000 & 4.0	& 340\,000 & 3700 & 1.7		& 1\,200\,000 & 27\,000 & 3.6	\\
512--1024 	& 540\,000 & 11\,000 & 1.5 	& 4\,200\,000 & 210\,000 & 4.0	& 1\,000\,000 & 10\,000 & 1.5	& 4\,000\,000 & 100\,000 & 4.0	\\
1024--1826 	& 1\,400\,000 & 32\,000 & 1.8 	& 5\,000\,000 & 190\,000 & 3.0	& 1\,900\,000 & 18\,000 & 1.5	& 10\,000\,000 & 190\,000 & 3.0	\\
\hline     
\end{tabular}
\end{table*}

%% file: Sec_Discussion.tex
\subsection{Comparison with earlier work \label{sec:earlierwork}}

\cite{Bouchy:2014aa} found that some 20\,000 astrometric BD detections could be expected for bright $G_{\text{RVS}}<12$ host stars having radial velocity time series in \gaia, assuming they have:
\begin{itemize}
\item a $0.6\%$ BD occurrence rate (as assumed in this paper);
\item 3.4 million main sequence and sub-giants stars (total count from \cite{Robin:2012aa} for these luminosity classes);
\item a 100\% detection efficiency.
\end{itemize}

As seen in Fig.~\ref{sect:resultsRv}, applying the limit of $G_{\text{RVS}}<12$ results in a sample of 1.8~million stars, which is only about half of those assumed in \cite{Bouchy:2014aa}. The difference is that in the latter, the selection is not constrained to be only FGK stars\footnote{We tried to extract the sample size of the FGK main sequence and subgiant with $G_{\text{RVS}}<12$ from the \cite{Robin:2012aa} catalog on Vizier, but unfortunately it does not contain the luminosity class.}. 
In \cite{Robin:2012aa} the fraction of FGK stars with $G_{\text{RVS}}<12$ is $71\%$ (20\% larger than expected from the discrepancy), though this is not limited to the aforementioned luminosity classes.

As mentioned in Sect.~\ref{sect:resultsAstro} we find around 5000 BDs with a detection efficiency of about $30-50\%$ (see Fig.~\ref{fig:results1page} around G 11--12.5).
Comparing this with the 100\% detection efficiency assumed in \cite{Bouchy:2014aa}, we can explain the discrepancy between their 20\,000 and our $\sim$5000~BDs: our sample is about half as large, and has a half as large detection fraction. Given that \cite{Bouchy:2014aa} did not constrain their estimate to FGK host stars, their BD number could be overestimated by a factor~2 when only considering the astrometric detection probability (which is closer to 50\% than 100\%). 

\cite{Bouchy:2014aa} also predicted a total of 60 transiting BDs for $G_{\text{RVS}}<12$, which is basically in line with the  35--50 BDs we predict in Table~\ref{tab:simResults}.




\cite{2014MmSAI..85..643S} made an order-of-magnitude estimate of thousands of BD companions with $G<16$~mag that can be astrometrically detected using their 3-sigma criterion, which is in line with the findings of this study.

\citet[][Sect.~2.10.5]{2018exha.book.....P} considered many tens of thousands of brown dwarfs to be detected in total, which matches our predictions. 

The most deviating prediction of several tens of BDs detectable by astrometry mentioned in \cite{2019ApJ...886...68A} seems to be explained by their used detection criterion: it focussed on tightly constraining the mass function, which requires more signal than a detection or even a well constrained orbital fit. A small investigation trying to reproduce their results indicated that a well constrained mass function correspond to a $\Delta \chi^2$ of several thousand or more, i.e. it is much more stringent and thus will lead to (much) more conservative rate estimates. 


 \subsection{Estimates from a uniformly sampled data set\label{sec:raimbaultBootstrap}}
In an attempt to estimate the number of expected BDs from a \textit{uniformly sampled dataset}, we used the Coralie survey dataset of \cite{raimbault19} in which about a hundred targets with companions having $2<M_{\text{J}} \sin i<433$ for a \textit{volume limited} sample are presented. Seventeen of these targets have $10<M_{\text{J}} \sin i<80$, of which two have $P\sim6$\,yr and one has $P\sim10$\,yr. As the latter will be hard to detect we remove it from the sample, but keep the two 6-year entries, i.e., a total sample of 16~BDs. We run additional simulation with this sample, but instead of sampling the companion mass, eccentricity, period, and period-eccentricity from our adopted CDF distributions, we bootstrap all our candidates from this sample (drawing samples with replacement). The comparison between the CDFs is illustrated in Figs.~\ref{fig:cdfHistPeriodFit}, \ref{fig:eccDist} and \ref{fig:pEcc}, showing that the \cite{raimbault19} distribution differs significantly from those derived from the \cite{2014MNRAS.439.2781M} and \cite{Diaz:2012aa} data. 

Note that we simply use $M_{\text{J}} \sin i$ as a proxy for $M_{\text{J}}$. The results are shown in Table~\ref{tab:RbsimResults}, indicating that the estimates would be 1.5--2 times higher than that using our adopted distributions (Table~\ref{tab:simResults}). Given that we adopted the \textit{minimum} mass estimate in our simulations, and the \gaia sensitivity is higher for more massive companions, this might suggest that if the Raimbault sample is representative for the real BD distribution, we could expect up to twice as many BDs as suggested in Table~\ref{tab:simResults}. 

\begin{table}[t]
\tiny
\caption{Simulation results from bootstrapping with a BD sample from \cite{raimbault19}. The number of detected BDs is tabulated, with in parentheses the number of host stars having $\geq3$ photometric transits (S/N>3) in \gaia data together with their mean number of transits. The period distributions `Raimbault' is shown in Fig.~\ref{fig:cdfHistPeriodFit}. All counts could be up to $50\%$ higher or lower due to the uncertainty on the assumed $0.6\%$ BD occurrence rate (see Sect.~\ref{sec:bdFractHostStars}).} 
\label{tab:RbsimResults}  
\centering    
\begin{tabular}{c@{\kern0.7em}  c@{\kern0.7em}  c | r @{\ }l | r@{\ } l  }          %
\hline\hline      
Magnitude & Astro & RV & \multicolumn{2}{c|}{5-yr nominal} & \multicolumn{2}{c}{10-yr extended} \\
selection & $\Delta \chi^2$ & $\Delta \chi^2$ & \multicolumn{2}{c|}{$\text{P}_\text{distr}$ = Raimbault} & \multicolumn{2}{c}{$\text{P}_\text{distr}$ = Raimbault} \\    
\hline                                 
--                 & -- 	        & --	& 780\,000 	&(\ 91: 3.8)    & 780\,000 &(\ 250: 3.9) \\
\hline
--                 & $>30$         & --	    & 60\,000 & ( \ \  2: 3.2)   & 100\,000 & ( \ \   11: 4.0)  \\
--                 & $>50$         & --	    & 42\,000 & ( \ \  1: 3.2)  & 74\,000 & ( \quad   3: 3.7)  \\
--                 & \ \ $>100$    & --	    & 26\,000 & ( \ \  1: 3.1)  & 48\,000 & ( \quad   1: 3.7)  \\
\hline
\hline
$G_{\text{RVS}}<12$ & -- 		    & --	& 12\,000 & ( \ \    5: 3.7)  & 12\,000 & ( \ \  14: 4.1)	 \\ 
\hline
$G_{\text{RVS}}<12$ & $>30$ 		& --	& 8200 & ( \ \  2: 3.7)  & 9700 & ( \quad  7: 4.0)	 \\ 
$G_{\text{RVS}}<12$ & $>50$ 		& --	& 7200 & ( \ \  1: 3.8)  & 8700 & ( \quad  5: 4.0)	 \\ 
$G_{\text{RVS}}<12$ & \ \  $>100$ 	& --	& 6000 & ( \ \ 0 \quad \ \ )  & 7500 & ( \quad 2: 3.9)	 \\ 
\hline
$G_{\text{RVS}}<12$ & --	        & $>30$     & 1800 & ( \ \ 4: 3.7)  & 3100 & ( \ \  12: 4.1) \\
$G_{\text{RVS}}<12$ & --	        & $>50$     & 1000 & ( \ \ 2: 3.8)  & 2000 & ( \ \   10: 4.2)  \\
$G_{\text{RVS}}<12$ & --	        & \ \ $>100$& 450 & ( \ \  1: 3.8)  & 1000 & ( \quad   7: 4.2) \\
\hline
$G_{\text{RVS}}<12$ & $>30$	        & $>30$     & 1100 & ( \ \ 1: 3.7)  & 2400 & ( \quad \   7: 4.1)  \\
$G_{\text{RVS}}<12$ & $>50$	        & $>50$     & 550 & ( \ \  1: 3.9)  & 1300 & ( \quad \  4: 4.0)  \\
$G_{\text{RVS}}<12$ & \ \ $>100$	& \ \ $>100$& 190 & ( \ \  0\quad \ \ )  & 560 & ( \quad \  1: 4.0)   \\

\hline                                             
\end{tabular}
\end{table}

%% file: Sec_Conclusions.tex

Based on detailed simulations, we find that \gaia is likely to detect tens of thousands of brown dwarfs from the astrometric data alone. From the \gaia radial velocity data alone there should be one- or two thousands of additional candidates, with a few hundred overlapping in the two domains. Systems with at least three photometric transits with S/N\,$>3$ are expected for perhaps a thousand systems, which would be  excellent targets for follow up. Among these, a few tens overlap with radial velocity detections and at most a hand full overlapping with astrometric detections. 
An extended mission of 10-years increases all of the detected numbers by 50--200\%, allowing the brown dwarf horizon to be extended significantly.

The numbers found in this study are ultimately limited in accuracy to about $\pm 50\%$ due to the present uncertain occurrence rates and period distributions of brown dwarfs around the considered FGK host stars. What is certain however it that the \gaia detections will enlarge the current brown dwarf sample by at least two orders of magnitude, allowing to investigate the BD fraction and orbital architectures as a function of host stellar parameters in greater detail than every before, leading to a better understanding of the 
formation scenarios of these substellar companions to FGK dwarfs and subgiants
that populate the $10-80M_{\text{J}}$ mass range, and the exploration of the (perhaps not completely dry) desert within it.



%% file: Sec_Acknowledgements.tex
\begin{acknowledgements}

\EDIT{We acknowledge financial support from the Swiss National Science Foundation (SNSF). This work has, in part, been carried out within the framework of the National Centre for Competence in Research PlanetS supported by SNSF.
}
LL is supported by the Swedish National Space Board.
We would like to thank the referee Alessandro Sozzetti for his detailed remarks and suggestions that much improved the quality and relevance of this paper. Thanks also to Lorenzo Rimoldini, Jeff Andrews, Danley Hsu, Laurent Eyer, Johannes Sahlmann and C\'eline Reyl\'e for helpful feedback on specific aspects of this paper. Thanks to David Katz for discussion on the per-FoV radial velocity error model and Shay Zucker for discussion on a useful photometric transit detection criterion.
This work has made use of data from the European Space Agency (ESA) mission
{\it Gaia} (\url{https://www.cosmos.esa.int/gaia}), processed by the {\it Gaia}
Data Processing and Analysis Consortium (DPAC,
\url{https://www.cosmos.esa.int/web/gaia/dpac/consortium}). Funding for the DPAC
has been provided by national institutions, in particular the institutions
participating in the {\it Gaia} Multilateral Agreement.
\end{acknowledgements}

%% file: Sec_GaiaDetectionLimits.tex

\section{Astrometric and radial velocity detection limits\label{sec:gaiaDetectionLimits}}

\label{sec:gaiaResponseComp}
This section is intended to provide insight in the astrometric and radial velocity detection limits by means of simulating a large-grid of models for unseen companions. Note that this data was not used in any of the computations of this paper, and just serves as illustration for the reader\footnote{For completeness we wish to convey that the astrometric sensitivity of \gaia to detect unseen companions has been assessed in various works in literature \cite[see e.g.][]{Casertano:2008aa,2010EAS....45..273S, 2010EAS....42...55S,2014EAS....67...93S, 2014MmSAI..85..643S}
 using the per-transit `S/N statistic' which compares the astrometric signature (Eq.~\ref{eq:astrometricSignal}) of the host star
against the (typical) accuracy of an individual epoch, $\sigma_{\text{fov,astro}}$.}. Though this paper focuses on the BD mass-range, the method is applicable to unseen companions of any mass like neutron stars, white dwarfs and black holes.


The \gaia  astrometric and radial velocity detection limits of a specific binary companion depend on the time sampling (Appendix~\ref{sec:gaiaTs}) and error models (Appendices~\ref{sec:gaiaError} and \ref{sec:gaiaErrorRv}).
Provided with these, we can construct the detection limits as a function of the companion orbital parameters. Below we detail the resulting behaviour.

\subsection{Simulating detection limits\label{sec:simAstroDetectLimits}}
For a given host mass ($M_{\text{*}}$ in $M_{\odot}$) and absolute magnitude ($G^{\text{abs}}_\text{*}$), and companion mass ($M_{\text{BD}}$), eccentricity ($e$), period ($P$ in days), and distance ($d$ in pc) we make $2\times 784$ simulations: for 784 HEALPix sky-uniform distributed time series (see Appendix~\ref{sec:gaiaTs}) we sample the time series for both a 5-yr and 10-yr mission, with each observation being subject to a random rejection probability of $\sim$10\% to simulate dead time. 

The reference time for the astrometric model, $t_{\textrm{ref}}$, is set to the middle of the selected time series. From the parallax ($\varpi$ in arcsec = 1/distance in pc) and $G^{\text{abs}}_\text{*}$ we compute the apparent G-band magnitude of the host star assuming zero extinction: $m_\text{*}= G^{\text{abs}}_\text{*} - 5 (\log10(\varpi) + 1)$, 
 which is used to extract the astrometric error on the observations using Eq.~\ref{eq:fovPrecision}. 
 
The companion-to-host semi-major axis ($\overline{a}$ in au) is computed from the period and sum of the masses using Kepler's third law: $\overline{a} = P^{2/3} (M_\text{*} + M_{\text{BD}} )^{1/3}$. The sky-projected semi-major axis ($a_\text{*}$~in~arcsec) of the host star with respect to the system barycentre is computed from $\overline{a}$, the parallax,  and mass ratio ($q=M_{\text{BD}}/M_{\text{*}}$) as:
 $a_\text{*}=\varpi \overline{a} q / (q+1)$. The orientation of the companion is randomly drawn in 3d space, i.e., inclination, argument of periapsis, position angle of the receding node, and mean anomaly at $t_{\textrm{ref}}$. For each astrometric simulation the centrality parameter $\lambda$ (Sect.~\ref{sec:gaiaDetectMethod}) is computed. 

In a next step, for the 5- and 10-yr simulations separately, we compute the fraction of the 768 simulations for which $\lambda>23, 43, 93$, i.e.\ the fraction of simulations corresponding to the $\Delta \chi^2>30,50,100$ astrometric detection thresholds. The 768 simulations uniformly cover the whole sky and each have a random orbit orientation, hence this fraction can be considered a `sky average' value. 

The radial velocity detection limits are computed in the same way 
using the radial velocity centrality parameter $\lambda$  (Sect.~\ref{sec:gaiaDetectMethodRv}) and the $\lambda>25, 45, 95$ thresholds corresponding to the $\Delta \chi^2>30,50,100$ radial velocity detection thresholds.
The RVS magnitude is approximated as: $G_\textrm{RVS} = G + 0.65$ for all three stellar types. Note that due to the random orientation of the systems in the simulation, the radial velocity limits will represent detection limits for a `typical' average inclination of 60$^\circ$.

Originally we intended to tabulate a comprehensive list of detection curve data for the reader that would like to re-use them for their own work. Due to the strong dependency on host type (mass and  absolute magnitude), relevant companion masses (which can be of any-mass unseen companion), and orbital parameters (mainly eccentricity and inclination)  this table would become very large and might still not cover the specific needs of the reader. Hence we suggest anyone interested in using our detection limits to either implement the different accuracy models and computation methods, or contact the main author 
to request specific detection curves of interest.

\begin{figure}[b!]
 \includegraphics[width=0.49\textwidth]{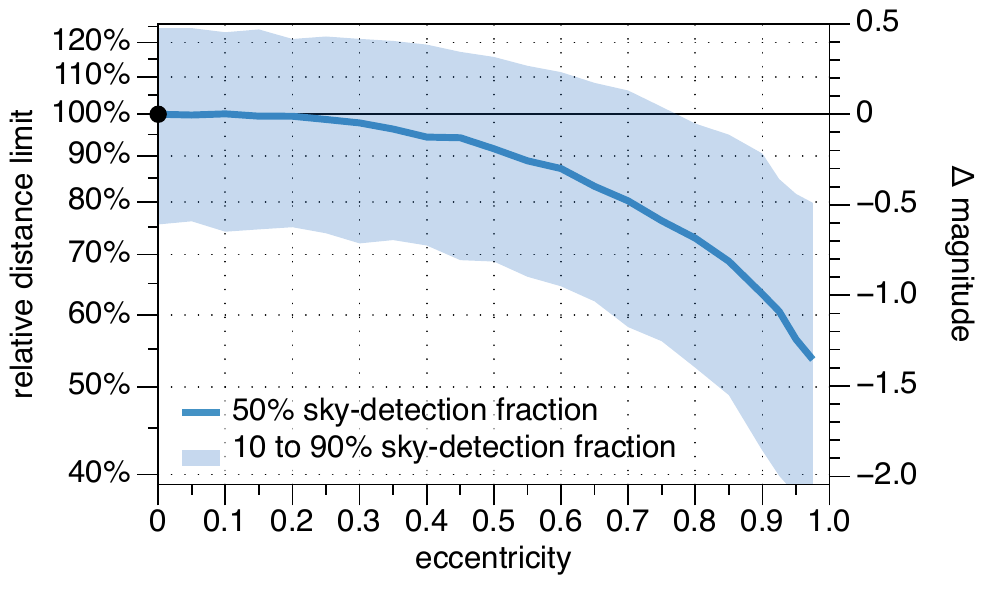}  
\caption{Astrometric eccentricity dependence of distance and magnitude detection limits for $\Delta \chi^2=100$ for a G2V star and a $10~\mathrm{M}_{\text{J}}$ BD with $P=4$\,yr. The black point at $e=0$ corresponds to the black point in the middle panel of Fig.~\ref{fig:gaiaDetectability}. Sky-detection fractions of 50 percentile (thick line) and 10--90 percentile (shaded region) illustrate that the detection threshold for such a (randomly oriented) system varies by $\pm25\%$, depending on sky position. 
}
\label{fig:gaiaDetecEcc}
\end{figure}

\subsection{Interpreting detection limits}

\begin{figure*}[ht!]
 \includegraphics[width=\textwidth]{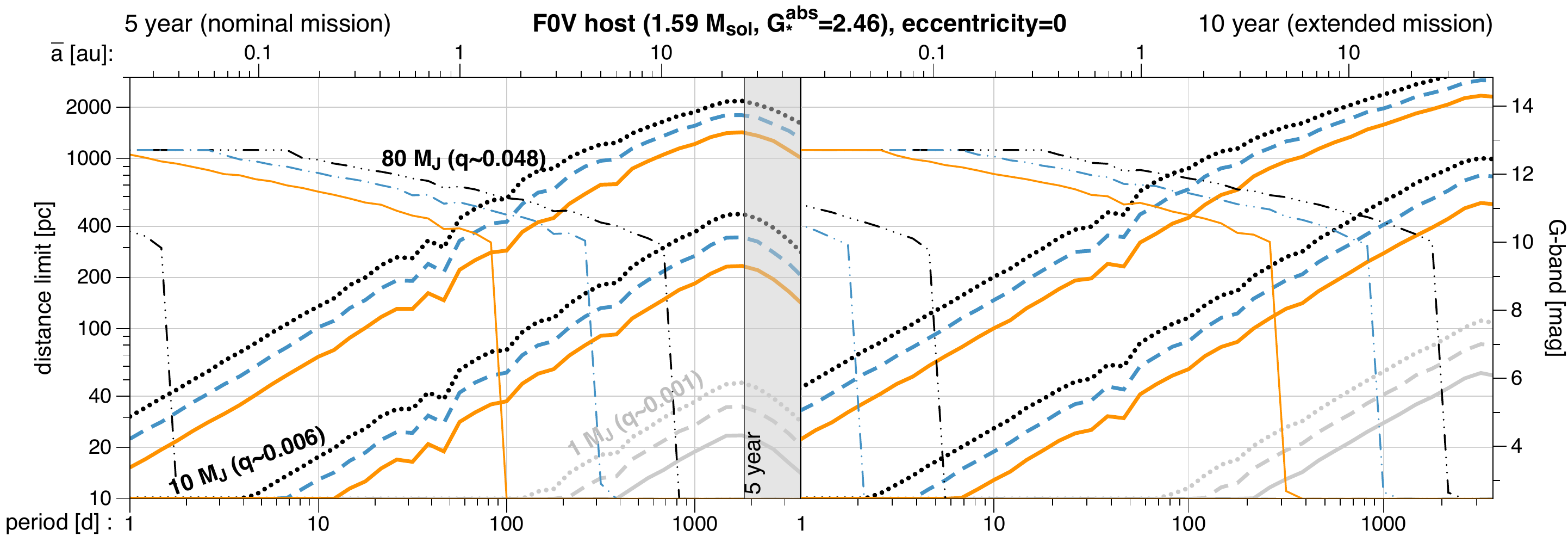}  
 \hspace{0.9cm}\\
 \includegraphics[width=\textwidth]{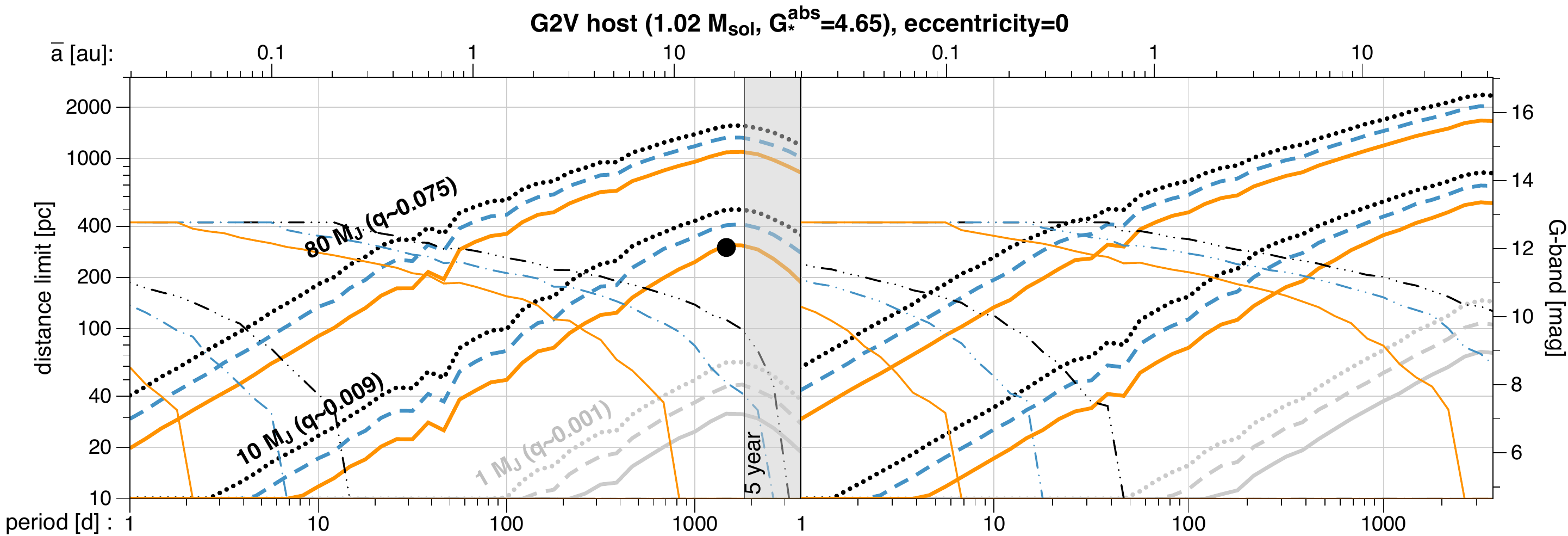} 
  \hspace{0.9cm}\\
 \includegraphics[width=\textwidth]{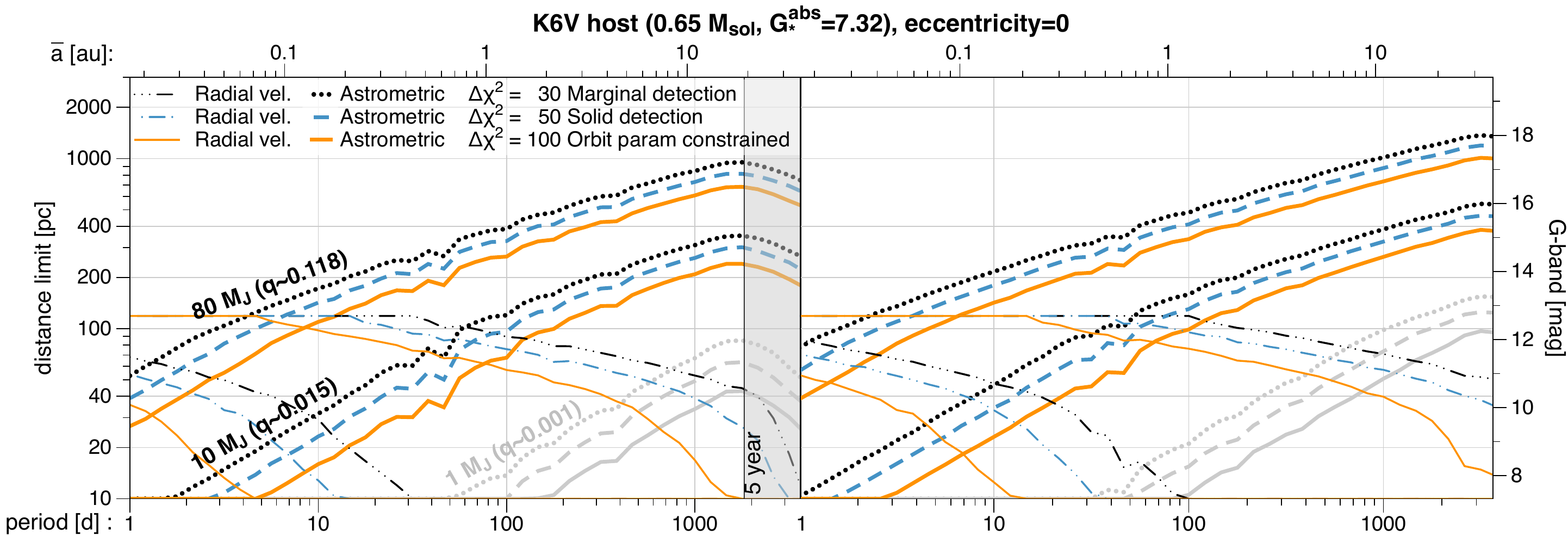} 
\caption{Illustration of the \gaia astrometric and radial velocity detection limits for the adopted BD mass range $10$-- $80$~M$_{\textrm{J}}$ 
around a F0V/G2V/K6V main sequence star (top/middle/bottom respectively) for  different $\Delta \chi^2$ detection thresholds (see Sect.~\ref{sec:gaiaDetectMethod}) resulting from a simulated 5-year nominal mission (left) and 10-year extended mission (right). For each $\Delta \chi^2$ detection threshold, the lines show for a given period (bottom axis) and semi-major axis (top axis), the zero-extinction limits in distance (left axis) and apparent G-band magnitude (right-axis) for which 50\% of the simulated sky positions stops to be detectable. The ascending thick lines are the astrometric limits, while the descending thin lines are the radial velocity limits. Note that the latter are truncated due to the $G_\textrm{RVS}=12$ limit (see text for details).  
For the dot in the middle panel, the astrometric eccentricity dependence and sky-position variation is expanded in Fig.~\ref{fig:gaiaDetecEcc}, see Sect.~\ref{sec:gaiaResponseComp} for more details. 
For comparison, in grey we include the astrometric detection thresholds for a 1~M$_{\textrm{J}}$ planet (it is not detectable using \gaia radial velocity).
}
\label{fig:gaiaDetectability}
\end{figure*}

To plot the response as function of period (or semi-major axis) as shown in Figs.~\ref{fig:gaiaDetecEcc} and \ref{fig:gaiaDetectability}, we compute these statistics for a range of logarithmically increasing distances and identify (interpolate) the distance of the 10\%, 50\%, and 90\% sky-detection fraction of the various $\lambda$ thresholds. 

We start by describing the astrometric detection limits.
The results of the astrometric 50\% sky recovery fraction is shown by the lines in Figs.~\ref{fig:gaiaDetecEcc} and \ref{fig:gaiaDetectability} , while the 10\% and 90\% fractions are used to draw the shaded range in Fig.~\ref{fig:gaiaDetecEcc}. For example: the black dot in both figures is a BD companion around a G2V star (1~$M_{\odot}$) with a zero eccentricity BD of mass $10~\mathrm{M}_\text{J}$, having $P=4$~yr, observed during a 5-yr nominal mission. It will have a good orbit recovery ($\Delta \chi^2=100$) for 50\% of the sky at 311\,pc. The distance range corresponding to the $10\%$ and $90\%$ sky coverage for this recovery is 381--234~pc, i.e.\ $\pm \sim 25\%$ visualised by the shaded range at eccentricity zero of Fig.~\ref{fig:gaiaDetecEcc}. 
As eccentricity increases, the detection efficiency slowly decreases, resulting in a distance limit about 10\% lower around $e=0.5$, and reaching more than 20\% lower for $e>0.7$. Given the significant eccentric distribution of BDs (see Sect.~\ref{sec:eccVsPeriod}) the overall effect is non-negligible. The solid line is similar for other periods of this system, though the  $10\%$--$90\%$ sky coverage range increases for shorter periods: e.g., to $\pm$45\% for $P=1$~yr. Comparing this to a 100~$M_\text{J}$ companion, the 50\% sky fraction solid line drops below 20\% for $e>0.8$, while the $10\%$--$90\%$ sky coverage range expands from $\pm 15\%$ to some $\pm25\%$ for $P=4$ to $P=1$\,yr.

Fig.~\ref{fig:gaiaDetectability} demonstrates that detectability increases for increasing period, which can be understood through Kepler's third law: the (projected) semi-major axis increases as a function of period, hence boosting detectability. The peak occurs around the mission length, and as discussed in Sect.~\ref{sec:gaiaDetectMethod}, periods (much) beyond the mission length are increasingly difficult to recover, and good parameter recovery is no longer connected to the adopted $\Delta \chi^2>100$. To make this clear, this region is shaded in the plots.

The adopted astrometric accuracy floor for $G<12$ (see Fig.~\ref{fig:FovAccLaw} of Appendix~\ref{sec:gaiaError}) leads to a different slope of the detectability curves of Fig.~\ref{fig:gaiaDetectability} for distance $<800$, $<300$, and $<85$~pc, for the shown F0V, G2V, and K6V host star, respectively. Thus, for the $G<12$ regime 
the difference in detectability between host stars is only dependent on $q$, which for a given $\mathrm{M}_{\text{BD}}$ is higher for late-type stars. 
As the detectability of a 1~$\mathrm{M}_{\textrm{J}}$ companion lies within the $G<12$ regime, this translates into the further distance limit for the 1~$\mathrm{M}_{\textrm{J}}$ companions for a K6V host compared to a F0V host, as seen in Fig.~\ref{fig:gaiaDetectability}.

For parts of the curves that are in the  $G$>$(\gg)12$ regime, the astrometric error is strongly dependent on apparent magnitude (Fig.~\ref{fig:FovAccLaw}), which effectively `outweighs' the smaller mass ratio for heavier stellar host mass, and hence produces a larger astrometric signal for earlier type stars, e.g.\ as seen in the further distance limit of the peak of a $100~\mathrm{M}_{\textrm{J}}$ companion for a F0V star compared to a K6V.


In summary of the astrometric detection limits: 
(1)~for a given host and BD mass the detectability increases with period and reaches maximum sensitivity for periods close to the mission length;
(2)~orbital parameters of systems with periods (well) beyond the mission length are not recovered;
(3)~increased eccentricity reduces detectability;
(4)~for a given companion mass, companions around earlier type stars are detectable at larger distances when $G>12$ 
due to the higher intrinsic brightness of the host star and the specific \gaia magnitude dependent accuracy, despite the smaller mass ratio.

A particular feature of the \gaia astrometric data sampling is that the detection sensitivity for short astrometric periods $P \lesssim 100$ days is very non-uniform and gives rise to a rather complex variation of detectability as function of sky position (i.e. time-sampling) and orbital parameters (mainly related to the sky orientation of the orbit). Even though Fig. A.2 provides sky and orbit averaged values along its lines, the effect is still mildly visible in the wiggles of the lines. An alternative illustration of this effect can be found in Fig.~5 of  \cite{2014MNRAS.437..497S}. A detailed explanation of this feature is beyond the scope of this paper and will be addressed elsewhere.

Regarding the radial velocity detection limits, we see the general expected negative slope from Eq.~\ref{eq:rvSignal}, though it is heavily affected by the radial velocity uncertainty curve (Appendix~\ref{fig:FovAccLawPhot}) that steeply increases for magnitudes above $G_\textrm{RVS}>9.5$, causing a rather `flat' response above about $G>9$ (see Fig.~\ref{fig:gaiaDetectability}).
Note that due to the adopted hard limit at $G_\textrm{RVS}<12$ for which we assume \gaia radial velocity time-series will be available, the radial velocity detection limits are truncated around $G\sim 12.7$, which only affects the most massive 80~M$_\textrm{J}$~BDs on orbits of several days. 

From Eq.~\ref{eq:rvSignal} one expects a lower response for the more massive F0V stellar type than the much lighter K6V, however, as seen above, the radial velocity uncertainty is strongly dependent on magnitude, which effective `outweights' the radial velocity signal dependency on the host mass star due to the much higher luminosity of the heavier stars. The only effect that remains is that for $G<9$ (where the radial velocity uncertainty is relatively flat) we see that detection limits do extend towards longer periods for the lighter host stars.

%% file: Sec_Appendices.tex
\section{\gaia time series\label{sec:gaiaTs}} 
\begin{figure}[b]
 \includegraphics[width=0.49\textwidth]{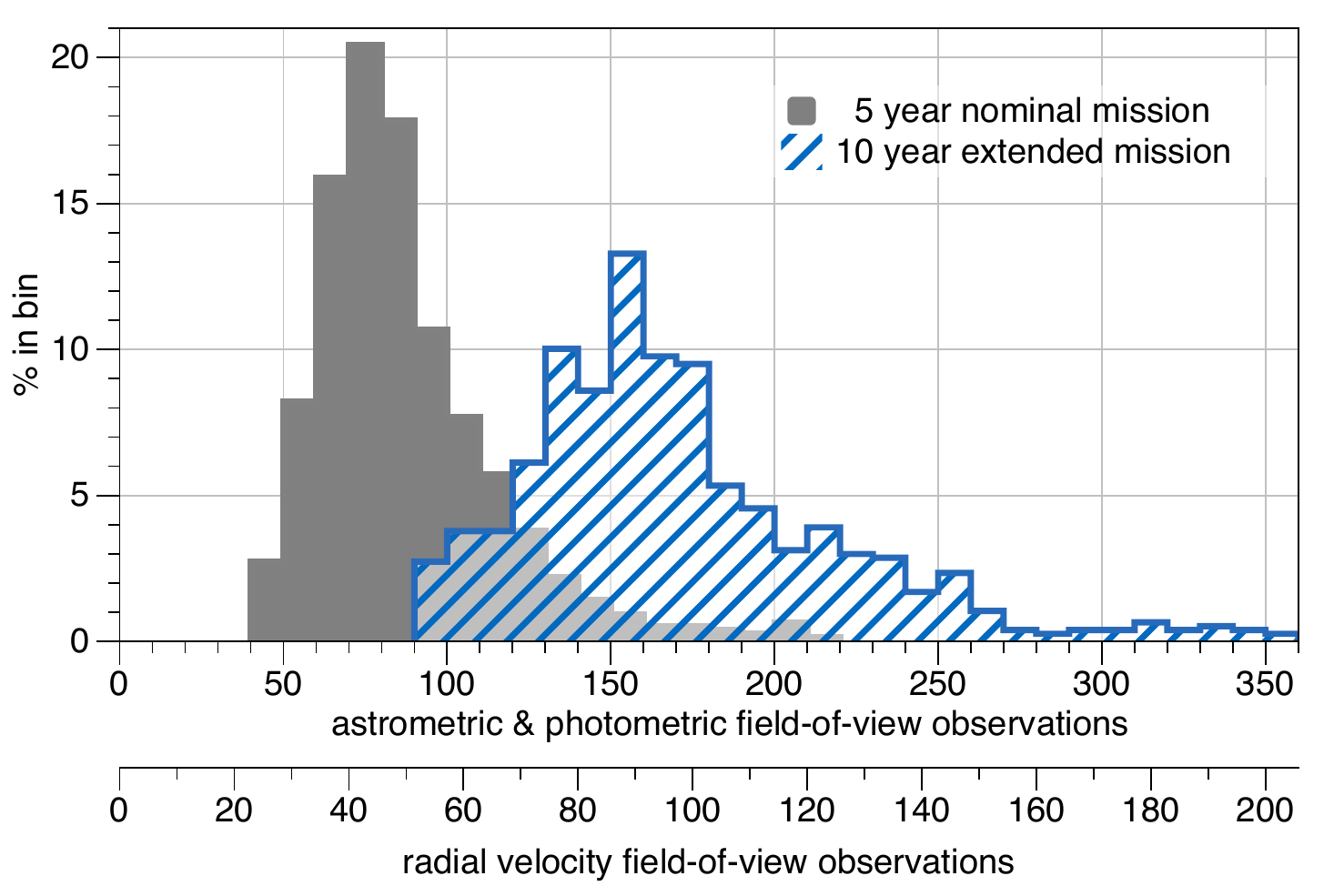} 
\caption{Counts of the simulated number of field-of-view (FoV) observations of the 196 sky positions used in this study for a 5- and 10-yr \gaia mission (bin size: 10), see top panels of Fig.~\ref{fig:resultsTypeRelated} for the corresponding sky-distribution. For radial velocity measurements the number of FoV observations is reduced by a factor of 0.57 (4/7, see Sect.~\ref{sec:gaiaDetectMethodRv}).
}
\label{fig:FovNumObs}
\end{figure}

We use AGISLab \citep{Holl:2012fk} to simulate the actual Gaia scanning law for the 5-yr nominal and 10-yr extended mission\footnote{From 
Julian year 2014.734 (2014-09-26, start of Nominal Scanning Law phase)  
to 2019.734 (2019-09-26) for the nominal 5-year mission, and up to 2024.734 (2024-09-25)  for the 10-yr extension.
}. 
We have used a uniformly distributed grid of 784 sky positions, these 
being the bin centres of a level 3 HEALPix map \citep{Gorski:2005qy}. 
A histogram of the number of field-of-view observation of all the simulated time series is shown in Fig.~\ref{fig:FovNumObs}.
The simulated Besan\c con star positions are then mapped to the time series of the nearest neighbour on this grid. 
For each simulated source, each observation of a time series has a 10\% chance to be randomly removed in order to account for the expected mission dead-time.

\section{\gaia astrometric FoV error model\label{sec:gaiaError}} 

\begin{figure}[hbt]
 \includegraphics[width=0.49\textwidth]{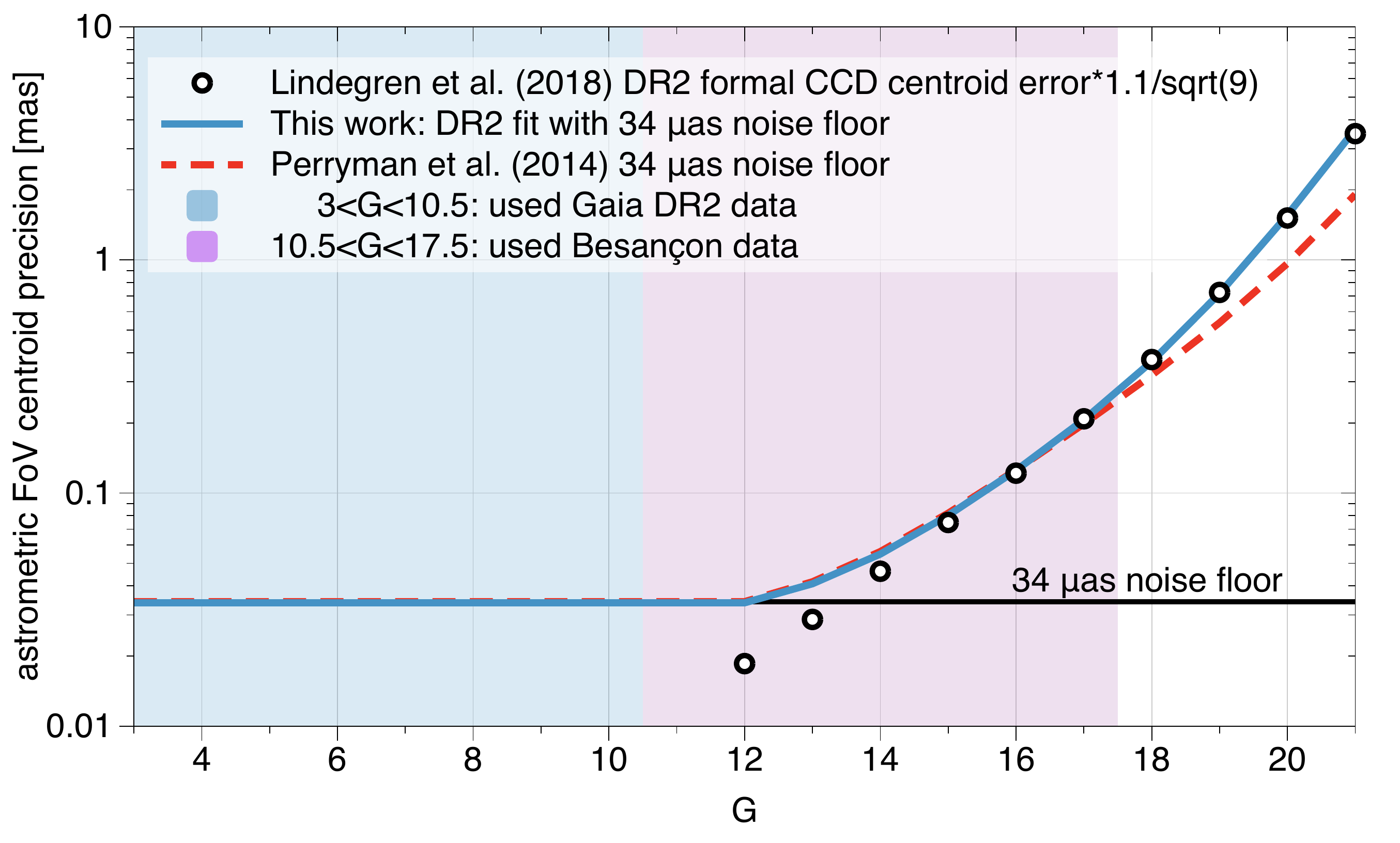} 
\caption{Adopted astrometric \gaia per-FoV precision model based on the expected \gaia precision derived from DR2 data, and compared with the  \cite{2014ApJ...797...14P} error model.
}
\label{fig:FovAccLaw}
\end{figure} 
With the \gaia mission and its calibration improvements still in progress, we adopt the same astrometric observation error model parameterisation described in Sect.~2.1 of \cite{2014ApJ...797...14P}.
This error model provides the per-field-of-view astrometric centroiding error 
$\sigma_{\textrm{fov,astro}}$ ($\mu$as) as a function of the host-star $G$-band magnitude: 
\begin{eqnarray} 
z &=& 10^{0.4(\textrm{max}[G,12]-15)} \nonumber \\
\sigma_\eta&=&( c_1\ z + c_2 \ z^2)^{0.5} \nonumber \\
\sigma_{\textrm{fov,\,astro}} &=& \left( \frac{ \sigma^2_\eta }{9} + \sigma^2_{\textrm{att}} +  \sigma^2_{\textrm{cal}}  \right)^{0.5}  \textrm{[$\mu$as]} \label{eq:fovPrecision}
\end{eqnarray}
\begin{eqnarray} 
 &\text{with}&  c_1= 53\,000 \text{ and } c_2= 310 \text{ in \cite{2014ApJ...797...14P}} \nonumber \\
 &\text{and}&  c_1= 49\,000 \text{ and } c_2= 1700 \text{ in this work (DR2 fit).} \nonumber 
\end{eqnarray}
Note that for bright stars with $G<12$ the observation times are reduced by CCD `gating' to avoid signal saturation, causing a more or less constant measurement precision for these stars. In this error model it is assumed to be strictly constant at a level of $\sigma_{\textrm{fov}}=  34 \ \mu$as per field-of-view when adopting a value of 20~$\mu$as per field-of-view transit for both  $\sigma^2_{\textrm{att}}$ and $\sigma^2_{\textrm{cal}}$ \citep{Risquez:2013aa, Lindegren:2012aa}, corresponding to a combined attitude and calibration noise level of 
85 $\mu$as  per CCD.  The DR2 per-CCD astrometric centroiding precision shown in Fig.~9 of \cite{Lindegren:2018aa} illustrates that the  systematic noise level at that time was about 250~$\mu$as per CCD. We here make the realistic assumption that this noise level will reduce by a factor of three for the final astrometric data release due to improved calibration models, which is already (partially) corroborated by the improvements shown in Fig.~A.1 of \cite{2021A&A...649A...2L}.

The free parameters $c_1$ and $c_2$ of the \cite{2014ApJ...797...14P} error model of Eq.~\ref{eq:fovPrecision} were adjusted to fit the pre-launch prediction of the end-of-mission astrometric performance. We have revised this fit based on the estimates from DR2, as shown in Fig.~\ref{fig:FovAccLaw}. The circles are DR2 median per-CCD formal uncertainties from Fig. 9 in \cite{Lindegren:2018aa}, converted to per-FoV estimates by dividing them by the square root of the 9 per-FoV CCDs, and increased by 10\% to match the actual robust scatter estimate observed, particularly for the fainter sources,  in the Astrometric Global Iterative Solution \citep[AGIS, ][]{Lindegren:2012aa, 2011ExA....31..215O, 2010ASPC..434..309L}. It is not expected that the calibration precision will further improve for $G>13$ where it is photon-limited. 

The \cite{2014ApJ...797...14P} and our DR2 derived error models agree within a few percent up to $G\sim17$, and start to deviate thereafter (8\% at $G=17.5$, 13\% at $G=18$, etc...), mainly due to the increased stray-light component in the real mission data. The \cite{2014ApJ...797...14P} predicted planet detections were basically brighter than $G=17$ and hence their conclusions would be unchanged based on this latest model. Our results extend to $G=17.5$, and hence this new error model will reduce the number of detections towards the faint end.




\section{\gaia radial velocity FoV error model\label{sec:gaiaErrorRv}} 

\begin{figure}[hbt]
 \includegraphics[width=0.49\textwidth]{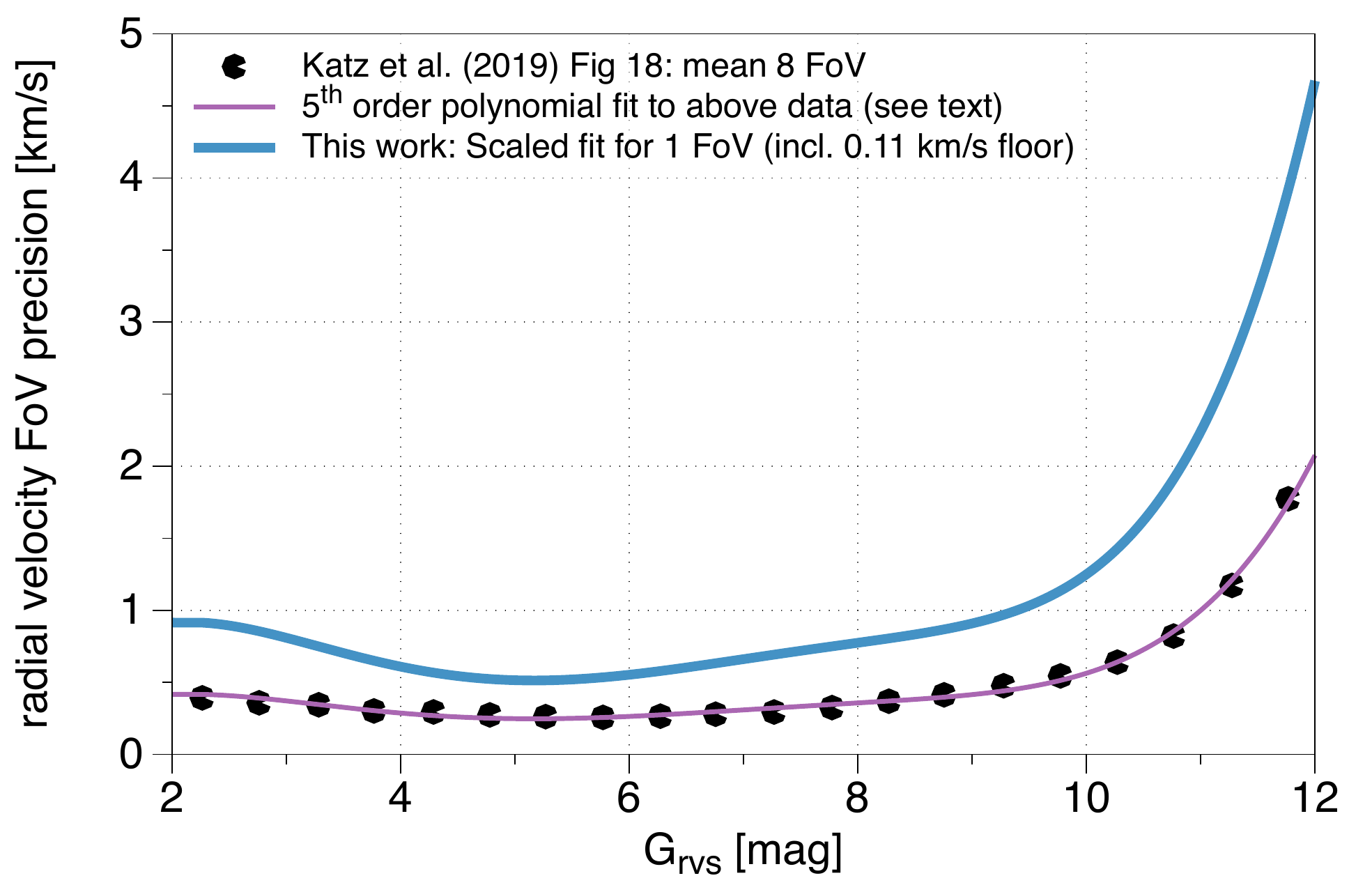} 
\caption{Adopted radial velocity \gaia per-FoV precision model valid for $G_{\text{RVS}}<12$, derived from \gaia DR2 radial velocity data.
}
\label{fig:FovAccLawRv}
\end{figure} 
To compute the detectability of BD companions using the \gaia RVS derived radial velocity measurements (Sect.~\ref{sec:gaiaDetectMethodRv}) we need an end-of mission estimate of the per-FoV precision of the radial velocity measurements. Due to changes from the pre-launch conditions \citep[e.g.\ the stray-light issue discussed by the][]{2016A&A...595A...2G} we estimated the expected end-of-mission performance from the currently available data, i.e., Fig.~18 of \cite{2019A&A...622A.205K}. We take the data points for the interval of [7,11] transits which have a mean at $N=8$, and using their Eq.~1 we fit a fifth degree polynomial model $P(G_{\text{RVS}})$ to the `formal' per-FoV transit uncertainty $\sigma_{\text{V}_\text{R}}$
\begin{equation}
\epsilon_{\text{V}_\text{R}} = \left[ \left( \sqrt{ \frac{\pi}{2\ N}} P(G_{\text{RVS}}) + 0.11^2 \right) \right] \ \ \text{with} \ \ P(G_{\text{RVS}}) \simeq \sigma_{\text{V}_\text{R}}	\ ,
\end{equation}
with the polynomial model
\begin{eqnarray}
P(G_{\text{RVS}}) &=& -1.430 + 2.944 x  -1.3084 x^2 + 0.25102 x^3\nonumber  \\
&&  -0.021959 x^4 + 0.00072645x^5 \\
\text{with} \ \ x&=& \text{max}\left(2.26, G_{\text{RVS}}\right) \nonumber
\end{eqnarray}
For the adopted per-FoV precision we include a 0.11\,km\,s$^{-1}$ noise floor, which is the current best estimate for the end-of mission precision (David Katz, private communication), resulting in
\begin{equation}
\sigma_{\text{fov,\,RV}} =  \sqrt{ P^2(G_{\text{RVS}}) + 0.11^2 } \ \ \textrm{[km\,s$^{-1}$]}	\ .
\end{equation}

\section{\gaia photometric FoV error model\label{sec:gaiaErrorPhot}} 

\begin{figure}[hbt]
 \includegraphics[width=0.49\textwidth]{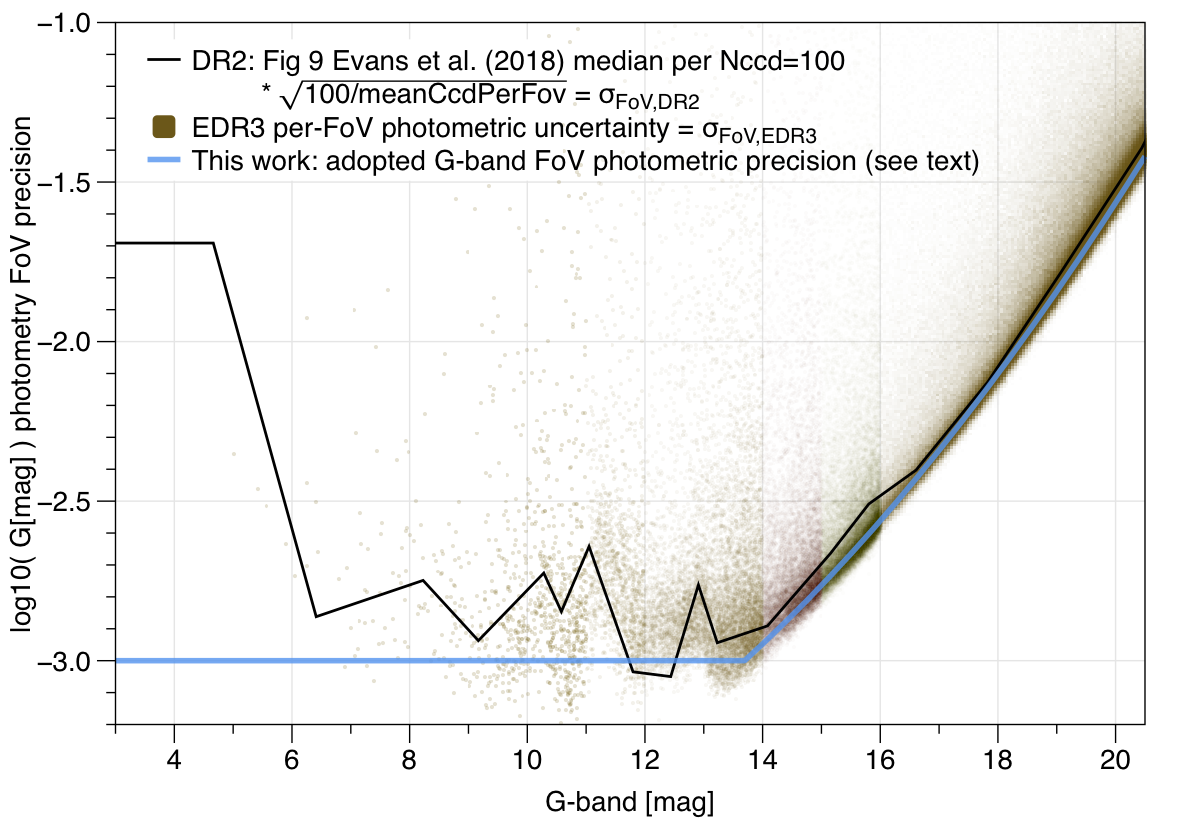} 
\caption{Adopted photometric \gaia G-band per-FoV precision model based on the approximate \gaia precision derived from EDR3 data.
}
\label{fig:FovAccLawPhot}
\end{figure} 
To compute the detectability of photometric brown dwarfs transits (Sect.~\ref{sec:gaiaDetectMethodTransit}) we require an end-of mission estimate of the per-FoV precision of the $G$-band photometry. As for the radial velocity precision, we have replaced the pre-launch estimates with those from currently available data. In Fig~\ref{fig:FovAccLawPhot} we estimate the per-FoV $G$-band photometric precision (blue solid lines) from the published uncertainties on the mean, for both DR2\footnote{This traced in a rough way the orange mode-line of the distribution in Fig.~9 of \cite{2018A&A...616A...4E}, and is rescaled to represent the corresponding precision per FoV (in absence of any noise floor). Note that \textit{meanCcdPerFov} represents the number: 8.857 = (6/7*9+1/7*8), i.e., the focal plane has 7 rows of CCDs of which one has 8 CCDs and the other have 9 CCDs that are used to derive G-band photometry.} (black line) and EDR3\footnote{We query a random sample of photometry of 1M EDR3 sources \citep{2020arXiv201201916R} with a minimum of 150 CCD observations (\texttt{phot\_g\_n\_obs}) and directly scale them to the (noise-floor free) per-FoV precision in magnitude: \texttt{fov\_uncertainty\_mag= 1.0857/sqrt((6/7*9+1/7*8))*phot\_g\_mean\_flux\_error *sqrt(phot\_g\_n\_obs)/ phot\_g\_mean\_flux}.} (scatter data). From this we see that above $G>16.5$ the improvement in EDR3 is relatively small. From this observation, and the fact that it extends down to about $G=13.5$ along a rather straight exponential line, we will assume that the EDR3 mode-line reached the photon limited precision and will no further improve in future releases. For the bright range the gating complicates the calibrations, and it is clear that this has substantially improved in EDR3, and should continue to improve in the coming releases. Therefore we adopted a simple noise floor of $\sigma_\text{FoV,\,G}=10^{-3}$\,mag\footnote{This coincides with the pre-launch estimate of 1\,mmag adopted by \cite{2012ApJ...753L...1D}. At that time it was assumed this floor would extend to $G=14$ or beyond, such that their predicted transit counts fainter than 14~mag should be considered as somewhat optimistic.}, which effectively extends up to $G\simeq13.5$. For $G>13.5$ we fit a second-order polynomial to the mode of the EDR3 data, resulting in our adopted model, illustrated by the blue line
\begin{eqnarray}
    \sigma_\textrm{fov,G} &=& 10^{\text{max}\left[-3, -3.56 \ -0.0857\, x \ + \ 0.00938\, x^2 \right]} \ \ \textrm{[mag]} \\
 \text{with }   x &=& G-0.15 \nonumber 	\ .
\end{eqnarray}

\section{Cumulative density functions\label{sec:cdfMass}} 

To generate $(M_{\mathrm{BD}})$ from a uniform distributed random number $U\sim U([0,1])$
we use the inverse cumulative density function. This is the inverse CDF of Eq.~\ref{eq:massDCF}:
\begin{equation}
    {\rm CDF}_\text{M}^{-1}(u)=\begin{cases}
    \frac{- b + \sqrt{b^2 -4a(c-u)}}{2a}, & \text{$\ \ \ \ \ 0\leq u <0.49$} \\
   (u-c)/b, & \text{$0.49\leq u<0.51$}\\
    \frac{- b + \sqrt{b^2 -4a(c-u)}}{2a}, & \text{$0.51 \leq u \leq 1$}
    \label{eq:invMassDCF}
  \end{cases}
\end{equation}
where the coefficients correspond to those presented in Eq.~\ref{eq:massDCF}.
Using the same technique the random sampling of the period and eccentricity distribution were performed, using the inverse functions of the CDF formulae given in Figs.~\ref{fig:cdfHistPeriodFit} and \ref{fig:eccDist}.

\section{Besan\c con model data for $G>10.5$ \label{appendix:besancon}}

To simulate the vast majority of the host star distribution we used the  latest 2016\footnote{\url{http://modele2016.obs-besancon.fr/}}  Besan\c con model, which has various improvements over the model of 2013, 
including a variable star formation rate over the thin disk life time \citep{2014A&A...564A.102C}.
The extinction distribution was set to the suggested  \cite{2006A&A...453..635M} model for $100\degr < l < 100 \degr $ and $-10\degr < b < 10\degr$, and diffuse extinction for higher latitudes and longitudes.


Simulations were generated in bins adjusted for three zones: (1)~the `outer' Galactic northern and southern hemispheres spanning $0\degr \leq l < 360 \degr$ in steps of $36\degr$, and $28\degr < |b|  \leq90\degr$ in steps of $6.2\degr$;
(2) the Galactic `centre' region spanning $-60\degr \leq l \leq 60 \degr$ in steps of $10\degr$, and $-28\degr \leq b  \leq28\degr$ in steps of $3.5\degr$; and (3) the Galactic `disk' region spanning $60\degr < |l| \leq 180 \degr$ in steps of $20\degr$, and $-28\degr \leq b  \leq28\degr$ in steps of $3.5\degr$.

The model includes the apparent \gaia $G$-band magnitude. Although the simulated $G$~passband uses the pre-launch transmission curves described in \cite{Jordi:2006aa, Jordi:2010aa}, any differences with the in-flight passband will be negligible for the outcome of this study.
The simulations were limited to spectral types from F0 to K9, and Luminosity class IV~(subgiants) and V (main sequence).  
A limit of $G<17.5$ was adopted based on the low detection-rates beyond this magnitude (see Appendix~\ref{sec:gaiaDetectionLimits}), and otherwise rather excessive simulation data volume.

The Besan\c con model output contains generated stars that are sampled uniformly in density for each of the sky-position bins. 
The total number of generated sources is: 130\,327\,900 of which 92\,863\,736~(71\%) are main sequence stars, and 37\,464\,164~(29\%) are sub-giants. Fig.~\ref{fig:bscFig1} shows the distribution of stars: on the sky, as function of magnitude, stellar type, and as function of effective temperature versus mass and bolometric magnitude. The bottom panel has some visual guides at $T_{\text{eff}}=7220$\,K and 3940\,K, and $M_{\text{bol}}=2.50$ and 7.59, corresponding to the main sequence F0 and K9 stars\footnote{\label{footnoteMamajek}V2019.3.22 (Eric Mamajek) of \url{http://www.pas.rochester.edu/~emamajek/EEM_dwarf_UBVIJHK_colors_Teff.txt}} \citep{Pecaut:2013aa}.

The Besan\c con model we extracted did not contain any stars with $G<10.2$\,mag. For this paper we therefore use \gaia DR2 derived data for $G<10.5$ (Appendix~\ref{appendix:gaiaGt10p5}) and use the Besan\c con data for $G>10.5$.

We also require $G_{\text{RVS}}$ to select sources which are assumed to have radial velocity time series (i.e. $G_{\text{RVS}}<12$). 
As our Besan\c con data uses the pre-launch \gaia magnitude system \citep{Jordi:2010aa}, we use those relations to compute $G_{\text{RVS}}$ from the $G$-band magnitudes and colours. 
A $G_\text{RVS}=12$ corresponds roughly to $G=12.65$ for our FGK main sequence and sub-giant sample. Applying the exact limit of $G_{\text{RVS}}<12$ on our Besan\c con data with $G>10.5$  results in a sample of 1.7~million stars.

\section{\gaia DR2 derived data for $G<10.5$ \label{appendix:gaiaGt10p5}}
Because the Besan\c con model data did not contain stars brighter than $G<10.2$, we derived the stellar distribution for $G<10.5$ from \gaia DR2 data. For the simulations in this study we need positions, apparent G-band magnitudes, distances, stellar host masses, radii, and $G_{\text{RVS}}$ magnitude estimates. 
The position \citep{Lindegren:2018aa}, apparent magnitude  \citep{2018A&A...616A...3R,2018A&A...616A...4E}, and estimate of the radius \citep{2018A&A...616A...8A} are available in DR2. 

For about 95\% of DR2 sources the \texttt{parallax\_over\_error>5} (i.e., relative error better than 20\%)
which means that a simple inverse of the parallax provides a meaningful distance, although it can still be biassed and have a large uncertainty. Since we are mainly interested in the \textit{local} stellar density, we used the bayesian estimated distances with Galactic prior of \cite{2018AJ....156...58B} to derive a meaningful distance estimate for the whole sample\footnote{\texttt{r\_est} from: \url{http://gaia.ari.uni-heidelberg.de/tap.html}}. Distances for our final sample typically deviate from the inverse parallax by less than 1\% (and by a maximum of 8\%).

To make an (initial) selection of main sequence and sub-giants stars we use the main sequence selection of FGK stars in \gaia DR2 described in Section~3.1 of \cite{Hsu_2019}, with their suggested $G_\text{BP}-G_\text{RP}$ range of [0.5, 1.7]. We made the following DR2 archive selection:
\texttt{lum\_val < 3*1.75*power(10, 2.62 - 3.74*bp\_rp + 0.962*bp\_rp*bp\_rp)}, where we added the factor~3 to confidently include the increase in luminosity from main sequence to sub-giants phase (which is around 3 for solar-type stars, and more like 1.8 for F-type stars), resulting in a preliminary selection of 204\,433 stars. In the \cite{Hsu_2019} paper the power of 10 was erroneously not mentioned (Danley Hsu, priv.\ comm.). 

Mass estimates were not yet included in DR2. In comparison with the brightest part of the Besan\c con data, masses of the current observable sample can be reasonably well estimated from the absolute magnitude alone, with a dispersion of about $0.2~M_{\odot}$. We fit this relation with a quadratic function for the zero-age main sequence masses of the FGK stars listed in \cite{Pecaut:2013aa}\textsuperscript{\ref{footnoteMamajek}}: $M /M_{\odot} = 2.4571 - 0.41272 G_{{\text{abs}}} + 0.022355 G_{{\text{abs}}}^2$, which correctly follows the core of the distribution of both the main sequence and sub-giants in Besan\c con mass versus $G  -5 \log10( \text{distance [pc]})+5$ (extinction correction does not have a large effect).
Because the most massive stars in our sample (F0) should not exceed $1.6~M_{\odot}$, we select only stars with $M<1.6~M_{\odot}$. This modifies our initial selection based on the luminosity as a function of $G_\text{BP}-G_\text{RP}$ such that the most luminous tip at the blue end gets clipped around a luminosity of $8.5~L_{\odot}$, which is similar to a F0V zero-age main sequence luminosity of $7.8~L_{\odot}$, and about 1.8 times less than $4.6~L_{\odot}$ for a F3V  zero-age main sequence star \citep{Pecaut:2013aa}\textsuperscript{\ref{footnoteMamajek}}, illustrating that such a cut should roughly retain the relevant sub-giants and not include too many giants. We estimate the absolute magnitude for the \gaia DR2 sources using \texttt{phot\_g\_mean\_mag - 5*log10(r\_est [pc]) + 5}. The extinction \texttt{a\_g\_val} is missing for about 25\% of the sources in our selection and for those who have it, it causes peculiar (non-physical) structures in our HR-diagram, hence we omit any extinction correction.
We then estimate the masses based on this derived absolute magnitude and apply the mass-cut of $M < 1.6~M_{\odot}$ which reduces our \gaia DR2 sample to a final 174\,122 stars with $G<10.5$.

Additionally, we would like to roughly separate the main sequence stars from the sub-giants, as in the Besan\c con data. The Besan\c con data shows that a cut of $R_{\odot}>1.7$ should contain the vast majority of the sub-giants, and below that the vast majority of the main-sequence stars. This makes sense given that this it about the zero-age main-sequence radius of our largest considered stars \citep[F0V-F1V,][]{Pecaut:2013aa}\textsuperscript{\ref{footnoteMamajek}}, Even though this separation is not perfect, the distributions of the Besan\c con and Gaia DR2 selected data sets overlap rather well in the colour-magnitude diagram, and are consistent with their expected location in this diagram. In this way we classify the selected 174\,122 DR2 stars in to 113\,832 (65\%) main sequence stars, and 60\,290 (35\%) sub-giants.


Finally, we need $G_{\text{RVS}}$ to select sources which are assumed to have radial velocity time series (i.e. $G_{\text{RVS}}<12$). 
\EDIT{We adopted the approximate 0.65\,mag mode offset derived from the Besan\c con data to convert DR2 photometry into $G_{\text{RVS}}$ for all DR2 stars irrespective of colour, i.e., $G_{\text{RVS,\,DR2}}= G_{\text{DR2}}-0.65$.} 
The standard deviation around this offset in the Besan\c con model was about 0.11~mag, hence we expect this fixed-value offset approximation to give meaningful approximations to the real $G_{\text{RVS}}$. 
So in conclusion, stars assumed to have radial velocity timeseries (i.e., $G_{\text{RVS}}<12$) were selected in the DR2 data using the cut $G<12.65$.
\EDIT{We note that a more accurate $G_{\text{RVS}}$ conversion for DR2 data is provided in Eq.~2 and 3 of \cite{2018A&A...616A...1G}.}